\documentclass[12pt]{article}
\usepackage{epsf,latexsym,amsmath,amssymb}
\makeindex

\newtheorem{theorem}{Theorem}[section]
\newtheorem{lemma}[theorem]{Lemma}
\newtheorem{proposition}[theorem]{Proposition}
\newtheorem{corollary}[theorem]{Corollary}

\newtheorem{definition}[theorem]{Definition}
\newtheorem{sublemma}[theorem]{Sublemma}

\newcommand{\mytwoindex}[2]{\mythreeindex{#1}{#1}{#2}}
\newcommand{\mythreeindex}[3]{\index{#1@[#2:] {#3}}}
\newcommand{\myindexaj}{\mythreeindex{aj}{$a_j$}{the number of $\pi_j$ and 
$\pi_j^{-1}$ appearing in a word or form}}
\newcommand{\myindexW}{\mythreeindex{WGamma}{$W_\Gamma,W_T,W_{T'}$}{the
number of potential walk classes or legal walks on a form, type, or
new type with various additional restrictions (such as irreducibility)}}
\newcommand{\myindexlambdaone}{\mythreeindex{lambda1}{$\lambda_1(G)$}{the
sup over all values $( c_G(v,v;k))^k$}}
\newcommand{\myindexEGamman}{\mythreeindex{EGamman}{E\char'133$\Gamma$\char'135$_n$}{The expected value of the number
of closed walks corresponding to a potential walk class (which depends
only the the form of the potential walk)}}

\newcommand{\mob}{\mu}

\newcommand{\newa}{P}

\newcommand{\dtreelike}{{$d$-Ramanujan}}
\newcommand{\opp}{{\rm opp}}

\newcommand{\sbd}{{\rm sbd}}
\newcommand{\walksum}[2]{{{\rm WalkSum}(#1,#2)}}
\newcommand{\supress}[2]{{#1\left[#2\right]_{\rm sup}}}
\newcommand{\realize}[2]{{{#1}|_{#2}}}
\newcommand{\ord}{{\rm ord}}
\newcommand{\from}{\colon}
\newcommand{\ignore}[1]{}

\newcommand{\El}{{E_{\rm long}}}
\newcommand{\Ef}{{E_{\rm fixed}}}
\newcommand{\kf}{k^{\rm fixed}}
\newcommand{\veckf}{{\vec k}^{\rm fixed}}

\newcommand{\ck}{{\cal K}}

\newcommand{\ce}{{\cal E}}
\newcommand{\ct}{{\cal T}}
\newcommand{\cg}{{\cal G}}
\newcommand{\ch}{{\cal H}}

\newcommand{\ci}{{\cal I}}
\newcommand{\cj}{{\cal J}}
\newcommand{\cjnd}{\cj_{n,d}}
\newcommand{\cind}{\ci_{n,d}}
\newcommand{\cinpd}{\ci_{n+1,d}}

\newcommand{\cgnd}{\cg_{n,d}}
\newcommand{\chnd}{\ch_{n,d}}
\newcommand{\cw}{{\cal W}}
\newcommand{\cl}{{\cal L}}

\newcommand{\fund}{{\rm fund}}
\newcommand{\taufund}{{\tau_\fund}}

\newcommand{\tang}{{\psi}}	
\newcommand{\tset}{{\Psi}}
\newcommand{\tseig}{{\tset_{\rm eig}}}
\newcommand{\tsmin}{{\tset_{\rm min}}}
\newcommand{\tsord}{{\tset_{\rm ord}}}

\renewcommand{\complement}[1]{#1^{\rm c}}

\newcommand{\nv}{{v}}	
\newcommand{\ned}{{e}}

\newcommand{\oddf}{!_{\rm odd}}

\newcommand{\Esymm}{{\rm E}_{\rm symm}}

\newcommand{\tree}{{\rm Tree}}
\newcommand{\tangle}{\ct}
\newcommand{\tanglefirst}{\tangle}

\newcommand{\irdsel}{{\rm IrSelTr}}
\newcommand{\sit}{{\rm SIT}}
\newcommand{\ssit}{{\rm SSIT}}

\newcommand{\irdtr}[2]{{\rm IrredTr}\left(#1,#2\right)}
\newcommand{\intn}{{\{1,\ldots,n\}}}

\newcommand{\cwstsl}{{\cw_{\rm STSL}}}

\newcommand{\cwsl}{{\cw_{\rm SL}}}

\newcommand{\E}[1]{\mbox{E}\left[#1\right] }
\newcommand{\prob}[1]{{\rm Prob}\left\{ #1 \right\} }

\newcommand{\proof}{{\par\noindent {\bf Proof}\space\space}}
\newcommand{\proofbox}{\begin{flushright}$\Box$\end{flushright}}

\newcommand{\ird}[1]{{\rm Irred}_{#1}}

\newcommand{\Trace}[1]{{\rm Trace}\left( #1 \right) }

\title{A Proof of Alon's Second Eigenvalue Conjecture and Related Problems}

\author{Joel Friedman\thanks{
        Departments of Computer Science and Mathematics,
        University of British Columbia, Vancouver, BC\ \ V6T 1Z4
        (V6T 1Z2 for Mathematics), CANADA.
        {\tt jf@cs.ubc.ca}.
        Research supported in part by an NSERC grant.}}

\date{March 27, 2002 \\ Third Revision: May 3, 2004}

\begin{document}           
\maketitle                 
\begin{abstract}
A $d$-regular graph has largest or first (adjacency
matrix) eigenvalue $\lambda_1=d$.
Consider for an even $d\ge 4$, a random $d$-regular
graph model
formed from $d/2$ uniform, independent permutations
on $\intn$.  We shall show that for any $\epsilon>0$ we have
all eigenvalues aside from $\lambda_1=d$ are bounded by
$2\sqrt{d-1}\;+\epsilon$ with probability $1-O(n^{-\tau})$,
where $\tau=\lceil \bigl(\sqrt{d-1}\;+1\bigr)/2 \rceil-1$.
We also show that this probability is at most $1-c/n^{\tau'}$,
for a constant $c$ and a $\tau'$ that is either $\tau$ or
$\tau+1$ (``more often'' $\tau$ than $\tau+1$).  We prove
related theorems for other models of random graphs, including
models with $d$ odd.

These theorems resolve the conjecture of Alon, that says that
for any $\epsilon>0$ and $d$,
the second largest eigenvalue of ``most'' random $d$-regular graphs
are at most $2\sqrt{d-1}\;+\epsilon$ (Alon did not specify precisely
what ``most'' should mean or what model of random graph one should
take).



\end{abstract}


\section{Introduction}

The eigenvalues of the adjacency matrix of a finite undirected graph, $G$, are
real and hence can be ordered
\mythreeindex{lambda}{$\lambda_i(G)$}{$i$-th largest eigenvalue of the
adjacency matrix of a finite graph}
$$
\lambda_1(G)\ge\lambda_2(G)\ge\cdots\ge\lambda_n(G),
$$
where $n$ is the number of vertices in $G$.  If $G$ is $d$-regular, i.e.,
each vertex is of degree $d$,
then $\lambda_1=d$.
In \cite{alon_eigenvalues}, Noga Alon conjectured that for any $d\ge 3$ and
$\epsilon>0$, $\lambda_2(G)\le 2\sqrt{d-1}+\epsilon$ for ``most'' $d$-regular
graphs on a sufficiently large number of vertices.
The Alon-Boppana bound shows that the constant
$2\sqrt{d-1}$ cannot be improved upon (see \cite{alon_eigenvalues,
nilli,friedman_geometric_aspects}).
The main goal of this paper is to prove this conjecture
for various models
of a ``random $d$-regular
graph.''

Our methods actually show that for ``most'' $d$-regular graphs,
$|\lambda_i(G)|\le 2\sqrt{d-1}+\epsilon$
for all $i\ge 2$, since our methods are variants of the standard
``trace method.''

Our primary interest in Alon's conjecture, which was Alon's motivation,
is that fact that graphs with $|\lambda_i|$ small for $i\ge 2$ have various
nice properties, including being expanders or magnifiers 
(see \cite{alon_eigenvalues}).  

For a fixed $n$ we can generate a random $d$-regular graph on
$n$ vertices
as follows, assuming $d$ is even (later we will give random graph models
that allow $d$ to be even or odd).
Take $d/2$ permutations on $V=\{1,\ldots,n\}$, $\pi_1,\ldots,\pi_{d/2}$,
each $\pi_i$ chosen uniformly among all $n!$ permutations
with all the $\pi_i$ independent.
We then form
$$
E = \Bigl\{ \bigl(i,\pi_j(i)\bigr),\bigl(i,\pi_j^{-1}(i)\bigr) \bigm| 
j=1,\ldots,
d/2, \quad i=1,\ldots,n  \Bigr\} \;,
$$
yielding a directed graph $G=(V,E)$, which we may view as undirected.  
We call this probability space of
random graphs $\cgnd$\mythreeindex{Gnd}{$\cgnd$}{space of random graphs
formed from $d/2$ permutations}.  
$G$ can have multiple edges and self-loops, and
each self-loop contributes $2$ to the appropriate diagonal entry of $G$'s
adjacency matrix\footnote{Such a self-loop is a {\em whole-loop} in the
sense of \cite{friedman_geometric_aspects}; see also Section~2 of this
paper.}.

The main goal of this paper is to prove theorems like the following,
which prove Alon's conjecture, for various models of a random $d$-regular
graph; we start with the model $\cgnd$.
\begin{theorem}\label{th:main}
Fix a real
$\epsilon>0$ and an even positive integer $d$.  Then there is a constant,
$c$, such that for a random graph, $G$,
in $\cgnd$ we have that with probability at least
$1-c/n^\tau$ we have for all $i>1$
$$
|\lambda_i(G)|\le 2\sqrt{d-1}\;+\epsilon,
$$
where
$\tau=\taufund=\lceil \bigl(\sqrt{d-1}\;+1\bigr)/2 \rceil-1$
\mythreeindex{taufund}{$\taufund$}{smallest order of a supercritical
tangle}.
Furthermore, for some constant $c'>0$ we have that $\lambda_2(G)>2\sqrt{d-1}$
with probability at least 
$c'/n^s$, where $s=\lfloor \bigl(\sqrt{d-1}\;+1\bigr)/2
\rfloor$.  (So $s=\taufund$ unless $\bigl(\sqrt{d-1}\;+1\bigr)/2$ is an
integer, in which case $\taufund=s-1$.)
\end{theorem}

Left open is the question of whether or not this theorem holds with
$\epsilon=0$ (which would yield ``Ramanujan graphs'')
or even some function $\epsilon=\epsilon(n)<0$.  
Calculations such as those in
\cite{friedman_geometric_aspects} suggest that it does, even for some
negative function $\epsilon(n)$.
Examples of ``Ramanujan graphs,'' i.e., graphs where
$|\lambda_i(G)|\le 2\sqrt{d-1}$ except $i=1$ (and, at times, $i=n$
when $\lambda_n=-d$)
have been given in 
\cite{lubotzky_phillips_sarnak,margulis,morgenstern} where $d$ is one
more than an odd prime or prime power.  Theorem~\ref{th:main} demonstrates
the existence of ``nearly Ramanujan'' graphs of any even degree.
We shall soon address odd $d$, as well.

Another interesting question arises in the gap between $\taufund$ and $s$ in
Theorem~\ref{th:main} in the case where
$\bigl(\sqrt{d-1}\;+1\bigr)/2$ is an integer; 
it is almost certain that one of them can be
improved upon.
In the language of Section~4 of this paper, 
$\taufund$\mythreeindex{taufund}{$\taufund$}{smallest order of a supercritical
tangle} is the smallest
order of a supercritical tangle, and $s$ that of a hypercritical tangle;
a gap between $\taufund$ and $s$ can only occur when there is a critical
tangle of order smaller than that of any hypercritical tangle.

Previous bounds of the form $\lambda_2\le f(d)+\epsilon$
include $f(d)=(2d)^{1/2}(d-1)^{1/4}$ of the author
(see \cite{friedman_relative}),
which is slight improvement over the Broder-Shamir bound of
$f(d)=2^{1/2}d^{3/4}$ (see \cite{broder}).  Asymptotically in $d$,
the bounds $f(d)=C\sqrt{d}$ of Kahn and Szemer\'edi (see
\cite{friedman_kahn_szemeredi}, here $C$ is some constant)
and $f(d)=2\sqrt{d-1}+2\log d + C$
of the author
(see \cite{friedman_kahn_szemeredi,friedman_random_graphs} and see
equation~(\ref{eq:rho}) for the more precise bound) are
improvements over the first two bounds.

The value of $\taufund$ in Theorem~\ref{th:main} depends on the particular 
model
of a random graph.  Indeed, consider the model 
$\chnd$\mythreeindex{Hnd}{$\chnd$}{random graph 
space formed by $d/2$ permutations that are
cycles of length $n$}
of a random graph,
which is like $\cgnd$ except that we insist that each $\pi_i$ be one of the
$(n-1)!$ permutations whose cyclic decomposition consists of one cycle of
length $n$.  The same methods used to prove Theorem~\ref{th:main} will show
the following variant.

\begin{theorem}\label{th:mainh}
Theorem~\ref{th:main} holds with $\cgnd$ replaced by
$\chnd$ and $\taufund
=\lceil \sqrt{d-1}\;\rceil-1$ and $s=\lfloor \sqrt{d-1}\rfloor$,
except that when $d=4$ we take $s=2$.
\end{theorem}
Once again, $\taufund=s$, unless a certain expression, in this case
$\sqrt{d-1}$ (excepting $d=4$),
is an integer.  Note that for $\chnd$, the value of
$\taufund$ is roughly twice as large as that for $\cgnd$ for $d$
large.

Next consider two more models of random $d$-regular graphs; in these two
models $d$ may be even or odd.  
Let $\cind$\mythreeindex{Ind}{$\cind$}{random graph model of $d$ perfect matchings ($n$ even)}, 
for positive
integers $n,d$ with $n$ even, be the model of a random $d$-regular graph
formed from $d$ random perfect matchings on $\intn$.

For an odd positive integer $n$, let a {\em near perfect matching} be a
matching of $n-1$ elements of $\intn$; such a matching becomes a $1$-regular
graph if it is complemented by a single half-loop\footnote{
	Readers unfamiliar
	with half-loops (i.e., self-loops contributing only $1$ to a
	diagonal entry of the adjacency matrix)
	can see Section~2 of this paper or 
	\cite{friedman_geometric_aspects}.
} at the unmatched
vertex.  Taking $d$ independent such $1$-regular graphs gives a model,
$\cjnd$\mythreeindex{Jnd}{$\cjnd$}{random graph model formed from
$d$ permutations each with exactly one fixed point ($n$ odd)}, 
of a $d$-regular graph on $n$ vertices for $n$ odd.

\begin{theorem}\label{th:maini}
Theorem~\ref{th:mainh} holds with $\chnd$ replaced by
$\cind$ and with no $d=4$ exception (i.e., $s=1$ for $d=4$).
Theorem~\ref{th:main} holds with $\cgnd$ replaced by
$\cjnd$.
\end{theorem}

We can assert the truth of the
Alon conjecture on more models of random graphs
by using results on continguity and related notions.
Consider
two families of probability spaces, 
$(\Omega_n,{\cal F}_n,\mu_n)_{n=1,2,\ldots}$ and
$(\Omega_n,{\cal F}_n,\nu_n)_{n=1,2,\ldots}$ over the same sets
$\Omega_n$ and sigma-algebras ${\cal F}_n$; denote
$\mu=\{\mu_n\}$
and $\nu=\{\nu_n\}$.  We say that $\mu$ {\em dominates} $\nu$ if
for any family of measurable events,
$\{E_n\}$ (i.e., $E_n\in{\cal F}_n$), we have $\mu_n(E_n)\to 0$ as
$n\to\infty$ implies $\nu_n(E_n)\to 0$ as $n\to\infty$.
We say that $\mu$ and $\nu$ are {\em contiguous} if $\mu$ dominates $\nu$
and $\nu$ dominates $\mu$.
\begin{corollary} Fix an $\epsilon>0$ and an integer
$d\ge 2$.  Let ${\cal L}_{n}$
be any family of probability spaces of $d$-regular graphs on $n$ vertices
(possibly defined for only certain $n$)
that is dominated by $\cgnd$, $\chnd$, $\cind$, or $\cjnd$.
Then
for $G$ in ${\cal L}_n$ we have that with probability $1-o(1)$ 
(as $n\to\infty$) for
all $i$ with $2\le i\le n$ we have
$$
|\lambda_i(G)|\le 2\sqrt{d-1}+\epsilon.
$$
\end{corollary}
There are a lot of results regarding contiguity and (at least
implicitly) domination; see \cite{greenhill,kim,wormald} and the references
there.  For example, if $\cgnd'$ is the restriction of $\cgnd$ to those
graphs without self-loops, then for $d\ge 4$ it is known that
(1) $\cgnd'$ and $\chnd$ are contiguous (by \cite{kim} and previous work), and
(2) $\cgnd$ dominates $\cgnd'$ (easy, since a self-loop occurs 
in $\cgnd$ with 
probability bounded away from $1$ for fixed $d$).  
Thus the Alon conjecture for
$\cgnd$ implies the same for $\chnd$ (but this contiguity and/or domination
approach does not give as tight a bound on the probablity that
$\lambda_2\le 2\sqrt{d-1}\;+\epsilon$ fails to hold as is given
in Theorem~\ref{th:mainh}).
Also, $\cgnd$ is contiguous
with the ``pairing'' or ``configuration'' model of $d$-regular (pseudo)graphs
(see \cite{greenhill}); it follows that the Alon conjecture holds for the
latter model, and thus
(see \cite{wormald}, especially the beginning of Section~2 and
Corollary~4.17)
the conjecture holds for $n$ (and $d$) even
for $\cind$ or the uniform
measure on all $d$-regular (simple) graphs on $n$ vertices.

Our method for proving Theorems~\ref{th:main}, \ref{th:mainh}, and
\ref{th:maini}
is a variant of the well-known ``trace method''
(see, for example
\cite{wigner,geman,komlos,broder,friedman_random_graphs}) originated
by Wigner,
especially the author's refinement in \cite{friedman_random_graphs}
of the beautiful Broder-Shamir style of analysis in 
\cite{broder}.
The standard trace method involves taking the expected value
of the trace of a reasonably
high power\footnote{In 
	\cite{broder,friedman_random_graphs} this power
	is roughly $c\log n$, where $c$ depends on $d$ and on aspects of the
	method.}
of the adjacency matrix.  In our situation we are
unable to analyze this trace accurately enough to prove
Theorem~\ref{th:main}, as certain infinite sums involved in our
analysis diverge 
(for example, the infinite sum involving $W(T;\vec m)$ and $P_{i,T,\vec m}$
just above the middle
of page 351 in \cite{friedman_random_graphs}, for types of order $>d$).
This divergence is due to certain ``tangles'' that
can occur in a random graph and can adversely affect the eigenvalues
(see Sections~2 and 4).
To get around these ``tangles'' we introduce a {\em selective trace}.
We briefly sketch what a selective trace is in the next paragraph.

Recall that a closed walk\mytwoindex{closed walk}{a walk beginning and
ending at the same vertex} about a vertex, $v$, is a walk in the graph
beginning and ending at $v$.
Recall that the trace of the $k$-th power of the adjacency matrix equals
the sum over all $v$ of the number of closed walks about $v$
of length $k$.  
The {\em $k$-th irreducible trace} (used in both \cite{broder} and
\cite{friedman_random_graphs}) is the same sum as the $k$-th power
trace, except that we
require the closed walks to be {\em irreducible}\mytwoindex{irreducible}{
a walk (resp.\ word) that has no consecutive steps of an edge (resp.\ letter) 
and its 
inverse}, i.e., to have no edge
traversed and then immediately thereafter
traversed in the opposite direction.
A selective trace is a sum like an irreducible trace, but
where we further require that
the walk have no
small contiguous piece that ``traces out''
a ``supercritical
tangle'' (the notions of ``tracing out'' and ``supercritical
tangles'' will be
defined later;
roughly speaking, a ``supercritical
tangle'' is a small graph with many cycles). 
Since these ``tangles'' occur with probability
at most proportional to $n^{-\tau}$, with $\tau=\taufund$ 
as in Theorem~\ref{th:main},
the selective trace usually agrees with the standard ``irreducible'' trace.

Analyzing the selective traces involves a new concept of the
``new type,'' which is a refinement of the ``type'' of 
\cite{friedman_random_graphs}.

We caution the reader about the notation used here.  In this paper
we work with only $d$-regular graphs.  In
\cite{broder} $2d$-regular graphs were studied; in 
\cite{friedman_random_graphs} the graphs are usually $2d$-regular, although
for a part of Section~3 the graphs are $d$-regular.
We also caution the reader that here we use the term ``irreducible''
(as used in \cite{broder,friedman_random_graphs} and, for example,
in the text \cite{godsil}) as opposed to ``reduced'' (which is quite
common) or
``non-backtracking'' (sometimes used in \cite{friedman_random_graphs})
in describing walks and related concepts.

We hope to generalize or
``relativize'' the theorems here to
theorems about new eigenvalues of random covers
(see \cite{friedman_relative} for a relativized Broder-Shamir theorem).  
In this paper we occasionally go out of our
way to use a technique that will easily generalize to this setting.

The rest of this paper is organized as follows.  In Section~2 we
review the trace method used in \cite{friedman_random_graphs} and
explain why it requres modification to prove Alon's conjecture; as a
byproduct we establish the part of Theorem~\ref{th:main} involving
$s$.  In Section~3 we give some background needed for some technical
details in later sections.  In Section~4 we formalize the notion of a
tangle, and discuss their properties; we prove the part
of Theorem~\ref{th:mainh} and \ref{th:maini}
involving $s$.  In Section~5 we describe
``types'' and ``new types,'' explaining how they help to estimate
``walk sums;'' walk sums are generalizations of all notions of ``trace''
used here.  In Section~6 we describe the ``selective trace'' used in
this paper; we give a crucial
lemma that counts certain types of selective closed walks in a graph.
In Section~7 we explain a little about ``{\dtreelike}'' functions, giving
a theorem to be used in Section~14 that also illustrates one of the main
technical points in Section~8.  
In Section~8 we prove that certain selective traces have an asymptotic
expansion (in $1/n$) whose coefficients are ``\dtreelike.''
In Section~9 we show that the expansion in Section~8 still exists when
we count selective traces of graphs not containing any finite set of
tangles of order $\ge 1$.
In Section~10 we introduce strongly irreducible traces, that simplify
the proofs of the main theorems in this paper.  
In Section~11 we prove a
crucial lemma that allows us to use the asymptotic expansion
to make conclusions about
certain eigenvalues; this lemma sidesteps the unresolved problem of
(even roughly) determining the coefficients of the asymptotic expansion
(in \cite{friedman_random_graphs} we actually roughly
determine the coefficients for the shorter expansion developed there).
In Section~12 we prove
the magnification (or ``expansion'') properties needed to apply the
sidestepping lemma of Section~11.
In Section~13 we complete the proof of Theorem~\ref{th:main}.
In Section~14 we complete the proof of Theorems~\ref{th:mainh} and
\ref{th:maini}, giving general conditions on a model of random graph
that are sufficient to imply the Alon conjecture.
In Section~15 we make some closing remarks.

We mention that the reader interested only in the Alon conjecture
for only $\cgnd$
(i.e., the first part of Theorem~\ref{th:main}) need not read
Sections~2, 4 (assuming a willing to
believe Lemma~\ref{lm:order_increases}), and 14 and
Subsections~3.7, 3.8, 5.4, and 6.4.
Section~2 explains the problems with the trace method encountered
in \cite{friedman_random_graphs}.  Subsections~3.7 and 3.8 and 
Section~4 concern
themselves with the second part of Theorem~\ref{th:main} (the close
to matching bound on how many graphs fail the $2\sqrt{d-1}+\epsilon$
bound).  Subsection~5.4 explains the new aspects in our approach to
the Alon conjecture; this subsection is not essential
to the exposition (but probably is helpful).  Subsection~6.4 and 
Section~14
involve the Alon conjecture for $\chnd,\cind,\cjnd$.

Throughout the rest of this paper we will work with $\cgnd$ unless
we explicitly say otherwise, and we understand $d$ to be a fixed
integer at least
$3$.  At times we insist that $d$ be even (for example, in
dealing with $\cgnd$ and $\chnd$).

\section{Problems with the Standard Trace Method}
\label{se:problem}

In this section we summarize the trace method used in
\cite{friedman_random_graphs}, and why 
this method cannot prove Alon's conjecture.
During this section we will review some of the ideas of
\cite{friedman_random_graphs}, involving asymptotic expansions of
various types of traces, which we modify in later sections
to complete our proof of Alon's
conjecture.

\subsection{The Trace Method}

We begin by recalling the trace method as used in
\cite{friedman_random_graphs}, and why it did not yield the Alon
conjecture.

The trace method (see \cite{wigner,geman,komlos,mckay,broder,
friedman_random_graphs}, for example) determines information on the
eigenvalues of a random graph in a certain probability space by computing
the expected value of a sufficiently high power of the adjacency matrix, $A$;
this expected value equals the expected value of the sum of that power of the
eigenvalues, since
$$
\Trace{A^k} = \lambda_1^k+\cdots+\lambda_n^k.
$$
Now $\Trace{A^k}$ may also be interpreted as the number of closed
walks (i.e., walks (see Section 3.1) in the
graph that start and end at the same vertex) of length $k$.
Now restrict our discussion to $\cgnd$.
A word, $w=\sigma_1\ldots\sigma_k$, of length $k$ over the alphabet
$$
\Pi=\{ \pi_1,\pi_1^{-1},\ldots,\pi_{d/2},\pi_{d/2}^{-1} \}
$$
(i.e. each $\sigma_i\in\Pi$), determines a random permutation, and the
$i,j$-th entry of $A^k$, is just the number of words, $w$, of length $k$,
taking $i$ to $j$.
But given a word, $w$, the probability, $P(w)$, that $w$
takes $i$ to $i$ is clearly independent of $i$.
Hence we have
$$
\E{\Trace{A^k}} = n \sum_{w\in\Pi^k} P(w)
$$

In \cite{broder}, Broder and Shamir estimated the right-hand-side of the
above equality to obtain an estimate on $\lambda_2$.  This analysis was
refined in \cite{friedman_random_graphs}.  We review the ideas there.

First, a word, $w$, is
said to be {\em irreducible} if $w$ contains no consecutive occurence
of $\sigma,\sigma^{-1}$.  
It is well-known that any word, $w$, has a unique 
{\em reduction}
to an irreducible 
word\footnote{
	In fact, the irreducible word has length which is its distance
	to the identity in the Cayley graph over the free group on
	$d/2$ elements (see \cite{figa-talamanca_picardello}, Sections~1
	and 7 of chapter 1).
	Alternatively, see Proposition~2.5 of \cite{dicks} or
	Theorem~1 of \cite{johnson} (this theorem says that a free
	group on a set, $X$, is in one-to-one correspondence with the set
	of reduced words, $X\cup X^{-1}$, which means that every word
	over $X\cup X^{-1}$ has a unique corresponding reduced word;
	here ``reduced'' is our ``irreducible'').
} (or reduced word), $w'$,  
obtained from $w$ by repeatedly discarding any
consecutive occurences of $\sigma$ and $\sigma^{-1}$ in $w$, and
$P(w)=P(w')$.  
Similarly a walk is said to be {\em irreducible} if it contains
no occurrence of a step along an edge immediately followed by the
reverse step along that edge\footnote{
In the case of an edge that is a half-loop (see Section 3), a half-loop
may not be traversed twice consecutively in an irreducible walk.
}.
Similarly, every irreducible walk has a unique reduction.
Let $\ird k$ be the set of irreducible words of
length $k$, 
and let $\irdtr{A}{k}$\mythreeindex{IrredTr}{$\irdtr{A}{k}$}{the number of
closed irreducible walks of length $k$ in the graph underlying $A$}
be the number of irreducible closed walks of
length $k$ in $G$\footnote{We have admittedly defined $\irdtr{A}{k}$ in
terms of $G$, but we shall soon see (Lemma~\ref{lm:chebyshev}) that
$\irdtr{A}{k}$ can be defined as a polynomial in $A$ and $d$.
}.
We shall see that to
evaluate the expected value of $\Trace{A^m}$ it suffices, in a sense
(namely that of equation~(\ref{eq:other_way}) below),
to evaluate
$$
\E{\irdtr{A}{k}} = n\sum_{w\in\ird k} P(w),
$$
for $k=m,m-2,\ldots$.
It is easy to see that for any fixed word, $w$, we have a power series
expansion
$$
P(w)=\newa_0(w)+\frac{\newa_1(w)}{n}+\frac{\newa_2(w)}{n^2}+\cdots
$$
(see, for example, Theorem~\ref{th:exp_polys}).

As examples, we note that for
a random permutation, $\pi$, on $\{1,\ldots,n\}$,
the probability that the sequence $1,\pi(1),\pi^2(1),\ldots$ first returns
to $1$ at $\pi^k(1)$ (i.e., the probability that $1$ lies on a cycle of
length exactly $k$) is $1/n$ for $k=1,\ldots,n$.  It follows that
$P(\pi_1^m) = \phi(m)/n$, where $\phi(m)$ is the number of positive integral
divisors of $m$, assuming $m\le n$.  
In this example $\newa_1(w)=\phi(m)$ involves number
theoretic properties of $m$.
For a second example, we first remark that $\pi^m$ maps a fixed vertex
to a different vertex with probability $1-\bigl(\phi(m)/n\bigr)$, and to
each of the $n-1$ different vertices with the same probability.  It is
then easy to see that
$$
P(\pi_1^{m_1}\pi_2^{m_2}) =  \frac{\phi(m_1)\phi(m_2)}{n^2} +
\frac{\bigl( n-\phi(m_1) \bigr) \bigl( n-\phi(m_2) \bigr) }{n^2(n-1)}
$$
provided that $m_1,m_2$ are at most $n$.
If $m_1,m_2$ are at least $2$, then the $\newa_i$ are non-zero for $i\ge 1$
and involve number theoretic functions of $m_1,m_2$ for $i\ge 2$.

So set
\begin{equation}\label{eq:g_idef}
g_i(k)=\sum_{w\in\ird k} \newa_{i+1}(w)
\end{equation}
(we easily see $\newa_0(w)=0$ for $w\in\ird k$ and $k\ge 1$ and so
$g_{-1}(k)=0$ for $k\ge 1$).
\begin{definition}\label{de:ram}
A function, $f(k)$, on positive integers, $k$, is
said to be {\em \dtreelike} if there is a polynomial $p=p(k)$ and a
constant $c>0$ such
that
$$
|f(k)-(d-1)^kp(k)|\le ck^c(d-1)^{k/2}
$$
for all $k$.  We call $(d-1)^kp(k)$ the {\em principal term} of $f$,
and $f(k)-(d-1)^kp(k)$ the {\em error term} (both terms are uniquely
determined if $d>2$).
\end{definition} 
In \cite{friedman_random_graphs} it was shown (among other things) that
for all $i\le \sqrt{d-1}/2$ we have that $g_i$ as above is {\dtreelike}.
This, it turns out, gives a second eigenvalue bound of roughly
$2\sqrt{d-1}+2\log d+C+O(\log\log n/\log n)$ for a universal constant, $C$.
We now explain why.

A standard counting and expansion argument is given in
\cite{friedman_random_graphs} (specifically Theorem~3.1 there)
to establish the following lemma.
\begin{lemma}\label{lm:counting}
For fixed even $d\ge 4$ there is an $\eta>0$ such that
with probability $1-n^{1-d}+O(n^{2-2d})$ we have that a
$G$ in $\cgnd$ has $max(\lambda_2,-\lambda_n)\le d-\eta$
(also with probability $n^{1-d}+O(n^{2-2d})$ we have that
$\lambda_2=d$).
\end{lemma}

Next to $\lambda_1=d$, one (or both) of $\lambda_2,\lambda_n$ is
the next largest eigenvalue in absolute value; Lemma~\ref{lm:counting},
by bounding the eigenvalues other than $\lambda_1$, will eventually be
used to show that the $g_i$ of equation~(\ref{eq:g_idef}) are essentially
determined, for small $i$, by $\lambda_1$'s ``contribution'' to
$\irdtr{A}{k}$ (see below).

Next we establish
the precise relationship between the traces of the $A^k$ and the
$\irdtr{A}{k}$.  Let $A_k$ be the matrix whose $i,j$-th entry is the
number of irreducible walks of length $k$ from $i$ to $j$.
\begin{lemma}\label{lm:chebyshev}
The $A_k$ are given by
$$
A_k=q_k(A), 
$$
where $q_k$ is the degree $k$ polynomial given via
\begin{equation}\label{eq:chebyshev}
q_k( 2\sqrt{d-1}\cos\theta) = \bigl( \sqrt{d-1}\bigr)^k
\biggl( \frac{2}{d-1} \cos k\theta + \frac{d-2}{d-1}\;\frac{\sin (k+1)\theta}
{\sin\theta}\biggr) 
\end{equation}
(which is a type of Chebyshev polynomial);
alternatively we have $q_1(x)=x$, $q_2(x)=x^2-d$, and for $k\ge 3$ we have
$$
q_k(x) = x\;q_{k-1}(x)-(d-1)q_{k-2}(x).
$$
Also
$$
\irdtr{A}{k}=\Trace{A_k}=\sum_{i=1}^n q_k(\lambda_i).
$$
\end{lemma}
The proof is given in \cite{lubotzky} and
\cite{friedman_random_graphs} (specifically Lemma~3.3, page 356, in
\cite{friedman_random_graphs}; the $F_k$'s there are the $A_k$'s here).
In Section~10 we
shall use the fact that for fixed $\lambda$, $q_k=q_k(\lambda)$ 
satisfy the recurrence
$$
\bigl(\sigma_k^2-\lambda\sigma_k+(d-1)\bigr) q_k=0,
$$
where $\sigma_k$ is the ``shift in $k$'' operator (i.e., 
$\sigma_kq_k=q_{k+1}$)

To go the other way
we note:
\begin{equation}\label{eq:other_way}
A^k = \sum_{i=k,k-2,k-4,\ldots} \;\; N_{k,i} A_i,
\end{equation}
where $N_{k,i}$ is the number of words of length $k$ that reduce to a given
irreducible word
of length $i$.
Thus
$$
\Trace{A^k} = \sum_{i=k,k-2,k-4,\ldots} \;\; N_{k,i} \;\irdtr{A}{i}.
$$
\begin{lemma}\label{lm:reducedwalks}
For $k,i$ even we have
$$
N_{k,i} \le \Bigl( 2\sqrt{d-1}\Bigr)^k (d-1)^{-i/2}\sqrt{(d-1)/d}
$$
if $i>0$ and
$$
N_{k,0} \le \Bigl( 2\sqrt{d-1}\Bigr)^k.
$$
\end{lemma}
An exact formula for $N_{k,i}$ is given in \cite{mckay}.  A weaker
estimate than the above lemma
was used in \cite{friedman_random_graphs}.  The proof of this
estimate is a simple spectral argument used by Buck (see \cite{buck,
friedman_relative}).
\proof 
Consider
the adjacency matrix, $A_T$, of the infinite $d$-regular tree, $T$.
Our proof requires the following sublemma.
\begin{sublemma}\label{lm:tree_norm}
$A_T$ has norm $\le 2\sqrt{d-1}$.
\end{sublemma}
Actually, it is well-known that the norm of $A_T$ is exactly $2\sqrt{d-1}$
(see, for example, page 9 of
\cite{woess}, and the theorems on $\lambda_1$ in Section~3
here).  However, the proof below is simple and generalizes to many other
situations (and can be used in many cases to determine the exact norm of
$A_T$).
\proof {\bf (of Sublemma~\ref{lm:tree_norm})}\space\space
Let $f$ be a function in $L^2(T)$, i.e., a function on the vertices
of $T$ whose sum of squares of values is finite.  Fix a vertex, $v_0$, of $T$,
to be viewed as the root of $T$; the {\em children} of a vertex, $v$, are
defined to be those vertices adjacent to $v$ and of greater distance 
than $v$ is to
$v_0$.
We have
$$
(A_Tf,f) = \sum_v\sum_{w\in\;{\rm children}(v)} 2 f(v)f(w)
$$
which, by Cauchy-Schwarz, is
$$
\le \sum_v\sum_{w\in\;{\rm children}(v)} \left( f^2(w)\sqrt{d-1} + 
\frac{f^2(v)}{\sqrt{d-1}} \right)
$$
$$
= f^2(v_0)d\Bigm/\sqrt{d-1}\;+\; \sum_{v\ne v_0} f^2(v)2\sqrt{d-1}.
$$
$$
\le \sum_v f^2(v) 2\sqrt{d-1} = 2\sqrt{d-1} \|f\|^2.
$$
Thus the norm of $A_T$ is $\le2\sqrt{d-1}$.
\proofbox
(To see that the norm of $A_T$ is exactly $2\sqrt{d-1}$ we find functions,
$f$, (of finite support)
for which the applications of Cauchy-Schwarz in the above proof are
``usually'' tight.  
Namely, we can take
$f(v)=(d-1)^{-{\rm dist}(v,v_0)/2}$ for ${\rm dist(v,v_0)}\le s$ and
$f(v)=0$ otherwise, where $s$ is a parameter which tends to $\infty$.
This technique works for some other graphs.)

We resume the proof of Lemma~\ref{lm:reducedwalks}.
Let $v$ be a vertex of $T$, and let $S$ be the vertices
of distance $i$ to $s$.  Then $|S|\,N_{k,i}$ is the dot product of 
$A_T^k\chi_{\{v\}}$
with $\chi_S$, where $\chi_U$ denotes the characteristic function of $U$,
i.e. the function that is $1$ on $U$ and $0$ elsewhere.
So by Cauchy-Schwarz
$$
|S|\,N_{k,i}=(A_T^k\chi_{\{v\}},\chi_S)\le \|A_T\|^k |\chi_{\{v\}}| \;
|\chi_S| = \Bigl( 2\sqrt{d-1}\Bigr)^k \sqrt{|S|}.
$$
We finish with the fact that $|S|=1$ if $i=0$, and otherwise
$|S|=d(d-1)^{i-1}$.
\proofbox
Notice that clearly $N_{k,k}=1$, and so for $i=k$
Lemma~\ref{lm:reducedwalks}
is off by a multiplicative factor
of roughly
$2^k$; according to \cite{mckay,figa-talamanca_picardello}, the
Lemma~\ref{lm:reducedwalks}
estimate of $N_{k,0}$ is off by roughly a factor of $k^{3/2}$.  The
roughness of Lemma~\ref{lm:reducedwalks} is unimportant for our purposes.

Now notice that by Lemmas~\ref{lm:counting} and \ref{lm:chebyshev} we have
$$
\E{\irdtr{A}{k}} = q_k(d)\bigl( 1+n^{1-d}+O(n^{2-2d})\bigr) + 
{\rm error},
$$
where
$$
|{\rm error}| \le
(n-1)\max_{|\lambda|\le d-\eta} |q_k(\lambda)|.
$$
It is easy to see (see \cite{friedman_random_graphs}) that
$q_k(d)=(d-1)^k$, and for some $\alpha>0$ we have
$$
\max_{|\lambda|\le d-\eta} |q_k(\lambda)| \le (d-1-\alpha)^k ck,
$$
for an absolute constant $c$ (with any $\eta$ as in
Lemma~\ref{lm:counting}).  We wish to draw some conclusions about
the principal term of the $g_i$'s.  We need the following
lemma:
\begin{lemma} For fixed $d,r$ there is a constant, $c$, such that
for $k\ge 1$ we have that in $\cgnd$
$$
\E{\irdtr{A}{k}} = g_0(k)+\frac{g_1(k)}{n} + \frac{g_2(k)}{n^2} + \cdots
+ \frac{g_{r-1}(k)}{n^{r-1}} + {\rm error},
$$
where
$$
|{\rm error}| \le  c (d-1)^{k-1}k^{4r+2}/n^{r}.
$$
\end{lemma}
\proof This follows from the calculations on page 352 of
\cite{friedman_random_graphs}; for each $i$, the $f_i$ in
\cite{friedman_random_graphs} is the polynomial in the principal
term of $g_i$ (and we mean $f_i$ corresponds precisely to $g_i$, not
$g_{i-1}$ or $g_{i+1}$).  (Actually we shall later\footnote{
	This stems from the fact that in \cite{friedman_random_graphs},
	the $e^{(r+1)k/n}k^{2r+2}$ just above equation~(21) (page 352)
	could have been replaced with $e^{rk/n}k^{2r}$.
} see that the
$4r+2$ in the error term estimate can be replaced by $4r$.)
\proofbox
We now take $k$ of order $\log^2n$ and use standard facts about
expansion and expansion's control on eigenvalues (namely our
Lemma~\ref{lm:counting})
to conclude, as in
\cite{friedman_random_graphs}, the following theorem.
\begin{theorem}\label{th:f_ivalues}
Let $g_0,g_1,\ldots,g_r$ be {\dtreelike}
for some $r\le d$.  Then the principal term of $g_i$ vanishes for 
$1\le i \le r$, and the principal term of $g_0$ is $d(d-1)^{k-1}$.
\end{theorem}
\proof See Theorem~3.5 of \cite{friedman_random_graphs}.
\proofbox

We next apply Lemma~\ref{lm:reducedwalks} to estimate the expected value
of the trace of $A^k$ where $k$ is roughly 
\begin{equation}\label{eq:hnrd}
h(n,r,d)=\frac{(r+1)\log n }{\log\Bigl( d/\bigl(2\sqrt{d-1}\,\bigr)\Bigr)} ,
\end{equation}
as in
\cite{friedman_random_graphs}, in order to obtain the following theorem.
\begin{theorem}\label{th:expansion_consequence}
With the same hypotheses as Theorem~\ref{th:f_ivalues},
we have (in $\cgnd$)
\begin{equation}\label{eq:expansion_consequence}
\E{  \sum_{i=2}^n \lambda_i^k} \le \rho^k
\end{equation}
for all $k\le h(n,r,d)$, with $h$ as above, where
\begin{equation}\label{eq:rho}
\rho = 2\sqrt{d-1}\; \Bigl( d/\bigl( 2\sqrt{d-1}\,\bigr) \Bigr)^{1/(r+1)}
\left( 1 + \frac{c\log\log n}{\log n} \right)
\end{equation}
and $c$ depends only on $r,d$.
\end{theorem}
\proof First we take $k=\lfloor h(n,r,d) \rfloor$ and find that $\rho$
can be taken as above.  For
smaller $k$ we appeal to Jensen's inequality.  See 
\cite{friedman_random_graphs} for details.
\proofbox
From Theorem~2.18 of \cite{friedman_random_graphs} we see that we
can take $r$ as large as $\lfloor \sqrt{d-1}\;/2\rfloor$.  With this
value of $r$, using equation~(\ref{eq:rho}), we conclude that
equation~(\ref{eq:expansion_consequence}) holds with
$\rho=2\sqrt{d-1}+2\log d + C+o(1)$ for an absolute constant, $C$, where
$o(1)$ is a quantity that for fixed $d$ tends to $0$
(proportional to $\log\log n/\log n$) as $n\to\infty$.

Whenever equation~(\ref{eq:expansion_consequence}) holds, 
then the expected value
of $\max(|\lambda_2|,|\lambda_n|)$ is bounded by $\rho$.  The Alon conjecture
would be implied if one could obtain $\rho\le 2\sqrt{d-1}\;+\epsilon$
for any $\epsilon>0$.


\subsection{Limitations of the Trace Expansion}

In this subsection we will show that for some $i\le O(\sqrt{d}\log d)$,
$g_i$ is not {\dtreelike}.  We similarly
show the part of Theorem~\ref{th:main} involving $s$.  Both these facts are
due to the possible occurrence of what we call {\em tangles}.
Tangles and avoiding them are the main themes in this paper.

We begin by describing an example of a {\em tangle}, and its effect
on eigenvalues and traces.
Consider $\cgnd$ for a fixed, even $d$ and a variable $n$ which we
view as large.  
Fix an integer $m$ with $1\le m\le d/2$ (assume $d\ge 4$).
Consider the event, $\tanglefirst$, that
$$
\mbox{$\pi_i(1)=1$ for $i=1,\ldots m$.}
$$
Clearly $\tanglefirst$ occurs with probability $1/n^m$.

Assume $\tanglefirst$ occurs in a fixed $d$-regular graph, $G=(V,E)$.
Let $W$ be the set of vertices of distance at least $2$ to the vertex $1$;
$W$ is a random set of vertices, but always of size at least $n-d-1$.
Consider the characteristic functions $\chi_{\{1\}},\chi_{W}$,
where $\chi_U$ is the function that is $1$ on the vertices in $U$, and 
$0$ elsewhere.  Let
$$
{\cal R}_A(v) = \frac{(Av,v)}{(v,v)}
$$
be the Rayleigh quotient associated to the adjacency matrix, $A$, of $G$.
The following lemma is well-known.
\begin{lemma}\label{lm:fix_bug}
Let $A$ be a real, symmetric matrix.
Let $u,v$ be nonzero vectors with $v$ orthogonal to $u$ and $Au$.  Then
$$
\lambda_2 \ge \min\bigl( {\cal R}_A(u), {\cal R}_A(v) \bigr).
$$
\end{lemma}
\proof Let $\mu$ denote the $\min$ on the right-hand-side of the above
inequality.  
By the hypothesis of the lemma, $(Au,v)=0$; along with the symmetry of $A$,
we have $(Av,u)=(v,Au)=0$.
If $w=\alpha u+\beta v$ with $\alpha,\beta$ scalars,
we have
$$
(Aw,w)= (Au,u)\alpha^2+(Av,v)\beta^2 \ge
\mu(u,u)\alpha^2+\mu(v,v)\beta^2 = \mu(w,w).
$$
It follows that the Rayleigh quotient of any vector in the span of $u$ and
$v$ is at least $\mu$.  Since this span is a two-dimensional subspace, the
max-min principle implies that $\lambda_2\ge \mu$.
\proofbox

We intend apply the above lemma with $u=\chi_{\{1\}}$ and $v=\chi_W$.
$$
{\cal R}_A(\chi_{\{1\}}) \ge 2m
$$
(since $(A\chi_U,\chi_U)$ counts twice the number of edges with both endpoints
in $U$).  Also
$$
(A\chi_{W},\chi_{W})
= (A\chi_V,\chi_V)-2(A\chi_V,\chi_{V\setminus W})+(A\chi_{V\setminus W},
\chi_{V\setminus W})
$$
$$
\ge (A\chi_V,\chi_V)-2(A\chi_V,\chi_{V\setminus W}) \ge dn-2d(d+1)
$$
so
$$
{\cal R}_A(\chi_{W}) \ge \frac{dn -2d(d+1)}{n-d-1}=d-O(1/n)
$$
viewing $m,d$ as fixed.
Since $\chi_{\{1\}}$ and $A\chi_{\{1\}}$ are supported in the neighbourhood
of distance at most $1$ from the vertex $1$, $\chi_W$ is orthogonal to
both of them.
Lemma~\ref{lm:fix_bug} now implies
$$
\lambda_2 \ge \min\bigl(2m,d-O(1/n)\bigr).
$$
Next consider the probability that 
$$
\mbox{$\pi_i(r)=r$ for $i=1,\ldots, m$}
$$
for at least one value of $r$.  Inclusion/exclusion shows that the
probability of this is at least
$$
\sum_r \prob{\mbox{$\pi_i(r)=r$ for $i=1,\ldots m$}} -
$$
$$
\sum_{r,s} \prob{\mbox{$\pi_i(r)=r$ and $\pi_i(s)=s$ for $i=1,\ldots m$}}
$$
$$
\ge n^{1-m}-\binom{n}{2} n^{-2m}.
$$
We summarize the above observations.
\begin{theorem}\label{th:weak}
For fixed integer $m$ with $1\le m\le d/2$,
we have that $\lambda_2\ge 2m$
for sufficiently large $n$ with probability at least
$n^{1-m}-(1/2)n^{2-2m}$.
\end{theorem}

The proof of the above theorem did not exploit the fact that aside from
having $m$ self-loops, the vertex $1$ is still adjacent to $d-2m$ other
vertices of a $d$-regular graph.
We seek a stronger theorem that exploits this fact.

\begin{theorem}\label{th:improved_bound}
For fixed integers $m\ge 1$ and $d\ge 4$, with $2m-1> \sqrt{d-1}$ and
$m\le d/2$,
we have that $\lambda_2>2\sqrt{d-1}$ 
for sufficiently large $n$ with probability at least
$n^{1-m}-(1/2)n^{2-2m}$.
\end{theorem}
(Notice that Theorem~\ref{th:weak} would require $m>\sqrt{d-1}$ for
the same conclusion.)
We are very interested to know if one can prove Theorem~\ref{th:improved_bound}
when
$2m-1=\sqrt{d-1}$ for integer $m$ and even integer $d$.  We expect not.
(The situation where $2m-1=\sqrt{d-1}$
gives rise to what we will call a ``critical tangle,'' and 
$2m-1\ge\sqrt{d-1}$ to a ``supercritical tangle,'' in Section~4.)
\proof
Note: in Theorems~\ref{th:remarkable} and \ref{th:key_to_s} we give
a proof of a generalization of this theorem requiring far less
calculation (but requiring more machinery).

Again, assume that $\pi_i(v_0)=v_0$ for $i=1,\ldots,m$ and some $v_0$.  
It suffices to show
$\lambda_2>2\sqrt{d-1}$ for sufficiently large $n$, under the assumption
that $2m-1>\sqrt{d-1}$.

Let
\begin{equation}\label{eq:alpha}
\alpha(m) = (2m-1)+\frac{d-1}{2m-1}.
\end{equation}
By Cauchy-Schwarz we have $\alpha>2\sqrt{d-1}$ (equality does not hold,
because $2m-1\ne (d-1)/(2m-1)$ since $2m-1>\sqrt{d-1}$).  

Let $\rho(v)$ denote $v$'s distance to $v_0$.
For a fixed $r$, let
$$
f(v) = \left\{ \begin{array}{ll} 
(2m-1)^{-\rho} & \mbox{if $\rho\le r$,} \\
0 & \mbox{otherwise,} \end{array}\right.
$$
where $\rho=\rho(v)$.  It is easy to check that $(Af)(v)\ge\alpha f(v)$
provided that $\rho(v)<r$ (this includes the case $v=v_0$, since $\rho(v_0)=0$,
but checking $v=v_0$ is a bit different from the other cases).
It follows that
\begin{equation}\label{eq:ray1}
\frac{(Af,f)}{(f,f)} \ge \frac{\alpha(f,f)_{r-1}}{(f,f)_r},
\end{equation}
where
$$
(f,f)_t = \sum_{\rho(v)\le t} f^2(v).
$$
But $1=f^2(v_0)\le (f,f)_{r-1}$ if $r\ge 1$, and also
$$
(f,f)_r\le (f,f)_{r-1}+(d-2m)(d-1)^{r-1}(2m-1)^{-2r}
$$ 
(since, by induction, the number of vertices at distance $r$ from
$v$ is at most $(d-2m)(d-1)^{r-1}$.)
So
\begin{equation}\label{eq:ray2}
\frac{(f,f)_r}{(f,f)_{r-1}}
\end{equation}
can be made arbitrarily close to $1$ by taking $r$ sufficiently large
(since $(d-1)(2m-1)^{-2}<1$).  

Let ${\cal R}$ be the Rayleigh quotient of $A$.
The last paragraph, especially equations~(\ref{eq:ray1}) and 
(\ref{eq:ray2}) implies that
for $\alpha'<\alpha$ there is an $r=r(\alpha')$ such that
${\cal R}(f)\ge \alpha'$.

So let $N$ be those vertices of distance $1$ or $0$ to
the support of $f$; the size, $|N|$, of $N$ is bounded as a function of 
$d$ and $r$.
The function $f$ is
orthogonal to $g=\chi_{V\setminus N}$ and $Ag$, 
and counting edges as before we see
$$
{\cal R}(g)\ge \frac{d|V|-2d|N|}{|V|-|N|}= d - O(|N|d/|V|).
$$
It follows that by taking $n$ sufficiently large, we can
make $\lambda_2\ge \alpha'$;  since $\alpha>2\sqrt{d-1}$, we can choose
$\alpha'>2\sqrt{d-1}$, making
$\lambda_2>2\sqrt{d-1}$.
\proofbox

Theorem~\ref{th:improved_bound} proves the part of Theorem~\ref{th:main}
involving $s$, by taking $s=m-1$ with $m$ as small as possible
(namely $m=\lfloor \bigl(\sqrt{d-1}\;+1\bigr)/2
\rfloor+1$).
The analogous parts of Theorems~\ref{th:mainh} and \ref{th:maini}
are slightly trickier, since the ``tangle'' involved has automorphisms;
we shall delay their proof (see Theorem~\ref{th:other_ss})
until we give a more involved discussion of
tangles in Section~4.

Notice that our proof is really computing the norm of $A_H$ where $H$ is
the $d$-regular graph with the vertex $1$ having $m$ self-loops, and which
is a (an infinite) tree when these loops are removed.  The function
$f$ as above
shows that $A_H$'s norm is at least $\alpha$.
The statement and proof of Sublemma~\ref{lm:tree_norm}
for the $d$-regular tree applies to the
above tree (with $2m-1$ replacing $\sqrt{d-1}$, and with $\alpha$
replacing $2\sqrt{d-1}$).
In this way our proof of Theorem~\ref{th:improved_bound}
is very much like one proof of
the Alon-Boppana theorem (see \cite{nilli,friedman_geometric_aspects}).

The discussion in this section leads to the following theorem.
\begin{theorem}\label{th:nottreelike}
There is an absolute constant (independent of $d$), $C$, such that the
$g_i$ of equation~(\ref{eq:g_idef}) cannot be {\dtreelike}
for all $i\le C\sqrt{d}\log d$.
\end{theorem}
\proof 
We fix an integer $s$ to be chosen later with 
$$
\Bigl( \sqrt{d-1} \; -1 \Bigr)/2 < s < d/2.
$$
Set $s+1=m$ and apply Theorem~\ref{th:improved_bound}.
Since $\alpha\ge 2s+1$ with $\alpha$ as
in equation~(\ref{eq:alpha}), for $k$ even we have that $\lambda_2\ge 2s$
with probability at least $n^{-s}+O(n^{-s-1})$.  Thus
$$
\E{ \lambda_2^k}^{1/k} \ge \bigl(n^{-s}+O(n^{-s-1})\bigr)^{1/k} 2s.
$$
According to Theorem~\ref{th:expansion_consequence}, if $g_0,\ldots,g_r$
are {\dtreelike} for some $r\le d$, then we have
$$
\E{ \lambda_2^k}^{1/k} \le \rho,
$$
with $\rho$ as in equation~(\ref{eq:rho}), provided that $k$ is even and
bounded by $h(n,r,d)$ as in Theorem~\ref{th:expansion_consequence}.
For some constant $c$ we have that for any $C$ and for
$r=C\sqrt{d}\log d$, equation~(\ref{eq:rho}) gives
$$
\rho =2\sqrt{d-1} \bigl(1 + cC^{-1}d^{-1/2}+c(\log n)^{-1}\log\log n\bigr).
$$
In other words,
\begin{equation}\label{eq:long}
\bigl(n^{-s}+O(n^{-s-1})\bigr)^{1/k} 2s \le
2\sqrt{d-1} \bigl(1 + cC^{-1}d^{-1/2}+c(\log n)^{-1}\log\log n\bigr).
\end{equation}
Take $k$ even and as close to $h(n,r,d)$ as possible;
note that by equation~(\ref{eq:hnrd}),
$$
\frac{\log n}{h}=
\frac{ \log\Bigl( d/\bigl(2\sqrt{d-1}\,\bigr)\Bigr)}{r+1} \le
\frac{\log d}{(C\sqrt{d}\log d)+1} \le 1/\Bigl(C\sqrt{d}\Bigr);
$$
hence, taking $n\to\infty$ in
equation~(\ref{eq:long}) implies that
for a universal constant, $c'$, we have
$$
e^{-sd^{-1/2}/C}2s\le 2\sqrt{d-1} \bigl(1 + cC^{-1}d^{-1/2}\bigr)
\le 2\sqrt{d-1}\bigl(1 + cC^{-1}\bigr) .
$$
Choosing $s=C\sqrt{d}$ and dividing by $2$ yields
$$
C\sqrt{d}\;/e \le \sqrt{d-1} \bigl(1 + cC^{-1}\bigr).
$$
Choosing $C$ large enough so that $C/e> 1+(c/C)$ makes this impossible.
\proofbox

We have proven that not all $g_i$ are {\dtreelike} for $i\le r$ where
$r=C\sqrt{d}\log d$.
Notice that in our terminology, Theorem~2.18 of \cite{friedman_random_graphs}
says that $g_i$ is {\dtreelike} for 
$i\le\lfloor \sqrt{d-1}\;\bigm/2 \rfloor-1$;
again, for each $i$ the $f_i$ in \cite{friedman_random_graphs} is the
polynomial in the principal part of our $g_i$.
This leaves the question of
whether Theorem~\ref{th:nottreelike} can be improved to an $r$ value closer
to $\lfloor \sqrt{d-1}\;\bigm/2 \rfloor-1$; we conjecture that it can
be improved to $r=\lfloor \bigl(\sqrt{d-1}\;+1\bigr)/2 \rfloor$, and
that the tangle with $m=\lfloor \bigl(\sqrt{d-1}\;+1\bigr)/2
\rfloor+1$ already ``causes'' this $g_r$ (or a lower one)
to fail to be {\dtreelike}.

This also leaves open the question of what can be said about the $g_i$
that are not {\dtreelike}.
Perhaps such $g_i=g_i(k)$ are a sum of $\nu^k p_\nu(k)$ over various $\nu$
with some added error term.  In this paper we avoid this issue,
working with a modified trace (i.e., ``selective'' traces)
for which the corresponding $g_i$ are {\dtreelike}.

\section{Background and Terminology}

In this section we review some ideas and techniques from the literature
needed here.  We also give some convenient terminology that is not
completely standard.

\subsection{Graph Terminology}

We use some nonstandard notions in graph theory, and we carefully
explain all our
terminology and notions here.

A directed graph, $G$, consists of a set of vertices, $V=V_G$, a set of
edges, $E=E_G$, and an incidence map, $i=i_G\from E\to V\times V$; if
$i(e)=(u,v)$ we will write $e\sim(u,v)$, say that $e$ is of {\em type}
$(u,v)$, and say that $e$ originates in $u$ and terminates in $v$.
(If $i$ is injective then it is usually safe to view $E$ as a subset of
$V\times V$, and we say that $G$ has no multiple edges.)  The adjacency
matrix, $A=A_G$, is a square matrix indexed on $V$, where $A(u,v)$ counts
the number of edges of type $(u,v)$.  The outdegree at $v\in V$ is the
row sum of $A$ at $v$, i.e., the number of edges originating in $v$; the
indegree is the column sum or number of edges terminating in $v$.

A graph\mytwoindex{graph}{a directed graph with a origin/terminal
reversing pairing of its edges}, $G$,
is a directed graph, $\widehat G$, such that each edge of type $(u,v)$
is ``paired'' with an
``opposite edge'' of type $(v,u)$; in other words, 
we have a map
$\opp=\opp_G\from E_{\widehat G}\to E_{\widehat G}$, such that
$\opp(\opp)$ is the identity, and if $e\in E_{\widehat G}$ has
$e\sim(u,v)$, then $\opp(e)\sim(v,u)$; in other words, the edges
$E_{\widehat G}$ come in ``pairs,'' except that a self-loop, i.e., an
$e\in E_{\widehat G}$ with $e\sim(v,v)$, can be paired with itself
(which is a ``half-loop\mytwoindex{half-loop}{a self-loop paired with itself
(in a graph)}'' in the terminology of 
\cite{friedman_geometric_aspects}) or paired with another self-loop at
$v$ (which is a ``whole-loop\mytwoindex{whole-loop}{two self-loops (about the same vertex) paired with each other (in a graph)}'').  Half-loops about $v$
contribute $1$ to the adjacency matrix entry at $v,v$ (i.e., contribute
$1$ to $A(v,v)$), and whole-loops
contribute $2$.  In this paper we primarily work with whole-loops,
needing half-loops only in the model $\cjnd$.
We refer to the {\em (undirected) edges}, $E_G$, of a graph, $G$,
as the set of ``pairs'' of edges, $\{e,\opp(e)\}$.  $G$'s vertex set
and adjacency matrix are just those of the directed graph, $\widehat G$,
i.e., $V_G=V_{\widehat G}$ and $A_G=A_{\widehat G}$.

A {\em numbering} of a set, $S$, is a bijection $\iota\from S\to
\{1,2,\ldots,s\}$, where $s=|S|$.  A {\em partial numbering} of a set,
$S$, is a numbering of some subset, $S'$, of $S$ (we allow $S'$ to be
empty, in which case none of $S$ is numbered).  We can speak of a graph,
directed or not, as having numbered or partially numbered vertices and/or
edges.  A numbering can be viewed as a total ordering.

Each letter $\pi\in\Pi=\{\pi_1,\pi_1^{-1},\ldots,\pi_{d/2}^{-1}\}$
has its associated inverse, $\pi^{-1}\in\Pi$,
and every word $w=\sigma_1\ldots\sigma_k$ over $\Pi$ has its associated
inverse, $w^{-1}=\sigma_k^{-1}\ldots\sigma_1^{-1}$.  If $\cw$ is any
set of words over $\Pi$, then a {\em $\cw$-labelling} of an undirected
graph, $G$, is a map or ``labelling'' $\cl\from E_{\widehat G}
\to\cw$ such that
$\cl\bigl( \opp(e)\bigr) = \bigl( \cl(e)\bigr)^{-1}$ for each $e\in 
E_{\widehat G}$.
For example, any graph $G\in\cgnd$ automatically comes with a
$\Pi$-labelling, namely $\bigl( i,\pi_j(i)\bigr)$ is labelled $\pi_j$, and
$\bigl( i,\pi_j^{-1}(i)\bigr)$ is labelled $\pi_j^{-1}$.

An {\em orientation} of an undirected graph, $G$, is the distinguishing
for each $e\in E_G$ of one of the two directed edges corresponding to $e$.

The following definition is special to this paper.
\begin{definition}\label{de:structural}
Fix sets $V,E$ and a set of words, $\cw$, over
$\Pi$, with $\cw^{-1}=\cw$.
A {\em structural map} is a map $s\from E\to\cw\times V\times V$.
A structural map defines a unique $\cw$-labelled, oriented graph,
$G$, with $V_G=V$ and $E_G=E$,
as follows: for each $e\in E$ with $s(e)=(\sigma,u,v)$,
we form a directed edge of type $(u,v)$ labelled $\sigma$ and declare
it distinguished, and pair it with a directed edge of type $(v,u)$
labelled $\sigma^{-1}$.
\end{definition}

\subsection{Variable-Length Graphs and Subdivisions}

In this paper we will work with graphs that have large or infinite
parts of them being paths or regular trees.  In this case we can 
easily eliminate
all the vertices in these parts 
by working with
``variable-length graphs.''
This leads to simpler calculations (in Theorem~\ref{th:infinite_edge}, that is
crucial to Lemma~\ref{lm:finiteness}, and in 
Theorem~\ref{th:crucial_cycle_count}).  This also gives us a notion
of regular tree of non-integral degree, in Theorem~\ref{th:remarkable}.

Recall (see \cite{shannon,adler,heegard,friedman_geometric_aspects}) that a
{\em VLG} or {\em variable-length graph}
(respectively, {\em directed VLG} or {\em directed variable-length graph})\mytwoindex{variable-length graph (VLG)}{a graph, directed or undirected,
with a positive integral ``length'' associated to each edge}
\mythreeindex{vlg}{VLG}{variable-length graph} 
is a graph (respectively, directed graph)
with an assignment of a positive
integer to each edge called the edge's 
{\em length}\footnote{In Shannon's terminology of \cite{shannon}, Chapter~1, 
Section~1, the edges have various ``times'' (as opposed to
``lengths'') such as a dot versus a dash
in Morse code.}.
The length of a walk in a VLG is the sum of the lengths of its edges
(each length is counted the number of times the edge appears in the walk).

A graph can be regarded as a VLG with all edge lengths $1$.
A VLG whose edge lengths are all $1$ can be identified with its underlying
graph.

A {\em bead}\mytwoindex{bead}{a vertex with indegree and outdegree $1$ (or, for undirected graphs, degree $2$) with no self-loops}
in a directed graph (respectively graph)
is a vertex with indegree and outdegree
$1$ (respectively, degree $2$)
and without a self-loop.  A {\em beaded path} is a path where every
vertex except possibly the endpoints are beads.
\begin{definition}\label{de:subdivision}
Let $G$ be a directed VLG.  To {\em subdivide} an edge, $e$, from $u$ to $v$
and of length $\ell$,
in $G$ is 
to replace $e$ with a beaded path of length $\ell$ from $u$ to $v$ in
$G$ (introducing $\ell-1$ new vertices).
A {\em subdivided form of $G$} is a graph, $G_\sbd$,
obtained by subdividing all edges
of $G$.
The same definition is made for VLG's, omitting the word ``directed''
everywhere, provided no half-loops are of length $2$ or greater\footnote{This
restriction will be explained just before
Proposition~\ref{pr:ird_invar}.  Actually, we can define a notion of
subdivision for all half-loops of odd length, but in this paper we
use half-loops only of length $1$.}.
\end{definition}
It should be clear that countings walks of certain types in a directed VLG,
$G$, should translate to an appropriate similar counting in $G_\sbd$,
and vice versa.  We next define an opposite of subdivision, supression,
and a vast generalization of supression, realization.

\begin{definition} Let $G$ be a strongly connected directed graph,
with $W\subset V_G$.  The {\em realization of $G$ to with vertex set $W$}
denoted $\realize{G}{W}$,
is the directed VLG on vertices $W$ with the following set of edges.
We create $\realize{G}{W}$
one edge from $u$ to $v$ (for $u,v\in W$) of length
$k$ for each walk from $u$ to $v$ in $G$ of length $k$ that
contains no $W$ vertices except as the first and last
vertices.  (So self-loops or edges in $G$ involving $W$ vertices
appear in $\realize{G}{W}$, since we regard self-loops or edges as
walks with no vertices except the first and last vertices.)
\end{definition}
The notion of realization appears constrained coding theory
(called ``fusion'' in \cite{heegard}, for example, and not given a name in
\cite{adler}; see also
\cite{friedman_geometric_aspects}).
We remark that if
$V_G\setminus W$ contains a cycle, then $\realize{G}{W}$ has
infinitely many edges.
\begin{definition}

Let $G$ be a strongly connected directed graph,
with $U\subset V_G$ a subset of beads in $G$ such that $U$ contains no
cycle.  The {\em supression of $U$
(in $G$)}, denoted $\supress{G}{U}$, is the realization of $G$ with
vertex set $V_G\setminus U$.
\end{definition}
The subdivision (by the supressed vertices)
of a supression returns the original directed graph.

We remark that if $G$ is a $\Pi$-labelled graph, then any supression
in $G$ has a natural $\Pi^{+}$-labelling, where $\Pi^+$ is the set
of words on $\Pi$ of length $\ge 1$.


\subsection{$\lambda_1$ of a VLG}

Let $G$ be a VLG (directed or undirected).  For $u,v\in V_G$ and a 
non-negative integer, $k$, let
$c_G(u,v;k)$\mythreeindex{cG}{$c_G(u,v;k)$}{the number of walks of 
  length $k$ from $u$ to $v$ in $G$}
denote the number of walks of length $k$ from $u$ to $v$ in
$G$.  We will use standard Perron-Frobenius theory
(see Sections~1.3 and 7.1 in \cite{kitchens} or
Chapters~1 and 6 in \cite{seneta}), which includes the rest of this
paragraph.
Assume that $G$ is strongly connected, i.e., for each $u,v\in V_G$ we
have $c_G(u,v;k)>0$ for some $k$.  Let $d=d_G$ be the period of $G$,
i.e., the greatest common divisor of the lengths of all closed walks in $G$.
Then all limits
$$
\limsup_{k\to\infty} \bigl( c_G(u,v;k) \bigr)^{1/k}, \quad
\lim_{k\to\infty} \bigl( c_G(v,v;kd) \bigr)^{1/kd}
$$
exist and are all equal (so independent of $u,v$); we define this
common limit to be $\lambda_1(G)$\myindexlambdaone, 
the {\em Perron value} of $G$.
It is easy to see that $c_G(v,v;k)\le \lambda_1^k$, using that
$c_G(v,v;k_1)c_G(v,v;k_2)\ge c_G(v,v;k_1+k_2)$.  If $G$ is a finite graph,
then $\lambda_1(G)$ is just the usual Perron-Frobenius (largest) eigenvalue
of $A_G$.

In directed graphs,
supression, realization, and subdivision preserve walk counts
(i.e., the $c(u,v;k)$'s) between appropriate vertices (those present
in the two graphs in question).  Therefore these operations also
preserve $\lambda_1$.

If $G$ is not strongly connected, we can define 
$\lambda_1(G)$\myindexlambdaone to be the supremum of $\lambda_1$ of
all the strongly connected components of $G$, or equivalently as the
supremum over all $\bigl( c_G(v,v;k)\bigr)^{1/k}$.

One can equivalently define $\lambda_1(G)$ with $\tilde c_G$ replacing
$c_G$, where $\tilde c_G(u,v;k)$ is the number of walks of length at
most $k$.  This is a sensible definition of $\lambda_1(G)$ when we
allow non-integral edge lengths\footnote{Why should a ``dash'' in 
  Morse code be precisely an integral
  multiple of a ``dot''?}.
One can also extend all these defitions to graphs with positively
weighted edges,
where the weight of a walk becomes the product of its edge weights, and
where $c_G$ or $\tilde c_G$ sums the weights of the walks.

\subsection{Shannon's Algorithm and Formal Series}

Shannon gives the following
algorithm (see \cite{shannon}, Chaper~1, Section~1)
for computing $\lambda_1(G)$ (or the ``valence'' or ``capacity'')
of a finite graph:
let $Z_G=Z_G(z)$ be the matrix whose $i,j$ entry is the sum of $z^\ell$ over
all edge lengths, $\ell$, of edges from $i$ to $j$,
with $z$ a formal
parameter.  Then $\lambda_1(G)$ is the reciprocal of the smallest real root 
in $z$ of
\begin{equation}\label{eq:shannon}
\det\bigl(I-Z_G(z)\bigr) = 0 .
\end{equation}
In this section we explain variants of this theorem/algorithm that hold
for infinite VLG's.  We first give some conventions that we will use
with formal power series.

By a {\em non-negative power series} we mean a series
$f(z)=\sum_{k=0}^\infty a_k z^k$, with $a_k$ non-negative reals.
We say that $f$ is the 
{\em generating function}\mytwoindex{generating function}{a power series, $\Sigma_k a_k z^k$, formed from coefficients $a_k$} of the $a_k$.
At times we view $f$ as a formal power series, but we will also have cause
to evaluate $f$ at non-negative reals (as a possibly diverging infinite
sum); it is easy to see that if $f(z_0)$ converges for a positive $z_0$,
then $f$ is continuous on $[0,z_0]$, and if $f(z_0)$ diverges then
$f(y_0)$ gets arbitrarily large as $y_0$ approaches $z_0$.

For such an $f$, $f$'s radius of convergence is
$$
\rho(f) = \limsup_{k\to\infty} a_k^{1/k}.
$$
The function
$f$ has a singularity at $z=\rho$, and $f(z_0)=\sum a_kz_0^k$ diverges
for $z_0>\rho$.  If the singularity at $z=\rho$ is a pole (e.g., when
$f(z)$ is a rational function), then $f(z)\to +\infty$ as $z\to\rho$
from the left.

N.B.: We do not identitfy a formal power series with any of its analytic
extensions unless we specifically say so.  For example,
$f(z)=1+z+z^2+\cdots$ has only the value $+\infty$ for real $z> 1$
unless we explicitly say to the contrary.

For a VLG, $G$, we set
$$
M_G(z) = I + Z_G(z) + Z_G^2(z) + \cdots 
$$
We have
\begin{equation}\label{eq:M}
\bigl( M_G(z) \bigr)_{u,v} = \sum_{k=0}^\infty c_G(u,v;k)z^k.
\end{equation}

We say that a non-negative power series, $f(z)=\sum a_k z^k$, {\em majorizes}
another one, $g(z)=\sum b_k z^k$, if
$$
a_1+\cdots+a_j \ge b_1+\ldots+ b_j
$$
for all $j\ge 1$.  If so, then $f(z_0)\ge g(z_0)$ for any $z_0\in[0,1]$
(with appropriate convensions on the value $+\infty$).  Given VLG's,
$G$ and $H$, we say that $G$ {\em majorizes} $H$ if $Z_G(z)$ majorizes
$Z_H(z)$ entry by entry;
equivalently, there is an endpoint preserving
injection from $E_H$ to $E_G$ that does not increase edge lengths.
If so, clearly $M_G(z)$ majorizes
$M_H(z)$ entry by entry.

\begin{theorem}\label{th:shannon_infinite}
Let $G$ be a strongly connected
VLG, directed or undirected, on a countable number of
vertices and edges.  The following are equal:
\begin{enumerate}
\item $1/\lambda_1(G)$, and
\item the radius of convergence of any entry
of $M_G(z)$.
\end{enumerate}
If $G$ has a finite number of vertices, then the above two numbers also
equal the following two:
\begin{enumerate}\addtocounter{enumi}{2}
\item the supremum of positive $z$ such that
$Z_G(z)$ converges (in each entry)
and has largest (i.e., Perron-Frobenius) eigenvalue
less than $1$, and
\item the supremum of positive $z$ such that 
$Z_G(z)$ converges and $\det\bigl(I-Z_G(y)\bigr)>0$
for all $y$ with $0<y<z$.
\end{enumerate}
If $G$ has a finite number of vertices, and if the entries $Z_G(z)$
are all rational functions of $z$, then the four above quantities equal
\begin{enumerate}\addtocounter{enumi}{4}
\item the first positive solution to $\det\bigl(I-Z_G(z)\bigr)=0$.
\end{enumerate}
Finally, all entries of $Z_G(z)$ will be rational functions of $z$ whenever
$G$ is finite (i.e., has finitely many vertices and edges)
or is a VLG realization of finite graph.
\end{theorem}
\proof (1)=(2): Clear from equation~(\ref{eq:M}).

(2)=(3): 
Let $B$ be a non-negative matrix.  According to Perron-Frobenius theory,
all the eigenvalues of $B$ are of absolute value at most $\lambda_1(B)$,
with equality only when the eigenvalue has the same algebraic and geometric
multiplicity.
It now follows from Jordan canonical
form that
a finite dimensional, non-negative matrix, $B$, has largest
(Perron-Frobenius) eigenvalue less than $1$ iff $I+B+B^2+\cdots$ converges.

(3)=(4): If $\lambda_1\bigl(Z_G(z_0)\bigr)<1$, then 
$\lambda_1\bigl(Z_G(y_0)\bigr)<1$ for all $y_0<z_0$, and hence
$\det\bigl(I-Z_G(y_0)\bigr)>0$ for such $y_0$.  Consider the first
$z_0$ for which $\lambda_1\bigl(Z_G(z_0)\bigr)<1$ fails to hold
(such a $z_0$ exists by the continuity of $\lambda_1$ as a function of
its entries).  Either $Z_G(z_0)$ does not converge, or else by continuity
of $\lambda_1$ we have $\lambda_1\bigl(Z_G(z_0)\bigr)=1$ and hence
$\det\bigl(I-Z_G(z_0)\bigr)=0$.

(4)=(5): By the above paragraph, if suffices to show that
$Z_G(z)$ cannot have a pole (in any of its entries)
at the first $z_0$ for which
$\lambda_1\bigl(Z_G(z_0)\bigr)<1$ fails to hold.  But if $Z_G(z)$ has a
pole in some entry at $z=z_0$, then for some $v\in V_G$ and positive
integer, $k$, $Z^k_G(z)_{v,v}$ has a pole at $z=z_0$, by the strong
connectivity of $G$.  But $\lambda_1^k\bigl( Z_G(z) \bigr)\ge
Z^k_G(z)_{v,v}$ for any $z$, and the latter tends to $+\infty$ as
$z\to z_0$ from the left.

Last part of the theorem:
Let $G'$ be a realization of a finite VLG, $G$, on the set $U\subset V_G$.
For $u_1,u_2\in U$, we shall calculate the $(u_1,u_2)$-entry of $Z_{G'}(z)$.
We claim this entry is the $(u_1,u_2)$-entry of $Z_{G}(z)$, plus the
$(u_1,u_2)$-entry
$$
v^{\rm T} (I+Z_{\bar G}(z)+Z_{\bar G}^2(z)+\cdots) w = v^{\rm T} 
\bigl(I-Z_{\bar G}(z)\bigr)^{-1}w,
$$
where $\bar U$ is the induced subgraph of $G$ on the vertex set
$\bar U=V_G\setminus U$, where $v$ is the vector whose entries correspond
to the edges from $u_1$ to the vertices of $\bar U$ and similarly for
$w$.  Since $v,w,Z_{\bar G}(z)$ all have polynomial entries,
we conclude that the $(u_1,u_2)$-entry of $Z_{G'}(z)$ is a rational
function.
\proofbox

Now we give two examples to show that Shannon's algorithm does not
literally apply as is to infinite graphs.  Let $G$ be a directed VLG with
one vertex and $a_k=\lfloor 2^k / (k+1)^2 \rfloor$ edges of length $k$
for all $k\ge 1$.
Then $Z_G(z)$ is a $1\times 1$ matrix with sole entry is
$f(z)=\sum_k a_k z^k$.  In this case we have $f(1/2)\le (\pi^2/6)-1<1$
and $f(z)$ diverges for any positive $z\ge 1/2$.  
It is not hard to see that $\lambda_1(G)=2$ (by
Theorem~\ref{th:shannon_infinite}, 
since $1+f+f^2+\cdots$ converges for $f<1$, which is the case
when $z\le 1/2$, and clearly diverges when $z>1/2$ where the series even
for $f$ diverges).
But the expression $\det\bigl(I-Z_G(z)\bigr)$ fails to have a zero at
$z=1/2$.

Next consider an undirected VLG, $G$, whose vertex set is the integers,
with each vertex having one self loop.  Then $\det\bigl(I-Z_G(z)\bigr)$
expanded in a power series has an infinite $z$ coefficient; if we
simply multiply the diagonals together (since this is a diagonal matrix),
we get the infinite product of $(1-z)$, which is $0$ for any $z>0$
(yet $\lambda_1(G)=1$).
While $G$ is not connected, we can add edges between $i$ and $i+1$ for
all $i$; this yields a connected graph with similar problems.

Finally we mention that Theorem~\ref{th:shannon_infinite} must be
modified when $G$ is not strongly connected.  Indeed, consider a directed
VLG
on three nodes, such that the first node has a single self-loop, and
the only other edges are edges from the second node to the third,
with $a_k$ such edges of length $k$.  Then the most natural way to
define $\lambda_1(G)$ is in terms of counting closed walks, so that
$\lambda_1(G)=1$.  But if $\sum a_k z^k$ diverges for any $z_0<1$, then
Theorem~\ref{th:shannon_infinite} fails.

\subsection{Limiting Graphs}

Let $G_i$ be a sequence of finite VLG's on the same vertex set, $V$, and the
same edge set, $E$.  Let $E$ be partitioned into two sets, $E_1,E_2$,
such that the following holds:
\begin{enumerate}
\item for each $e\in E_1$, the length of $e$ in $G_i$ is independent of $i$,
and
\item for each $e\in E_2$, the length of $e$ in $G_i$ tends to infinity
as $i\to\infty$.
\end{enumerate}
The {\em limit of the $G_i$} is the graph, $G_\infty$, which is any
$G_i$ with its $E_2$ edges discarded.

This simple remark is crucial for an important finiteness lemma
(Lemma~\ref{lm:finiteness}).
\begin{theorem}\label{th:infinite_edge}
With notation as above, 
$$
\lim_{i\to\infty} \lambda_1(G_i) = \lambda_1(G_\infty).
$$
\end{theorem}
\proof By counting closed walks we see that
$\lambda_1(G_i)\ge\lambda_1(G_\infty)$; this establishes the theorem
with ``$\ge$'' replacing ``$=$''.  To see ``$\le$'' replacing ``$=$'', 
assume that
$\lambda_1\bigl(A_{G_\infty}(z_0)\bigr)<1$ for some $z_0\in [0,1]$.  Clearly
$A_{G_i}(z_0)\to A_{G_\infty}(z_0)$ as $i\to\infty$.  So the continuity
of $\lambda_1$ on its entries implies $A_{G_i}(z_0)<1$ for $i$ sufficiently
large, and so $1/z_0\ge \limsup\lambda_1(G_i)$.  Now take a supremum over
$z_0$ with $\lambda_1\bigl(A_{G_\infty}(z_0)\bigr)<1$.
\proofbox

\subsection{Irreducible Eigenvalues}
\label{sb:lambda_irred}
Let $G$ be an undirected graph with corresponding directed graph 
$\widehat G$.
Let $G_{\ird{}}$\mythreeindex{GIrred}{$G_{\ird{}}$}{edge graph of $G$ with 
edges joined only when they form an irreducible path}
be the graph with vertices $E_{\widehat G}$ and an edge
from $e_1$ to $e_2$ iff $e_1e_2$ forms an irreducible path in $G$;
i.e., $e_1$ and $e_2$ are not opposites
(i.e., paired) in $G$, and $e_1$ terminates in
the vertex where $e_2$ originates.  (Therefore, if $e$ is a half-loop
in $G$, then
there is no edge from $e$ to itself in $G_{\ird{}}$.)
Then walks in $G_{\ird{}}$ give ``irreducible''
(or ``reduced'' or ``non-backtracking'') walks
in $G$.
A closed walk of length $k$ in $G_{\ird{}}$ gives a closed walk in $G$ (with specified
starting vertex) that is {\em strongly irreducible}, meaning that the
closed walk is irreducible and the last step in the closed walk is not the inverse of
the first step.
We define the {\em irreducible eigenvalues of $G$} to be those of
$G_{\ird{}}$, and we define the largest or Perron-Frobenius eigenvalue
of $G_{\ird{}}$ to be $\lambda_{\ird{}}=\lambda_{\ird{}}(G)$,\mythreeindex{lambdaIrred}{$\lambda_{\ird{}}(G)$}{the largest eigenvalue of $G_{\ird{}}$} the 
{\em largest irreducible eigenvalue of $G$}.

For $G$ an undirected VLG, we may define
$G_{\ird{}}$ as a VLG and hence define $\lambda_{\ird{}}(G)=
\lambda_1(G_{\ird{}})$.
In any graph,
an irreducible walk that enters a beaded-path must directly traverse this
path to its end (any backward step makes the walk reducible).
It easily follows that if we subdivide edges in a VLG that are not
half-loops
(i.e., whole-loops or edges between distinct vertices),
then counts of irreducible walks of a given length between $V_G$ vertices
remain the same\footnote{We do not know any simple or very natural way
to subdivide half-loops of even length
in a VLG while keeping $\lambda_{\ird{}}$
invariant.  A half-loop of length $\ell$ should be traversable zero
or one time in a row (but not twice or more in a row) in an irreducible
walk.  For odd lengths, $\ell$, this can be achieved by adding a path 
(of length $(\ell-1)/2$) with a
half-loop at the end.  
Morally speaking, subdividing an edge in a VLG, $G$,
can be
viewed as subdividing the corresponding edge pair in
underlying directed graph, $\widehat G$, and then gluing them
together with an ``opposite'' pairing (that glues the newly introduced
vertices together); for half-loops, we should glue a single directed 
self-loop of length $\ell$ to itself, via reflection about the middle);
for even length half-loops, that fact that reflection leaves the middle
vertex fixed creates problems if we wish to remain in the catergory of
undirected graphs$\ldots$
}.
As a 
consequence we get the following simple but important proposition.
\begin{proposition}
\label{pr:ird_invar}
We have
$\lambda_{\ird{}}$ of a VLG
is invariant under the subdivision of any set of edges devoid of
half-loops; in particular, $\lambda_{\ird{}}$ is invariant under
passing to a subdivided form of any VLG with all half-loops of length one.
Similarly, 
$\lambda_{\ird{}}$ of a graph is invariant under any supression.
\end{proposition}

We now state a theorem for use later in this paper; the theorem requires
a definition.
\begin{definition} A connected graph, $G$, is {\em loopy} if
$|E_G|\ge |V_G|$, or equivalently if $G$ contains an irreducible closed walk, 
or
equivalently if $G$ is not a tree.  $G$ is {\em $1$-loopy} if $G$ is
connected and the removal of any edge from $G$ leaves a graph each of
whose connected components are loopy.
\end{definition}

\begin{theorem}\label{th:loopy}
For a connected graph, $G$, the following are equivalent:
\begin{enumerate}
\item $G$ is $1$-loopy,
\item $G_{\ird{}}$ is strongly connected, and
\item $G$ is not a cycle and all vertices in $G$ have degree at least $2$.
\end{enumerate}
In particular, if $G$ is connected
and $d$-regular for $d\ge 3$, then $G_{\ird{}}$ is strongly connected.
\end{theorem}
Condition~(3) in the above theorem was pointed out to us by a referee.
\proof 
(2)$\Rightarrow$(1):
Consider a directed edge, $e\sim(u,v)$, of $G$, and let $e'$ be
$e$'s opposite (we permit $e=e'$, i.e., the case of a half-loop).
If the removal
of $e$ leaves the connected component of $v$ being a tree, then there is
no irreducible walk from $e$ to $e'$.  On the other hand, if this
connected component is not a tree, then a cycle in this component
about $v$ of minimum
length is irreducible, which then extends to an irreducible walk from
$e$ to $e'$.  To summarize, if $G$ is not $1$-loopy, then $G_{\ird{}}$
is not strongly connected; otherwise, each edge has an irreducible
path to its opposite, i.e., each edge is connected to its opposite
in $G_{\ird{}}$.

(1)$\Rightarrow$(2):
Assume $G$ is $1$-loopy.
If $e_1,e_2$ are two distinct, unpaired
directed edges originating in the same vertex,
$v$, then an irreducible path from $e_1$ to its opposite followed by
$e_2$ shows that $e_1$ and $e_2$ are connected by an irreducible path.
Thus any two edges that share a vertex are strongly
connected in $G_{\ird{}}$.

Finally if $e_1,e_2$ are two undirected edges without a common vertex,
then a shortest path connecting them gives an irreducible path
from some orientation of $e_1$ to some of $e_2$.  By the above we can
follow them with irreducible paths to the edge of opposite orientation.
Thus any two edges that do not share a vertex are strongly
connected in $G_{\ird{}}$.
So $G_{\ird{}}$ is strongly connected.

(1)$\Rightarrow$(3)
Since an isolated vertex is loopy, a $1$-loopy graph has all vertices
of degree at least $2$.  Also a $1$-loopy graph cannot be a single cycle,
since removing any one edge would give a tree.

(3)$\Rightarrow$(1)
Any tree either (i) is an isolated vertex, or (ii) is a path with two
vertices of degree one, or (iii) has at least three vertices of degree
$1$ (this follows by considering two vertices, $u,v$, of maximum distance in
a tree, and then taking a vertex of maximum distance to the unique
path joining $u$ to $v$).  
Consider the tree, $T$, obtained as the
non-loopy connected component by removing an edge, $e$, from a graph
$G$ that is non-loopy.  If $T$ is the only connected component, $e$
has both its endpoints in $T$, and if not then $T$ has only one endpoint
in $T$; in either situation, considering the cases (i)--(iii), it is
easy to check that $G$ is a cycle or has at least
one vertex (a vertex in $T$) of degree $1$.
\proofbox

\subsection{$\lambda_1$ and Closed Walks for Infinite
Graphs}

In this subsection we recall some facts about $\lambda_1$ and closed walks of
graphs, either indicating the proofs or giving references.  This section
is geared to infinite graphs, since most of the facts are very easy when
the graph is finite.

Let $G$ be a graph of bounded degree, i.e. there is an $r$ such that
the degree of each vertex is at most $r$.
Then $A_G$, $G$'s adjacency matrix, is a bounded linear operator (bounded
by $r$) on $L^2(G)$, the square summable functions on $G$'s vertices.
The next theorem shows that $\|A_G\|$, the norm of the operator $\|A_G\|$,
equals $\lambda_1(G)$ as was defined by counting closed walks about a 
vertex of at most some given length.
\begin{theorem} Assume $G$ is connected.  For any vertex, $v$, of $G$ and
positive integer, $k$, recall that
$c(v,k)$ is the number of closed walks of length $k$ in $G$ about $v$.  Then
$c(v,k)\le \|A_G\|^k$, and
$$
\lim_{r\to\infty} [c(v,2r)]^{1/(2r)}
$$
exists and equals $\|A_G\|$.
\end{theorem}
Since the limit above equals $\lambda_1(G)$, we see that 
$\lambda_1(G)=\|A_G\|$.
\proof See \cite{buck}.
\proofbox

We also need the following simple fact.
\begin{theorem} For every $\epsilon>0$ there is an $f\ne 0$ such that
$\|Af\|\ge(\lambda_1-\epsilon)\|f\|$, where $f$ has finite support.
\end{theorem}
\proof By definition of norm, there is a $g\ne 0$ with
$\|Ag\|\ge\bigl(\lambda_1-(\epsilon/2)\bigr)\|g\|$.  For any $\nu>0$
we may write $g=g_1+g_2$ where $g_1$ is of finite support and $\|g_2\|\le
\nu$.  It is easy to see (using the fact that $A$ is bounded) that
if $\nu$ is sufficiently small we can take $f=g_1$ to satisfy the
above theroem.
\proofbox

\subsection{A Curious Theorem}

\begin{figure}
    \centering
        \epsfysize=2.5in
        \epsfbox[0 0 745 370]{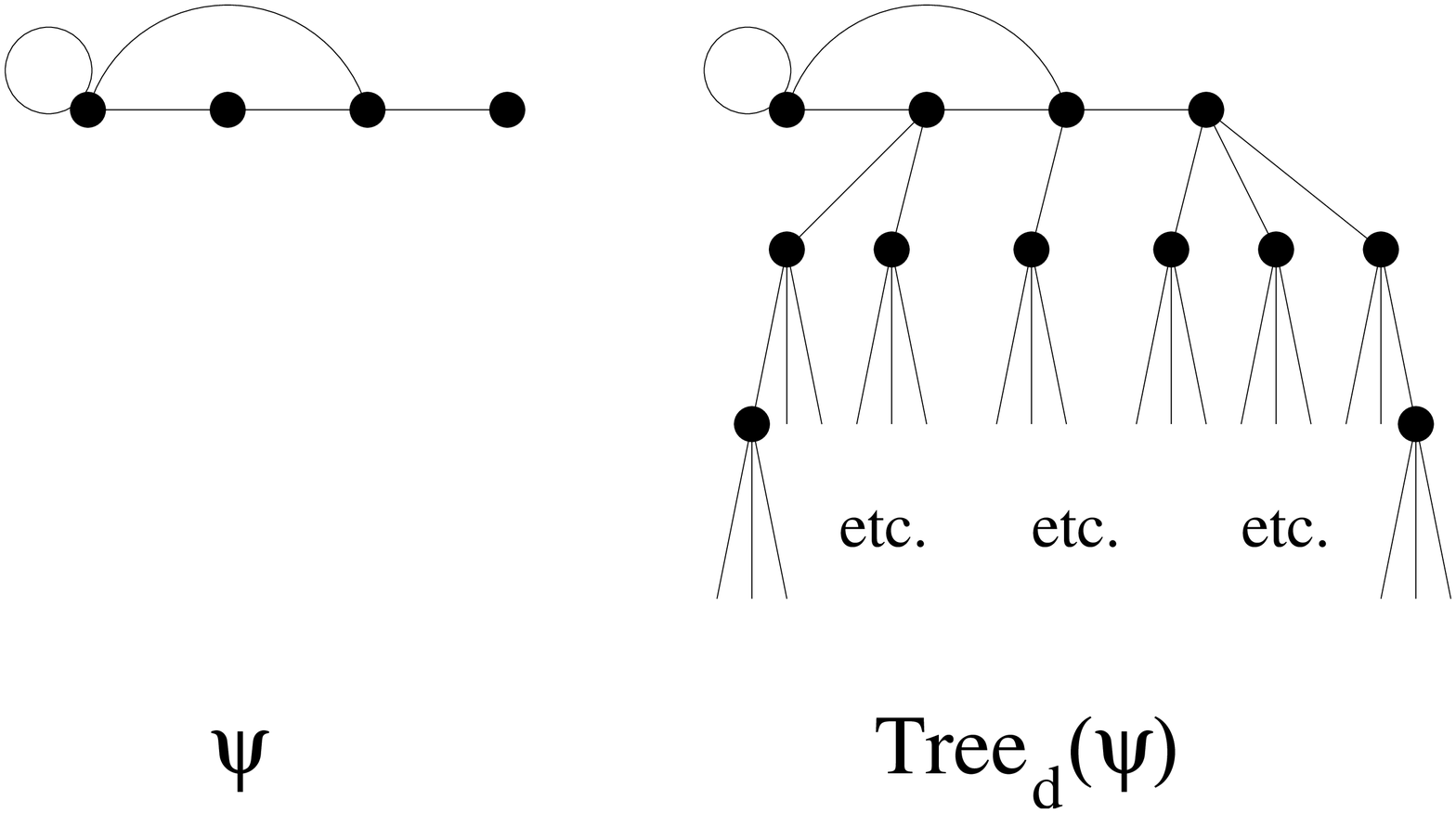}
    \caption{A graph, $\tang$, and ${\rm Tree}_d(\tang)$
	with $d=4$.}
    \label{fg:dtree}
\end{figure}

\begin{definition} Let
$\tang$ be a finite connected graph with each vertex of degree at
most $d$ for
an integer $d>2$. 
By $\tree_d(\tang)$ we mean the unique
(up to isomorphism) undirected graph, $G$,
that has an inclusion $\iota\from \tang\to G$
such that $G$ is $d$-regular
and such that $G$ becomes a forest when we remove (the
image under $\iota$ of) $\tang$'s edges.
\end{definition} 
We have seen an example of this construction in Section~2, where
$\tang$ is one vertex with $m$ self-loops, in the proof of
Theorem~\ref{th:improved_bound}.  See Figure~\ref{fg:dtree} for
another example.

In the category of $d$-regular graphs with a $\tang$ inclusion,
$\tree_d(\tang)$ is none other than the universal cover.

The methods used to prove Theorems~\ref{th:main}, \ref{th:mainh},
and \ref{th:maini}
suggest the following curious theorem.

\begin{theorem}\label{th:remarkable}
Let $d\ge 3$, and let $\tang$ be a finite connected
graph with each vertex of degree
$\le d$.  Then
\begin{eqnarray*}
\lambda_1\bigl( \tree_d(\tang) \bigr)=2\sqrt{d-1} & \Longleftrightarrow &
\lambda_{\ird{}}(\tang)\le\sqrt{d-1}, \\ 
\lambda_1\bigl( \tree_d(\tang) \bigr)>2\sqrt{d-1}  & \Longleftrightarrow &
\lambda_{\ird{}}(\tang)
>\sqrt{d-1}.
\end{eqnarray*}
The same is true for any real $d>2$, provided that
$\lambda_1\bigl( \tree_d(\tang) \bigr)$ is interpreted with an appropriate
analytic continuation in $d$ (described below).
\end{theorem}

Before proving the theorem, we describe the analytic continuation to which
we refer.

Let $\widetilde T$ be an undirected rooted tree,
where every vertex has $d-1$ children, 
with $d>1$ an integer (for now).  ($\widetilde T$ has degree $d-1$ at the
root and degree $d$ elsewhere, so $\widetilde T$ is not regular.)
Let $a_n$ for $n=2,4,\ldots$ be the number of walks in $\widetilde T$
from the root to
itself that never pass through the root except at the beginning and end.
It is easy to see that
$$
S = S_d(z)= \sum_{n=2}^\infty a_n z^n,
$$
satisfies the recurrence $S=z^2(d-1)(1+S+S^2+\cdots)$ (since the walks
counted by $S$ take one step, in $d-1$ possible ways, to a child
of the root, followed by some number (possibly zero)
of $S$ walks, followed by the step back to the root); so
$S(1-S)=z^2(d-1)$, so that near $z=0$
\begin{equation}\label{eq:S_quadratic}
S=S_d(z) = \frac{1-\sqrt{1-4(d-1)z^2}}{2}.
\end{equation}

One can alternatively say that if $H$ is the realization of $\widetilde T$
with vertex set $\{v_0\}$, where $v_0$ is the root, then the $1\times 1$
matrix $Z_H(z)$ has entry $S(z)$.  Similarly, if $\widetilde T$ were
modified to have $d-r$ children at the root (but every other vertex with
$d-1$ children, as before), then $Z_H(z)$ would have entry $(d-r)S(z)/(d-1)$.

By $\lambda_1\bigl( \tree_d(\tang) \bigr)$ for $d>1$ real
(assuming each degree in $\tang$ is $\le d$)
we mean $\lambda_1$
of the VLG formed from $\tang$ with an additional self-loop about
each vertex of degree $r$ that has ``formal weight''
$(d-r)S_d(z)/(d-1)$ (this can be viewed as adding an infinite set of self-loops
of given weights and lengths corresponding to this power series\footnote{
	In other words, we view a power series $\sum a_nz^n$ as the sum
	of terms representing
	$a_n$ edges of length $n$ (even when $a_n$ is not an integer).
}, or can simply be viewed as a term to add
to the diagonal of $Z_G(z)$).
When $d$ in an integer, the realization of $\tree_d(\tang)$ with vertex
set $V_\tang$ is exactly this VLG, and by Proposition~\ref{pr:ird_invar}
we know $\lambda_{\ird{}}$ remains the same.
\proof
By \cite{godsil}, exercise 13 page 72, we know that
$\lambda_{\ird{}}(A)$ is given by
$1/y$ of the smallest root, $y$, of
$$
\det\bigl(I-yA+y^2(D-I)\bigr) = 0,
$$
where $D$ is the diagonal matrix whose entry at vertex $v$ is the degree of 
$v$, and $I$ is the identity matrix.  Now
$$
Z_G(z) = zA + \frac{dI-D}{d-1}\; S(z).
$$
It is then easy to verify that
$$
I - Z_G(z) = \bigl( 1-S(z)\bigr) \bigl(I-yA+y^2(D-I)\bigr)
$$
for $y=y(z)=z/\bigl(  1-S(z)\bigr) $ (note that $y(z)$ is an increasing
function in $z$ for as long as $z(1+S+S^2+\cdots)$ converges).  
But $S(z_1)\le 1/2$ for all
positive real $z_1\le z_0$, where $z_0=1/\bigl( 2\sqrt{d-1}\bigr)$,
and also $y(z_0)=1/\sqrt{d-1}$.  So $I-y_1A+y_1^2(D-I)$ becomes non-invertible
for a $y_1<y(z_0)$ precisely when $Z_G(z_1)$ has eigenvalue $1$ for a
$z_1\in [0,z_0]$.  Now apply the equality of quantities~(1) and (3)
in Theorem~\ref{th:shannon_infinite}.
\proofbox
Note that the proof shows that if $1/y_1=\lambda_{\ird{}}(A)$ 
and $1/z_1$ is $\tree_d(\tang)$, then $y_1=y(z_1)$ provided
$1/z_1 > 2\sqrt{d-1}$.

\section{Tangles}

In Section~2 we saw that a vertex with $m$ self-loops in a
$\cgnd$ graph, with $m$ ``large,'' gives rise to a ``large'' second
eigenvalue (i.e., larger than $2\sqrt{d-1}$ for sufficiently large $m$,
as $n\to\infty$).  Here we generalize this observation to what we call
a ``tangle.''  Our proof of the Alon conjecture via a trace method must
somehow overcome all ``hypercritical'' tangles.

\begin{definition} 
Given two $\Pi$-labelled graphs, $G$ and $H$, we say $G$ {\em contains}
$H$ (or $H$ occurs in $G$) if there is an
inclusion\footnote{By an inclusion we mean a graph homomorphism that is
a injection on the vertices and on 
the edges.} $\iota\from H\to G$ that preserves
the labelling;
the {\em number of times} a graph, $H$, {\em occurs} in $G$ is
the number of distinct\footnote{
	For example, if $H=G$ consists of one edge joining two distinct
	vertices, $u,v$, then the identity is considered distinct from
	the morphism interchanging the vertices.  By the same principle,
	if $H$ has exactly $k$ automorphisms, then the number of times
	$H$ occurs in a graph is always a multiple of $k$.
} such $\iota$.
A {\em tangle} (or {\em $\cgnd$-tangle}) is a $\Pi$-labelled
connected graph, $\tang$, that is contained in some element of $\cgnd$.
\end{definition}
For example, in Section~2 we studied the
tangle with one vertex and $m$ self-loops labelled $\pi_1,\ldots,\pi_m$.
We define $\chnd$- and $\cind$- and $\cjnd$-tangles similarly, with
the following modifications to the meaning of the $\pi_i$'s.  For
$\chnd$, each of the $d/2$ independent $\pi_i$'s is uniform over all
permutations whose cyclic decomposition consists of a single cycle.
For $\cind$, each of $d$ independent $\pi_i$'s are uniform over all
perfect matchings, i.e., over all permutations that are involutions
without fixed points.
For $\cjnd$, each of $d$ independent $\pi_i$'s are uniform over all
involutions with exactly one fixed point.

\begin{theorem}\label{th:key_to_s}
Fix a positive integer, $d\ge 3$, and a graph $\tang$.  Any graph, $G$,
on $n$ vertices,
that contains $\tang$ has second eigenvalue at least
$\rho-o(1)$, where $o(1)$ is a function of $n$ tending to $0$ as $n\to\infty$,
and where $\rho$ is the norm of the adjacency matrix of $\tree_d(\tang)$.
\end{theorem}
This generalizes Theorem~\ref{th:improved_bound}.
\proof Fix $\epsilon>0$; we will show that for $n$ sufficiently large any
such $G$ has $\lambda_2(G)\ge\rho-\epsilon$.  First, there is a finitely
supported $f\ne 0$ on $\tree_d(\tang)$ with $\|Af\|\ge \|f\|(\rho-\epsilon)$,
where $A$ is the adjacency matrix of $\tree_d(\tang)$.
If $V_\tang$ is the set of vertices on which $f$ is non-zero, then replacing $f$
with the non-negative first Dirichlet eigenfunction on $V_\tang$ (see 
\cite{friedman_geometric_aspects}) we may assume $f$ is non-negative,
nonzero, and that $Af\ge f(\rho-\epsilon)$ (with equality everywhere except
at the boundary of $V_\tang$).
There is a covering map (see \cite{friedman_geometric_aspects} and
\cite{friedman_relative}) $\pi\from\tree_d(\tang)\to G$; set $\pi_*f$ to be
the function on $G$ defined by
$$
(\pi_* f)(v) = \sum_{\pi(w)=v} f(w).
$$
If $A_G$ is the adjacency matrix of $G$, then clearly
$$
A_G (\pi_* f) \ge (\rho-\epsilon)(\pi_* f).
$$
So, on the one hand,
${\cal R}(\pi_* f)\ge \rho-\epsilon$, where ${\cal R}$ is 
the Rayleigh quotient for $A_G$.
On the other hand, the support, $N$, of $\pi_* f$ is of size no greater than
that of $f$, and this size is bounded (independent of $G$ and $n$).
So the same reasoning as in the proof of Theorem~\ref{th:improved_bound}
shows that 
$$
{\cal R}(g) \ge d-o(1), \qquad\mbox{where}\quad
g=\chi_{V\setminus \widetilde N},
$$
where $\widetilde N$ is the set of vertices of distance $0$ or $1$ to $N$.
Since $\pi_* f$ is orthogonal to both $g$ and $A_G g$,
we are done (by Lemma~\ref{lm:fix_bug}).
\proofbox

\begin{definition} A tangle, $\tang$, is {\em critical}\mytwoindex{critical}{
a tangle with $\lambda_{\ird{}}=\sqrt{d-1}$} (respectively,
{\em supercritical}\mytwoindex{supercritical}{
a tangle with $\lambda_{\ird{}}\ge\sqrt{d-1}$}, 
{\em hypercritical}\mytwoindex{hypercritical}{
a tangle with $\lambda_{\ird{}}>\sqrt{d-1}$}) if $\lambda_{\ird{}}(\tang)$
equals (respectively, is at least, exceeds) $\sqrt{d-1}$.
\end{definition}
According Theorems~\ref{th:remarkable} and \ref{th:key_to_s},
a fixed
hypercritical tangle can only occur in a graph with sufficiently many vertices
if the graph has $\lambda_2>2\sqrt{d-1}$.

Now that we know how tangles affect eigenvalues, we want to know how often
the tangles occur.  This discussion, and the particular application to
$\chnd$, $\cind$, and $\cjnd$, will take the rest of this section.

\begin{definition} A {\em leaf} on a graph is a vertex of total degree $1$
(whether the graph is directed or not).
We say that a graph is {\rm pruned} if it has no leaves.
A {\em simple pruning} is the act of removing one leaf and its incident
edge from a graph; {\em pruning} is the repeated performance of some
sequence of simple prunings; {\em complete pruning} is the act of pruning
until no more pruning can be done.
\end{definition}
For example, completely pruning a tree results in a single vertex with
no edges;
completely pruning a cycle leaves the cycle unchanged.
\begin{proposition}\label{pr:lies_on_irred}
Given a graph, $G$, there is a unique pruned graph $H$ 
obtainable from completely pruning $G$.  Furthermore, $H$ is completely
pruned iff each edge of $H$ lies on an irreducible cycle.
\end{proposition}
\proof Let $e_1,\ldots,e_t$ denote the edges pruned in
one pruning of $G$, in the
order in which they are pruned.  Let $G'$ be a different complete pruning
of $G$, which we assume does not contain all the $e_i$.  Let $j$ be the
smallest integer such that $e_j$ lies in $G'$.  On the one hand,
the removal of $e_1,\ldots,e_{j-1}$ from $G$, or any subgraph
of $G$, allows $e_j$
to be pruned from $G$, or any subgraph of $G$, including $G'$.  On the
other hand, the prunability of $e_j$ from $G'$ contradicts the completeness
of the pruning that formed $G'$.  It follows that any complete pruning
contains all the edges of any other, and so any two are the same.

We now address the last statement of the theorem.
If $H$ has a leaf, then the edge incident upon this leaf does
not lie in an irreducible cycle.  Conversely, if $H$ has no leaves,
and if $e$ is an edge with endpoints $u,v$, consider the graph, $H'$,
obtained by removing $e$ from $H$.  If $u$ and $v$ are connected in $H'$,
then a minimal length path that joins them, along with $e$, gives an
irreducible cycle containing $e$.  Otherwise, since $u$'s connected 
component is not a tree (or else $H$ would have leaves), this 
component has an irreducible cycle, and a shortest walk from $u$ to
this cycle, once around the cycle, and back to $u$ gives an irreducible
cycle beginning and ending at $u$.  Similarly there is such a cycle about
$v$.  The cycle about $u$, followed by $e$, followed by the $w$ cycle, and
back through $e$ (in the other direction), gives an irreducible cycle
containing $e$.
\proofbox

A morphism of tangles is a morphism of $\Pi$-labelled graphs, i.e., a
graph morphism that preserves
the edge labelling.

\begin{definition}\label{de:tangle_order}
The {\em order} of a tangle, $\tang$, is 
${\rm ord}(\tang)=|E_\tang|-|V_\tang|$ (so while a whole-loop is counted as one
edge, for the model $\cjnd$, to
be considered soon, a half-loop is also counted as one edge\footnote{
  Rougly speaking, 
  the reason for this is that a whole-loop and half-loop are both
  $1/n+O(1/n^2)$ probability events.}).  More generally,
for a graph, $G$, or any structure with an underlying graph, $G$ (such as
a ``form'' or ``type'' to be defined in Section 5), its order is
${\rm ord}(G)=|E_G|-|V_G|$.
\end{definition}

\begin{theorem}\label{th:tangle_count}
Let $\tang$ be a tangle of non-negative order.
Then the expected number of occurrences
of $\tang$ in an element, $G$, of $\cgnd$
is $n^{-r}+O(n^{-r-1})$, where $r$ is the order of $\tang$.  
The probability that at least
one occurrence occurs is at least
$n^{-r}/c - n^{-2r}/(2c^2)+O(n^{-r-1})$, with $r$ as
before and where
$c$ is the number of automorphisms of the complete pruning of $\tang$.
\end{theorem}
Notice that the number of automorphisms of the complete pruning of a tangle,
$\tang$, is at least as many as that of $\tang$, and it may be strictly
greater\footnote{Indeed, a structural induction argument (i.e., by
pruning one leaf) shows that any
automorphism of the complete pruning of $\tang$
has at most one extension to $\tang$.
On the other hand, if $\tang$ is a cycle of length $q$ with all edges in one
``direction'' labelled $\pi_1$, then $\tang$ has $q$ automorphisms; yet if
we add one edge labelled $\pi_2$
to $\tang$ at any vertex, the new graph
has only the trivial automorphism.}.
The proof following will imply that the probability of $\tang$'s
occurrence is also
at most $n^{-r}/c+O(n^{-r-1})$, which matches the lower bound to first order,
provided that $r\ge 1$.
\proof 
Let $V_\tang=\{u_1,\ldots,u_s\}$, and for a tuple $\vec m=(m_1,\ldots,m_s)$
of distinct integers between $1$ and $n$, let $\iota_{\vec m}$ denote the event
that the map, $\iota$, mapping $u_i$ to $m_i$, is an occurrence of $\tang$
in $G$.
If $a_i$ is the number of $\tang$'s edges labelled $\pi_i$, then
each event $\iota_{\vec m}$ involves
setting $a_i$ values of
$\pi_i$, all of which occur
with probability
\begin{equation}\label{eq:tangle_expecteds}
\frac{(n-a_1)!}{n!}\cdots \frac{(n-a_{d/2})!}{n!}.
\end{equation}
Since the sum of the $a_i$ is $|E_\tang|$, this probability is 
$n^{-|E_\tang|}$.  
Since there are $n!/(n-|V_\tang|)!=n^{|V_\tang|}+O(n^{|V_\tang|-1})$ different 
$\iota_{\vec m}$'s,
the expected number of occurrences is $n^{-r}+O(n^{-r-1})$, where
$r={\rm ord}(\tang)$.

Next notice that if $\tang'$ is a pruning of a tangle, $\tang$, then
the probability that $\tang$ occurs is $1+O(n^{-1})$ times the probability
that $\tang'$ occurs (adding each pruned edge adds a condition that occurs
with probability between $1$ and $(n-c_1)/(n-c_2)$ with $c_1,c_2$
constants).  Hence we may assume $\tang$ is pruned.

An automorphism of $\tang$ can be viewed as a permutation on
$V_\tang$, which is the same as a permutation, $\sigma$, on
$\{1,\ldots,s\}$ (identifying a $u_i\in V_\tang$ with $i$).
Such a permutation, $\sigma$, acts by permuting the components
of the $\vec m$'s.  Say that $\iota_{\vec m}$ is equivalent to
$\iota_{\vec k}$ if $\vec m$ and $\vec k$ differ by a permutation,
$\sigma$, associated to an automorphism of $\tang$; i.e., if
$\iota_{\vec m}$ and $\iota_{\vec k}$ correspond to the same subgraph
of $G$.  Let $R$ be a set
of representatives in the equivalence classes of all $\vec m$'s.

By inclusion/exclusion, the probability that $\tang$ occurs at least once is
at least
\begin{equation}\label{eq:incexc}
\sum_{{\vec m}\in R} \prob{\iota_{\vec m}} - 
\frac{1}{2}\sum_{\substack{ {\vec k}\ne {\vec m} \\ {\vec k},{\vec m}\in R}}
\prob{\iota_{\vec m}\cap\iota_{\vec k}}.
\end{equation}
The first summand is $(1/c)n^{-r}+O(n^{-r-1})$, by the argument given
for the expected number.  For the second summand, we may write
$\iota_{\vec m}\cap\iota_{\vec k}$ as $\iota_{\vec q}(\tang')$, where
$\vec q$ is a vector comprised of the
distinct components of $\vec m$ and $\vec k$, and where $\tang'$ is the tangle
obtained by gluing two copies of $\tang$ along certain vertices (corresponding
to where the components of $\vec m$ and $\vec k$ coincide).  If 
$\vec m$ is disjoint from (i.e., nowhere coincides with)
$\vec k$, then $\tang'$ is two disjoint
copies of $\tang$; for fixed $\vec k\in R$ we have
$$
\sum_{\substack{ {\vec m}\in R\\ {\vec m}\;{\rm disjoint\;from\;}\vec k}}
\prob{\iota_{\vec m}\;|\;\iota_{\vec k}} = n^{-r}/c + O(n^{-r-1}),
$$
the summation being over conditional probabilities, since the conditioning
of $\iota_{\vec k}$ and summing over $\vec m$ disjoint only affects
equation~(\ref{eq:tangle_expecteds}) by changing $n!/(n-a_i)!$ terms
into $(n-c_i)!/(n-c_i-a_i)!$ terms for constants $c_i$, which is a
second order change.
Hence
$$
\sum_{\substack{{\vec k},{\vec m}\in R \\ {\vec k},{\vec m}\;{\rm disjoint}}}
\prob{\iota_{\vec m}\cap\iota_{\vec k}}
=
\sum_{ \vec k\in R} \prob{\iota_{\vec k}}
\sum_{\substack{ {\vec m}\in R\\ {\vec m}\;{\rm disjoint\;from\;}\vec k}}
\prob{\iota_{\vec m}\;|\;\iota_{\vec k}}
$$
$$
= \sum_{ \vec k\in R} \prob{\iota_{\vec k}} \bigl( n^{-r}/c + O(n^{-r-1})
\bigr)
= n^{-2r}/c^2 + O(n^{-2r-1}).
$$
To understand the situation where $\vec m$ and $\vec k$ overlap somewhere,
we pause for some lemmas.

\begin{lemma}\label{lm:inclusion_order}
Let $\iota\from \tang\to G$ be an inclusion of graphs, with
$G$ connected.
Then the order of $G$ is at least that of $\tang$.
\end{lemma}
\proof Let $G_\tang$ be $G$ with the vertices of $\iota(\tang)$ identified, and
all $\iota(\tang)$ edges discarded.  Then $G_\tang$ is connected, and so has
order at least
$-1$; on the other hand, clearly the order of $G_\tang$ is the
order of $G$ minus that of $\tang$ minus $1$ (for the vertex that is the
identification of all $\iota(\tang)$ vertices).
Hence
$$
{\rm ord}(G) = {\rm ord}(G_\tang) + {\rm ord}(\tang) +1 \ge
-1+{\rm ord}(\tang)+1={\rm ord}(\tang).
$$
\proofbox

\begin{lemma}\label{lm:edge_removal}
Let $G$ be a pruned graph and let $e\in E_G$.  If $G\setminus\{e\}$
(i.e., $G$ with $e$ removed) has two connected components, then each
connected component has order at least $0$.
\end{lemma}
\proof
Consider a connected component, $G'$, of $G\setminus\{e\}$.  If $G'$ did not
contain a cylce, then $G'$ would be a tree, and 
then $G$ would not be completely pruned.
So $G'$ contains a cycle, and we may apply Lemma~\ref{lm:inclusion_order}
to deduce that $G'$ has order at least $0$
\proofbox

\begin{lemma}
\label{lm:order_increases}
Let $\iota\from \tang\to G$ be an inclusion of pruned graphs.
Then the order of $G$ is at least that of $\tang$, and $G$'s order is strictly
greater than $\tang$'s if $\iota(\tang)$ is properly
contained in $G$.
\end{lemma}
\proof A connected component of a graph that is pruned (and non-empty)
has non-negative order.  So we may assume $\iota(\tang)$ meets every connected
component of $G$.  It suffices to prove the case where $\iota(\tang)$ meets
one connected component of $G$, i.e., the case where $G$ is connected.
Choosing an edge, $e$, that is missed by $\iota(\tang)$, we have that
$\iota$ includes $\tang$ into $G\setminus\{e\}$ (i.e., $G$ with
$e$ removed); we apply Lemma~\ref{lm:inclusion_order} to those components
of $G\setminus\{e\}$ containing part of $\iota(\tang)$, and to the 
possibly one other component of $G\setminus\{e\}$ we apply
Lemma~\ref{lm:edge_removal}.
It follows that the order of $\tang$ is at most that of 
$G\setminus\{e\}$; but the order of $G\setminus\{e\}$ is one less than
that of $G$.
\proofbox

\begin{lemma}\label{lm:order_greater}
Let $\iota_1,\iota_2\from \tang\to H$ be two inclusions
of a tangle, $\tang$, in a connected (labelled) graph,
$H$, such that $\iota_1(\tang)\cup\iota_2(\tang)=H$.
Assume that $\iota_1(\tang)\ne H$.  Then the order of $H$ is greater
than that of $\tang$.
\end{lemma}
\proof Ignoring $\iota_2$, the preceeding lemma applies to $\iota_1$
to immediately yield this lemma.
\proofbox

Lemma~\ref{lm:order_greater} shows that
$$
\sum_{\substack{ {\vec k},{\vec m}\;{\rm not\;disjoint}
\\ {\vec k},{\vec m}\in R,\;\;{\vec k}\ne{\vec m}}}
\prob{\iota_{\vec m}\cap\iota_{\vec k}} = O(n^{-r-1}),
$$
since the summation can be broken down into a finite number of sums
over tangles of order at least $r+1$.  Thus
$$
\sum_{{\vec m}\in R} \prob{\iota_{\vec m}} - 
\frac{1}{2}\sum_{\substack{ {\vec k}\ne {\vec m} \\ {\vec k},{\vec m}\in R}}
\prob{\iota_{\vec m}\cap\iota_{\vec k}}.
$$
$$
= n^{-r}/c+O(n^{-r-1})- n^{-2r}/(2c^2) + O(n^{-2r-1})+O(n^{-r-1}),
$$
which completes the proof of Theorem~\ref{th:tangle_count}.
\proofbox

It is easy to see that the above proof of Theorem~\ref{th:tangle_count}
uses very little about the model of random graph, and therefore generalizes
as follows.
\begin{theorem} Let $\ck_n$ be a model of $d$-regular random graphs on
$n$ vertices labelled $\intn$,
defined for some values of $n$.  Further assume that
(1) $\ck_n$ is invariant under renumbering $\intn$, and (2) any
tangle, $\tang$, has expected number of occurrences $n^{-r}+O(n^{-r-1})$
where $r$ is the order of $\tang$.  Then Theorem~\ref{th:tangle_count} holds
for $\ck_n$.
\end{theorem}

\begin{theorem}\label{th:other_ss}
For $G$ drawn from $\chnd$ or
$\cind$, we have that $\lambda_2(G)>2\sqrt{d-1}$
with probability at least $n^{-s}/2+O(n^{-s-1})$ where
$s=\lfloor \sqrt{d-1}\rfloor$, except for when $d=4$ in $\chnd$, where
we may take $s=2$.  The same holds for $\cjnd$ with probability
$n^{-s}+O(n^{-s-1})$, where
$s=\lfloor \bigl(\sqrt{d-1}\;+1\bigr)/2
\rfloor$ (and no exceptional values of $d$).
\end{theorem}
\proof 
For $\cjnd$, the tangle consisting of $1$ vertex with a number of self-loops
(in this case half-loops), proves the theorem for $\cjnd$ just as it
did for $\cgnd$ in Theorem~\ref{th:improved_bound}.

Consider the tangle, $\tang$,
with two vertices and $m$ edges joining the two vertices
labelled $\pi_1,\pi_2,\ldots$; $\tang$ is an $\chnd$-tangle
provided that $m\le d/2$, and an $\cind$-tangle if $m\le d$.  
$\lambda_{\ird{}}(\tang)$
is clearly $m-1$, $s=\ord(\tang)=m-2$, and the automorphism group of
$\tang$ is of order $2$.
So if $m-1>\sqrt{d-1}$ and if $G$ contains
this tangle, then we have 
$\lambda_2(G)>2\sqrt{d-1}$ for $n$ sufficiently large.
If we take $s=\lfloor \sqrt{d-1}\rfloor$, then $m-1>\sqrt{d-1}$; we
require $m\le d$ for $\cind$, amounting to
$$
\lfloor \sqrt{d-1}\rfloor + 2 \le d,
$$
which is satisfied for all $d\ge 3$.  For $\chnd$ we require
$$
\lfloor \sqrt{d-1}\rfloor + 2 \le d/2,
$$
which is satisfied for all $d\ge 7$.  Since $d$ is even and $\ge 4$ in
$\chnd$, we finish by examining the cases $d=4$ and $d=6$.

For $d=4$ consider the tangle, $\tang$, with vertices $v_1,v_2,v_3,v_4$,
edges labelled $\pi_1,\pi_2$ from $v_1$ to $v_2$, from $v_2$ to $v_3$,
and from $v_3$ to $v_4$.
Then ${\rm ord}(\tang)=2$ and $\lambda_{\ird{}}(\tang)>
\lambda_{\ird{}}(\tang')$ where $\tang'$ is the subgraph of $\tang$
induced on $v_1,v_2,v_3$.
But in the proof of Theorem~\ref{th:taufund_chnd} we compute
$\lambda_{\ird{}}(\tang')=\sqrt{3}$.  Hence $\tang$ is hypercritical
of order $2$, so we may take $s=2$ in the theorem in the case of $d=4$
and $\chnd$.

For $d=6$, the proof of Theorem~\ref{th:taufund_chnd} gives a tangle
of order $2$ with $\lambda_{\ird{}}>\sqrt{5}$
(see equation~(\ref{eq:d6tangle}) and the discussion around it).  
So we may take
$s=2$ in our theorem when $d=6$ in $\chnd$.

\proofbox

\section{Walk Sums and New Types}
\label{se:newtypes}

\begin{figure}
    \centering
        \epsfysize=2.5in
        \epsfbox[0 0 389 426]{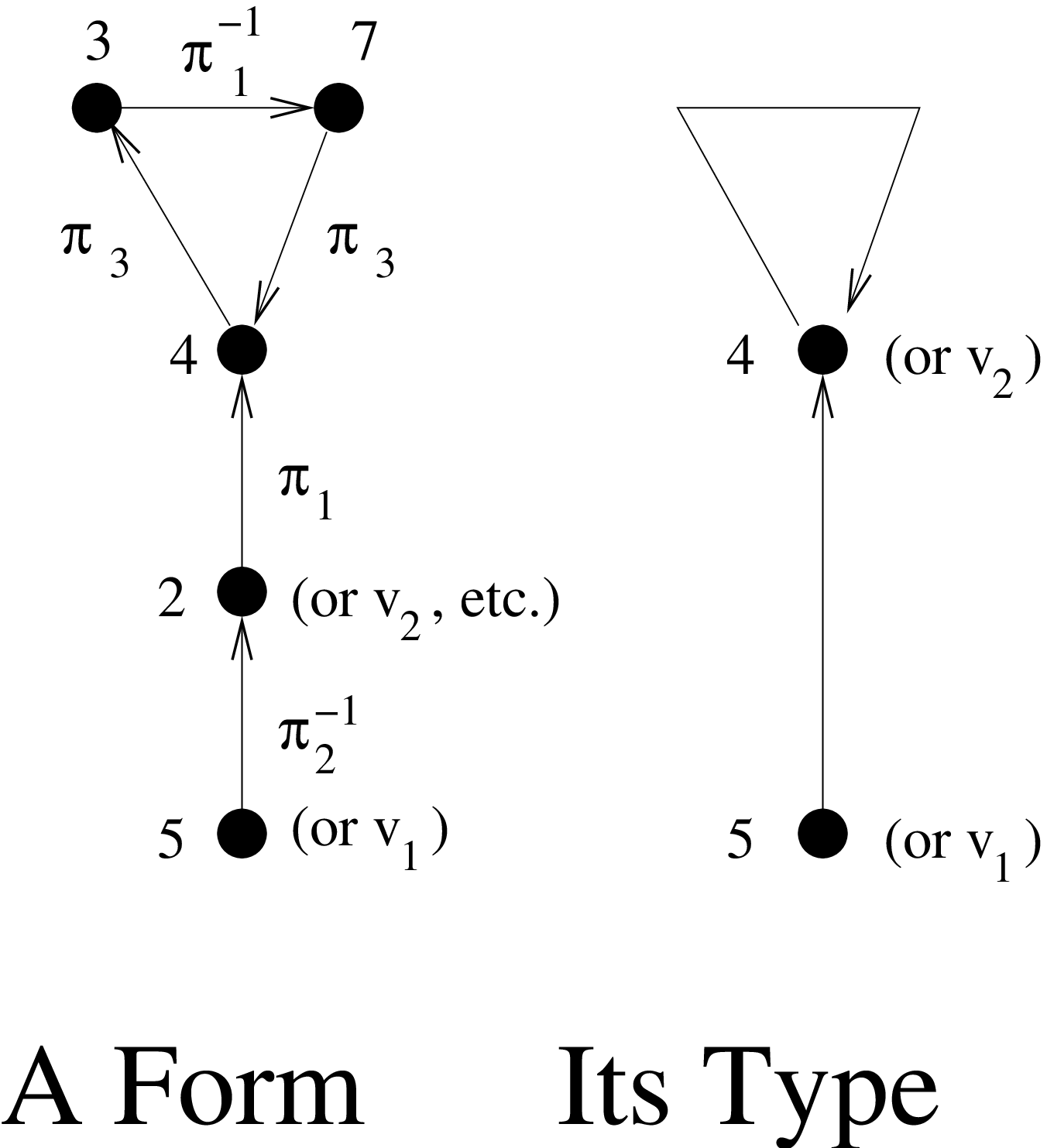}
    \caption{A form and its associated type.}
    \label{fg:form1}
\end{figure}

In this section we give some general techniques that are used in
estimating the expected values of
all the various traces that are used in this paper.
The main idea, originated in \cite{broder} and strengthened in
\cite{friedman_random_graphs}, is to group contributions to the trace 
in the following way.
Consider the word, $w$, and vector, $t$
$$
w=(\sigma_1,\ldots,\sigma_{10})=
(\pi_2^{-1},\pi_1,\pi_3,\pi_1^{-1},\pi_3,\pi_3,\pi_1^{-1},\pi_3,\pi_1^{-1},
\pi_2),
$$
$$
\vec t = (t_0,\ldots,t_{10})=(5,2,4,3,7,4,3,7,4,2,5)
$$
(see Figure~\ref{fg:form1}).
This represents a possible or potential irreducible closed walk of length 10,
from $5$ to $2$ along the edge
labelled $\pi_2^{-1}$, from $2$ to $4$ along $\pi_1$, etc.
Graphically we depict this potential walk by the subgraph it traces out, called
its ``form'' (see Figure~\ref{fg:form1}). 
The pair $(w;\vec t\>)$ represents a possible contribution to
$\irdtr{A}{10}$.  For any word, $w$, of length $k$ and vector,
$\vec t$, of length $k+1$, let $P(w;\vec t\>)$ be the probability that
$\sigma_i(t_{i-1})=t_i$ for all $i$, i.e., that the cycle does occur.
Then the expected value of $\irdtr{A}{10}$ is just a sum of appropriate
$P(w;\vec t\>)$'s, what we will call a ``walk sum.''

In the form of Figure~\ref{fg:form1}, we see that the numbers of the 
vertices $5,2,4,3,7$, are irrelevant, since our random graph model is
``symmetric,'' or invariant under renumbering the vertices.  So we
may replace the vertex $5$ by an abstract symbol $v_1$, $2$ by $v_2$,
etc. (for reasons to made clear later, we do want to remember the order
in which the vertices were traversed); this replacement
is not necessary, but
makes clear that there is no particular preference for any numbering
of the vertices.

In our example of Figure~\ref{fg:form1},
$(w;\vec t\>)$, a loop about the vertex $4$ is traversed
twice before we return to the starting vertex $2$.
Broder and Shamir realized that the other vertices, those of degree $2$
that are not the starting vertex (here the vertices $2,3,7$) are less
interesting features of the ``form.'' 
By supressing these ``less interesting'' vertices we get the ``type''
of the form (see Figure~\ref{fg:form1}); in \cite{friedman_random_graphs},
walk sums were grouped by their form, and the sums for each form were
grouped by the type of the form.
A type is a graph with certain
features, but when a form gives rise to a type then the edges
of the type inherit $\Pi^+$ labels from the form, and inherit ``lengths,''
which are the lengths of the $\Pi^+$ labels (recall that $\Pi^+$ is the
set of nonempty words over $\Pi$).
For example, the edge $(5,4)$ of the type in Figure~\ref{fg:form1} inherits
a length of $2$ and a label of $\pi_2^{-1}\pi_1$ from the form next to it;
edge $(4,4)$ inherits a length of $3$ and a label of $\pi_3\pi_1^{-1}\pi_3$.

This paper introduces a ``new type,'' used in analyzing walk sums 
corresponding to selective traces.  A new type is a type with
some additional information, primarily fixing the lengths of certain
type edges and requiring all other lengths to be sufficiently large.

\subsection{Walk sums}

By a {\em weakly potential $(k,n)$-walk}, $(w;\vec t\>)$, we mean a pair 
consisting of
a word $w=\sigma_1\ldots\sigma_k$ of length $k$
over $\Pi$,  and a vector,
$\vec t=(t_0,\ldots,t_k)$, with each $t_i\in\intn$; we sometimes refer
to $k$ as the {\em length} and $n$ as the {\em size} of the weakly potential
walk.  Given such a
$(w;\vec t\>)$, let
$\ce(w;\vec t\,)$ denote the event that the $\pi_i$ are chosen so that
$\sigma_i(t_{i-1})=t_i$
for all $i=1,\ldots,k$.
Let $P(w;\vec t\>)$ denote the probability that $\ce(w;\vec t\>)$ occurs.

A {\em potential $(k,n)$-walk} is a weakly potential $(k,n)$-walk that
is {\em feasible}, meaning
that $P(w;\vec t\>)\ne 0$, or equivalently that the 
following two feasibility
conditions hold: (1) for any $i,j$ such that $\sigma_i=\sigma_j$, we have
$$
t_{i-1}=t_{j-1} \Leftrightarrow  t_i=t_j,
$$
and for any $i,j$ such that $\sigma_i=\sigma_j^{-1}$, we have
$$
t_{i-1}=t_{j} \Leftrightarrow  t_i=t_{j-1}.
$$
For example, if $\vec t=(1,1,2)$ and $w=\pi_1\pi_1$, then
$(w,\vec t\>)$ is a weakly potential
walk but not a potential walk.
A {\em potential $(k,n)$-closed walk} is a potential walk, $(w,\vec t\>)$ as 
above with $t_k=t_0$.

All our variants of traces, irreducible and not irreducible, selective
and not, can be viewed as sums of $P(w;\vec t\>)$ over appropriate $w$'s
and $\vec t\;$'s.  In this subsection we formalize this notion and make
preliminary remarks about
such sums and asymptotic expansions.

\begin{definition} A {\em walk collection, $\cw=\cw(k,n)$,} is a collection,
for any two positive integers $k$ and $n$,
of $(k,n)$-potential walks $(w;\vec t\>)$ as above, 
i.e. $w$ is a word over $\Pi$ of length
$k$, and $\vec t=(t_0,\ldots,t_k)$ is a $k+1$ dimensional vector over $\intn$,
and $(w;\vec t\>)$ is feasible.
The {\em walk sum} associated to $\cw$ is
$$
\walksum{\cw}{k,n}=\sum_{(w;\vec t\>)\in\cw(k,n)} P(w;\vec t\>).
$$
\end{definition}

The main goal of this paper is to organize the various $(w,\vec t\>)$ pairs
into groups over which we can easily sum $P(w;\vec t\>)$.  One simple
organizational remark is that symmetries in the $(w;\vec t\>)$ pairs
often simplify matters.  Specifically, given an permutation, $\tau$,
of the integers, and a vector $\vec t$ as above, let
$$
\tau(\vec t\>) = \bigl(\tau(t_0),\ldots,\tau(t_k)\bigr).
$$
Say that $\vec s$ and $\vec t$ {\em differ by a symmetry} if $\vec s=\tau
(\vec t\>)$ for some $\tau$; in this case clearly $P(w,\vec s\>)=P(w,\vec t\;)$
for any word $w$ of length $k$.  We use $\vec s\sim\vec t$ to denote
that $\vec s$ and $\vec t$ differ by a symmetry.

\begin{definition} A walk collection, $\cw$, is {\em SSIIC} if it is
\begin{enumerate}
\item {\em symmetric}, i.e. $(w,\vec t\>)\in\cw(k,n)$ implies that
$(w,\vec s\>)\in\cw(k,n)$ for all $\vec s\sim\vec t$ such that
$s_i\le n$ for all $i$,
\item {\em size invariant}, i.e. if $(w,\vec t\>)$ is a potential
$(k,n)$-walk, then for any $n'>n$, $(w,\vec t\>)\in\cw(k,n)$ iff
$(w,\vec t\>)\in\cw(k,n')$,
\item {\em irreducible}, meaning that $(w,\vec t\>)\in\cw$ implies that
$w$ is irreducible, and
\item {\em closed}, meaning that $(w,\vec t\>)\in\cw$ implies that
$t_0=t_k$.
\end{enumerate}
\end{definition}
The walk sums of interest here, namely traces that are irreducible
or strongly irreducible and possibly selective,
will all be SSIIC.
We now make a series of remarks about walk sums that apply to all
SSIIC walk sums; some of the remarks apply more generally.

\begin{definition}
Let $\vec t$ be as above, i.e., a positive integer valued vector.
Define the {\em equivalence class of $\vec t$} 
to be
$$
[\vec t\>] = \{ \vec s \mid \vec s\sim\vec t\},
$$
i.e., the set of all positive integer values vectors differing from
$\vec t$ by a symmetry.
Define the {\em $n$-th equivalence class of $\vec t$}
$$
[\vec t\>]_n = \{ \vec s \mid \vec s\sim\vec t \quad\mbox{and all}\quad
s_i\le n\}
$$
(we may omit the $n$ subscript if $n$ is understood).
\end{definition}
Let $n$ be fixed, and set
$$
\Esymm(w;\vec t\>) = \Esymm(w;\vec t\>)_n
= \sum_{\vec s\in[\vec t\>]_n} P(w;\vec s\>)
$$
$$
= n(n-1)\cdots(n-v+1)P(w;\vec t\>),
$$
where $v$ is the number of distinct elements of $\vec t$.  A symmetric
walk sum is just the sum of certain $\Esymm(w,\vec t\>)$'s, and we can
write
$$
\walksum{\cw}{k,n}=\sum_{(w;[\vec t\>])\in\cw(k,n)}\Esymm(w;\vec t\>),
$$
where 
the summation over $(w;[\vec t\>])\in\cw(k,n)$ means that
we sum over one $\vec t$
in each equivalence class and over all $w$.

Each $\vec t=(t_0,\ldots,t_k)$
has an $\vec s\sim\vec t$ where the size of $\vec s\;$'s
entries are at most $k+1$.  So for each $k$ there are a finite number
of equivalence classes, $\cw(k)$, of $(w;\vec t\>)$ such that $w$ is
of length $k$ and $\vec t$ is of some finite size.  We refer to
$\cw(k)$ as the set of {\em potential walk classes} of length $k$
(or {\em potential closed walk classes} when we restrict to those $\vec t\;$'s
with $t_k=t_0$).  So if $\cw$ is
size invariant we may write
$$
\walksum{\cw}{k,n}=\sum_{(w;[\vec t\>])\in\cw(k)}\Esymm(w;\vec t\>)_n,
$$
where the right-hand-side summation has a fixed, finite number of summands
independent of $n$
(for fixed $k$).

Our next step is to comment about $\Esymm(w,\vec t\>)_n$.  Notice that if
the conditions $\sigma_i(t_{i-1})=t_i$ involve determining 
$a_j$\myindexaj values
of $\pi_j$, then
\begin{equation}\label{eq:probability}
P(w;\vec t\>) = \prod_{i=1}^{d/2} \frac{1}{n(n-1)\cdots(n-a_j+1)}
=\prod_{i=1}^{d/2}\frac{(n-a_j)!}{n!}.
\end{equation}
Let $e=a_1+\cdots+a_{d/2}$.
We have
$$
\Esymm(w,\vec t\>)_n = n(n-1)\ldots (n-v+1) \prod_{i=1}^{d/2}
\frac{1}{n(n-1)\cdots(n-a_j+1)}.
$$
Notice that in the power series expansions about $x=0$ of
$$
(1-x)(1-2x)\ldots(1-mx) \quad\mbox{and}\quad
(1-x)^{-1}(1-2x)^{-1}\ldots(1-mx)^{-1},
$$
the $x^i$ coefficient
is a polynomial (of degree at most $2i$) in $m$.
It follows that 
there exist polynomials $p_0,p_1,\ldots$ in the
variables $a_1,\ldots,a_{d/2},v$ such that
\begin{equation}\label{eq:expansion_polynomials}
\Esymm(w,\vec t\>)_n = n^{v-e} \sum_{i=0}^\infty n^{-i}
p_i(a_1,\ldots,a_{d/2},v)
\end{equation}
for $n$ sufficiently large.
\begin{definition}
\label{de:expansion_polynomials}
We define the {\em expansion polynomials}\mytwoindex{expansion polynomials}{The
polynomials $p_i=p_i(a_1,\ldots,a_{d/2},v)$
giving the $1/n$ expansion of $\Esymm(w,\vec t\>)$},
$$
p_i=p_i(a_1,\ldots,a_{d/2},v)
\mythreeindex{pi}{$p_i=
p_i(a_1,\ldots,a_{d/2},v)$}{the expansion polynomials},
$$
to be the polynomials that give the
expansion in equation~(\ref{eq:expansion_polynomials}).
Throughout the paper, $a_i$\myindexaj 
denotes the number of $\pi_i$ values determined
by the relevant structure (in this case the potential walk,
$(w,\vec t\>)$).
\end{definition}
We remark that since $e=a_1+\cdots+a_{d/2}$, we have that $v$ is determined
from the $a_i$ if $v-e$ is fixed and known; in such situations, the
expansion polynomials may be regarded as functions of the $a_i$ alone.

\begin{theorem}\label{th:exp_polys}
For any $w,\vec t$ and any integer $q\ge 0$ we have
\begin{equation}\label{eq:simple_expansion}
\Esymm(w,\vec t\>)_n = n^{v-e} \biggl( p_0+\frac{p_1}{n}+\cdots+\frac{p_q}{n^q}
+ \frac{{\rm error}}{n^{q+1}} \biggr)
\end{equation}
where
$$
|{\rm error}| \le \exp\bigl((q+1)k/(n-k)\bigr)\; k^{2q+2},
$$
and the $p_i$ are the expansion polynomials.
\end{theorem}
The proof is contained between
Lemma~2.7 and Corollary~2.10 of \cite{friedman_random_graphs}, although
the proof there has a minor error.  We will correct it here and review 
the entire proof,
since we will later 
need variants of this theorem for $\chnd,\cind,\cjnd$.
If
\begin{equation}\label{eq:g_rational}
g(x)=(1-b_1x)\cdots(1-b_rx)(1-c_1x)^{-1}\cdots(1-c_sx)^{-1},
\end{equation}
where the $b_i$ and $c_j$ are positive constants,
then $g$'s $i$-th derivative satisfies the bound
$$
|g^{(i)}(x)|/i!\le (1-xc_{\max})^{-i}\left(\sum b_j+\sum c_j\right)^i
$$
where $c_{\max}$ is the maximum of the $c_j$
(by equation~(6) in \cite{friedman_random_graphs} on page~339).
This estimate, using Taylor's theorem, expanding in $x=1/n$ about $x=0$,
gives the error term for Theorem~\ref{th:exp_polys} of
$$
(1-\zeta c_{\max})^{-q-1} \left(\sum b_j+\sum c_j\right)^{q+1},
$$
for some $\zeta\in[0,1/n]$.
Since
$$
-\log(1-\zeta c_{\max})=(\zeta c_{\max})+(\zeta c_{\max})^2/2+\cdots
\le \zeta c_{\max}/(1-\zeta c_{\max})
$$
$$
\le (k/n)/\bigl(1-(k/n)\bigr) = k/(n-k),
$$
we conclude that the error term is at most
\begin{equation}\label{eq:error_term}
\exp\bigl( (q+1)k/(n-k) \bigr) \left(\sum b_j+\sum c_j\right)^{q+1}.
\end{equation}
In the case of Theorem~\ref{th:exp_polys}, the $\sum b_j$ represents
the sum of $0,1,\ldots,v-1$, which is at most $\binom{k}{2}$, and
the $\sum c_j$ represents the sum over $j$ of all sums of $0,1,\ldots,a_j-1$,
which is at most $\binom{k}{2}$.  Since $2\binom{k}{2}\le k^2$, we
get an error term at most $\exp\bigl((q+1)k/(n-k)\bigr)k^{2q+2}$.
(See \cite{friedman_random_graphs} for more details.)
\proofbox

\begin{definition} Given a pair, $(w,\vec t\>)$, as above, its {\em order}
is $e-v$, with $e,v$ as above.
\end{definition}

\begin{lemma}\label{lm:big_order}
Given a word, $w$, of length $k$ over $\Pi$, we have
$$
\sum_{\vec t\;{\rm such\;that}\;(w,\vec t\>)\;{\rm is\;of\;order} \; \ge r} 
P(w,\vec t\>)
\le n\binom{k}{r+1}\left(\frac{k}{n-k}\right)^{r+1},
$$
which for $k\le n/2$ is at most $ck^{2r+2}n^{-r}$ for some constant $c$
depending only on $r$.
\end{lemma}
\proof 
This can be found in
\cite{friedman_random_graphs}, second displayed equation and
discussion before on page 352; this is the same idea used in the $r=1$
case proven in \cite{broder}.  (The extra factor of $n$ appears here
but not in \cite{friedman_random_graphs}, since
we do not fix the initial vertex of the walk.  Also note that the ``order,''
used here, is one less than the ``number of coincidences,''
used in \cite{friedman_random_graphs}.)  For the ease of reading, we shall
discuss these ideas in our notation.

We may evaluate $P(w,\vec t\>)$ by considering the steps of the walk
$w=\sigma_1\ldots\sigma_k$ one by one; i.e., we fix a $v_0\in\intn$
and consider the random walk $v_1=\sigma_1(v_0)$, $v_2=\sigma_2(v_1)$,
etc.  We inductively consider the probability that $(v_0,\ldots,v_s)$
is equivalent to $(t_0,\ldots,t_s)$ for $s=1,\ldots,k$.
Assuming $(v_0,\ldots,v_s)$
is equivalent to $(t_0,\ldots,t_s)$, and assuming we wish to have
$(v_0,\ldots,v_{s+1})$
equivalent to $(t_0,\ldots,t_{s+1})$,
one can divide the choice and outcome
of the random variable $v_{s+1}=\sigma_{s+1}(v_s)$ into three cases:
(1) $v_{s+1}$ has already been determined by previous information
(i.e., the value $\sigma_{s+1}(v_s)$ has been determined in a previous
step); 
(2) $v_{s+1}$ has not been determined and
(2a) $v_{s+1}$ must occur as one of the $v_0,\ldots,v_s$
(i.e., when $t_{s+1}$ occurs previously
in $t_0,\ldots,t_s$), or (2b) $v_{s+1}$ must
must be different from $v_0,\ldots,v_s$ (i.e., when $t_{s+1}$ does
not occur previously).
We call case (1) a forced choice, case (2)
a free choice, with case (2a) a coincidence, and case (2b) a generic
choice.

For example, if $w=\pi_1\pi_1\pi_2\pi_3$ and $\vec t\> = (1,1,1,2,1)$,
then $v_1$ is a coicidence (since $t_1$ occurs previously as $t_0$ and
$\pi_1(v_0)$ has not been determined), $v_2$ is
a forced choice (since $\pi_1(v_0)=v_0=v_1$ has been determined),
$v_3$ is a generic choice ($t_3$ does not occur in $t_0,t_1,t_2$),
and $v_4$ is a coincidence.

A choice $v_{s+1}=\sigma_{s+1}(v_s)$ is deterministic if it is a forced
choice;  a choice occurs with probability $1 - O(s/n)$ if it is
a generic choice, and with probability $1/n + O(s/n^2)$ if it is a 
coincidence.
We see that a word of order $t$ has exactly $t+1$ coincidences.
If follows that each word of order at least $r$ and length $k$ has some
$r+1$ positions of $k$ that can be marked as coincidences.  Fixing a 
marking, each coincidence of the marking occurs with probability
at most $k/(n-k)$ (since $v_{s+1}$ must assume one of the at most
$s+1\le k$ values
$v_0,\ldots,v_s$, and at most $k$ values of the permutation $\sigma_{s+1}$
could have been determined up to that point).  
Finally $v_0$ can be chosen in $n$ ways.  We conclude
the lemma.
\proofbox

\begin{lemma}\label{lm:equiv_classes}
For any irreducible word, $w$, over $\Pi$, of length $k$,
there are at most 
\begin{equation}\label{eq:equiv_classes}
\sum_{j=0}^r \binom{k}{j} k^j \le c k^{2r}
\end{equation}
equivalence classes $[\vec t\>]$ whose order with $w$ is $\le r-1$.
\end{lemma}
\proof See the third displayed equation of page 352 of
\cite{friedman_random_graphs} and the discussion preceding.  For
completeness and ease of reading, we repeat the argument here.

If $w$ has order $j-1$, then it has $j$ coincidences occurring in $k$
places.  Again, in each coincidence, $v_{s+1}=\sigma_{s+1}(v_s)$,
$v_{s+1}$ is being chosen from at most $k$ values.  This gives the
left-hand-side of equation~(\ref{eq:equiv_classes}).  For the
right-hand-side we notice that
$$
\sum_{j=0}^r \binom{k}{j} k^j \le \sum_{j=0}^r k^{2j}\le
k^{2r}(1+k^{-2}+k^{-4}+\cdots) \le k^{2r} (4/3)
$$
for $k\ge 2$.
\proofbox

\begin{theorem}\label{th:SSIICexpansion}
Let $\cw$ be SSIIC and let $r\ge 1$.  Then for all $k\le n/2$ we have
$$
\walksum{\cw}{k,n} =   f_0(k)+\frac{  f_1(k)}{n}+\cdots+
\frac{  f_{r-1}(k)}{n^{r-1}}+\frac{{\rm error}}{n^r},
$$
where 
$$
f_i(k) = \sum_{j=0}^{r-1}\;\; \sum_{(w;[\vec t\>]){\rm \;order\;}j,\in\cw(k)}
p_{i-j}(w;[\vec t\>])
$$
(with $p_i$ the expansion polynomials, which can be viewed as a
function of $(w;[\vec t\>])$ since $(w;[\vec t\>])$ determines the
$a_i$ and $v$)
and where for some $c$ depending only on $r$,
$$
|{\rm error}| \le ck^{4r}(d-1)^k.
$$
\end{theorem}
\proof By Lemma~\ref{lm:big_order}, we introduce an error of at most
$ck^{2r+2}n^{-r}$ per word by ignoring potential walks of order at least
$r$.
Each word, $w$, has at most $ck^{2r}$ associated potential walk classes
of order at most $r-1$
(by Lemma~\ref{lm:equiv_classes}), and truncating the associated 
asymptotic expansion, as in equation~(\ref{eq:simple_expansion}),
of each associated potential walk class
results in an error of at most $ck^{2r}$
(by Theorem~\ref{th:exp_polys}).
So each word, $w$, of length $k$ involved in $\cw$
contributes an error of at most
$ck^{4r}$, and there are at most $d(d-1)^{k-1}$ such words
(for $k\ge 1$) since $\cw$ consists of only irreducible words.
\proofbox

\subsection{The Loop}

Here we analyze walk sums associated with simple loops.  This gives
some ideas and a lemma to be used in Section 8.

\subsubsection{The Singly Traversed Simple Loop}
\label{sb:stsl}
Let $w$ be an irreducible
word over $\Pi$ of length $k$ and $\vec t$ a $k+1$ tuple
over $\intn$ as in the previous subsection.
If $t_0=t_k$ and the $t_i$'s are otherwise
distinct, we say $(w;\vec t\>)$
is a {\em singly traversed simple loop}
or {\em STSL} for short; we define $\cwstsl$ to be the collection of all
STSL's.
When $\ce(w;\vec t\>)$ occurs and $(w;\vec t\>)\in\cwstsl$,
the closed walk
from $t_0$ following $w$ traces
out a ``simple loop'' once,
that begins and ends at $t_0$, moving through distinct edges
and vertices throughout the closed walk.

Clearly $\cwstsl$ is SSIIC, so according to Theorem~\ref{th:SSIICexpansion}
we have an asymptotic expansion in $1/n$ with coefficients $f_i(k)$
for the associated walk sum.  We now briefly indicate why the
$f_i$ are {\dtreelike}.  This is a mildly tedious
exercise, covered (in much greater generality)
in \cite{friedman_random_graphs}.  We quote the main points
here.

First, note that there is exactly one equivalence class $[\vec t\>]$
of $\vec t\;$'s that appear in $\cwstsl$.  So we may write $\Esymm(w)$
for $\Esymm(w;\vec t)$ for any $\vec t$ of the equivalence class, and
we may write $p_i(w)$ for the $p_i(w;\vec t\>)$ in
Theorem~\ref{th:SSIICexpansion} or \ref{th:exp_polys}
(note also that $v=k$ in the notation
of Theorem~\ref{th:exp_polys}).

Let for $\sigma,\tau\in\Pi$, let
$\ird{k,\sigma,\tau}$ denote the irreducible words of length $k$
beginning with $\sigma$ and ending with $\tau$.
For $w\in\Pi^k$ let $a_i(w)$\myindexaj denote the number of $\pi_i$ and 
$\pi_i^{-1}$ occurring in $w$.  Since an STSL has no forced choices,
the definition of $a_i$ here agrees with that in
Definition~\ref{de:expansion_polynomials}.

\begin{lemma}\label{lm:irdeigens}
Let $p=p(a_1,\ldots,a_{d/2},k)$ be a polynomial.  
For every $\sigma,\tau$ there are polynomials $Q_1,Q_2,Q_3$ of $k$
of degree at most the degree of $p$ such that
$$
\sum_{w\in\ird{k,\sigma,\tau}} p\bigl( a_1(w),\ldots,a_{d/2}(w),k\bigr)
= (d-1)^k Q_1(k)+(-1)^k Q_2(k) + Q_3(k).
$$
\end{lemma}
\proof This is immediate from Lemma 2.11 of \cite{friedman_random_graphs} or
Corollary 2.12 (note that
our $d$ is $2d$ in \cite{friedman_random_graphs}).
\proofbox
Note that the formula for the above lemma comes about from the fact that
$d-1,-1,1$ are the eigenvalues of $G_{\ird{}}$ where $G$ is the graph with
one vertex and $d/2$ whole-loops (with $G_{\ird{}}$ defined
as in Subsection~\ref{sb:lambda_irred}).

\begin{corollary}
\label{cr:stsl_dtreelike}
We have each $f_i(k)$ as in Theorem~\ref{th:SSIICexpansion}
for ${\cal W}=\cwstsl$
is {\dtreelike}.
\end{corollary}
\proof 
Recall that an $(w,\vec t\>)\in\cwstsl$ is of order $0$, and there is only
one equivalence class of $\vec t$ for STSL's.
So we have
$$
f_i(k)=\sum_{\sigma\ne\tau^{-1}} \sum_{\ird{k,\sigma,\tau}}
p_i\bigl( a_1(w),\ldots,a_{d/2}(w),k\bigr),
$$
where the $p_i$ are the expansion polynomials.
The result now follows from Lemma~\ref{lm:irdeigens}, with error term
bounded by a polynomial in $k$.

\subsubsection{Simple Loops}

Consider any irreducible $(w,\vec t\>)$ that traces out a simple loop,
i.e., the vertices and edges visited form one closed walk, but now we don't
require that the loop is traversed only once.
Corresponding to this geometric picture of a simple loop we can form
the associated walk collection of {\em simple loop closed walks}; such closed walks,
being irreducible, must traverse the loop traced out some number of
times.

So let $\cwsl$ be the set of $(w,\vec t\>)$ pairs with 
\begin{enumerate}
\item $w=\sigma_1\raise1.6pt\hbox{\ldots}\sigma_k$ irreducible,
\item $\sigma_1\ne\sigma_k^{-1}$,
\item $t_0,\ldots,t_{r-1}$
distinct for some $r$ dividing $s$,
\item $t_{i+r}=t_i$ for $0\le i\le k-r$, and 
\item $w=u^s$ for some word $u$
with $rs=k$.
\end{enumerate}
$\cwsl$ is the walk collection of simple loop walks.

Clearly we have
\begin{equation}\label{eq:multiply_traversed}
\walksum{\cwsl}{k,n}=\sum_{s|k} \walksum{\cwstsl}{s,n}.
\end{equation}
We easily conclude the following theorem.
\begin{theorem} The $f_i(k)$ corresponding to $\cwsl$ are {\dtreelike},
and have the same principal term as the $f_i(k)$ corresponding to
$\cwstsl$.
\end{theorem}

\proof Consider the $f_i(k)$ corresponding to $\cwstsl$.  By 
Corollary~\ref{cr:stsl_dtreelike}, all the $f_i$ are {\dtreelike}.  By
equation~(\ref{eq:multiply_traversed}), it suffices to show that
$$
\widetilde f_i(k) = \sum_{s|k} f_i(s)
$$
are also {\dtreelike}.  Fixing an $i$ and setting 
$f_i(k)=p(k)(d-1)^k + r(k)$ as the decomposition of $f_i$ into principal
and error terms, we see
$$
\widetilde f_i(k) = p(k)(d-1)^k + \widetilde r(k),
$$
where
$$
|\widetilde r(k)| \le \left| \sum_{s\le k/2} p(s)(d-1)^s \right| +
\sum_{s\le k} cs^c (d-1)^{s/2}
$$
for some $c$ as in Definition~\ref{de:ram}.  It is clear that
the right-hand-side above is bounded by $(d-1)^{k/2}$ times a polynomial
in $k$.
\proofbox

\subsection{Forms, Types, and New Types}

In this subsection we will classify potential walks, $(w;\vec t\>)$,
or more generally potential walk classes, according to some characteristics
of the subgraph
that the walk traces out.

\begin{definition} A {\em form}\mytwoindex{form}{an oriented, $\Pi$-labelled
graph with edges and vertices numbered that is meant to represent the
graph traced out by a potential word, forgetting about the particular
choices in $\intn$ of the vertices}, $\Gamma$, is an oriented, $\Pi$-labelled
graph, $G_\Gamma=(V_\Gamma,E_\Gamma)$, with edges and vertices numbered.
A {\em specialization} of a form, $\Gamma$, is an 
injection $\iota\from V_\Gamma\to \intn$.
\end{definition}

With each potential walk, $(w;\vec t\>)$, we can associate a form, $\Gamma=
\Gamma(w;\vec t\>)$, with a 
specialization, $\iota$, as follows.  (The form, $\Gamma$, is not unique, but
is unique up to unique isomorphism, as described below; the
specialization is unique given the form.)
\begin{enumerate}
\item Set $V_\Gamma=\{v_1,\ldots,v_r\}$ to be any numbered ($v_i$ numbered
$i$)
set of size $r$, where
$r=|V_\Gamma|$ is the number of distinct elements among the $t_i$
(where $\vec t=(t_0,\ldots,t_k)$)
\item $\iota(v_i)$ is the $i$-th distinct element of the sequence
$t_0,t_1,\ldots,t_k$,
\item Set $E_\Gamma=\{e_1,\ldots,e_m\}$ to be any numbered set of size $m$
($e_i$ numbered $i$),
where $m=|E_\Gamma|$ is the number of distinct triples 
$\{ (\sigma_i,t_{i-1},t_i) \}_{i=1,2,\ldots,k}$, where we identify a
triple $(\sigma,s,t,)$ with $(\sigma^{-1},t,s)$,
\item if $(\sigma_j,t_{j-1},t_j)$ is the $r$-th distinct tuple in
$\{ (\sigma_i,t_{i-1},t_i) \}_{i=1,2,\ldots,k}$ (with the previous
identification), then $s(\ned_r)=\bigl(\sigma_j,\iota^{-1}(t_{j-1}),
\iota^{-1}(t_j)\bigr)$ defines the structural map 
(see Definition~\ref{de:structural})
of $G_\Gamma$.
\end{enumerate}
(see the example in Figure~\ref{fg:form1} explained at the beginning
of this section).

In other words, the form is the subgraph traced out by $(w,\vec t\>)$,
with some additional information (we remember the order in which
the vertices and edges are visited, and the direction each edge is
first traversed).  We say that forms $\Gamma_1$ and $\Gamma_2$ are 
{\em isomorphic} if they are isomorphic as oriented, numbered,
$\Pi$-labelled graphs.  Because of the numbering, there is at most
one isomorphism between any two forms (or a form and itself).  We
say that $(w,\vec t\>)$ is {\em of form $\Gamma$} or
{\em associated to $\Gamma$}, written
$(w,\vec t\>)\in\Gamma$, if one (or any) of the forms associated
to $(w,\vec t\>)$ is
isomorphic to $\Gamma$.  Given $(w,\vec t\>)$, there is always an
associated form, $\Gamma$, with $V_\Gamma=\{t_0,\ldots,t_k\}$ and
associated specialization, $\iota$, being the identity; however, we
usually view $V_\Gamma$ as any numbered set of the right size, since
all of our random graph models are symmetric.

If $(w,\vec t\>)\in\Gamma$, then define
$$
\E{\Gamma}_n = \Esymm(w,\vec t\>)_n
$$\myindexEGamman
which depends only on $\Gamma$, and not on the particular $(w,\vec t\>)$ to
which $\Gamma$ is associated; indeed,
\begin{equation}\label{eq:form_prob}
\E{\Gamma}_n
 = \frac{n!}{(n-v)!}\prod_{i=1}^{d/2} \frac{(n-a_i)!}{n!},
\end{equation}
where $v=|V_\Gamma|$, and $a_i$\myindexaj
is the number of edges in $\Gamma$ labelled
with $\pi_i$ and $\pi_i^{-1}$ (this is exactly the $a_i(w)$ of
equation~(\ref{eq:probability}) of any word, $w$, associated to $\Gamma$).

Hence, if $\cw$ is symmetric and size invariant, we may write
\begin{equation}\label{eq:fundamental}
\walksum{\cw}{k,n} = \sum_\Gamma W_\Gamma(\cw,k)\E{\Gamma}_n,
\end{equation}
where $W_\Gamma(\cw,k)$\myindexW
is the number of potential walk classes in $\cw(k)$
associated to $\Gamma$, and we sum over one $\Gamma$ in each isomophism
class of forms.

\begin{definition} A {\em legal walk} in a form, $\Gamma$, is a walk
starting in $\nv_1$ that visits all the vertices of $G_\Gamma$ in order
(of their numbering),
all the edges in order, and any edge is first traversed
in the direction of its orientation.  Each legal walk
of length $k$ generates a walk class in the natural way.
\end{definition}

The following easy proposition is worth stating formally; it follows
from the definitions.
\begin{proposition} $W_\Gamma(\cw,k)$ is the number of legal walks on
$\Gamma$ of length $k$.
\end{proposition}

\begin{definition} The {\em order} of a form, $\Gamma$, is $\ord(\Gamma)
=|E_\Gamma|-
|V_\Gamma|$, subject to the Definition~\ref{de:tangle_order} convention
that any self-loop, whole-loop or half-loop, counts as a single
edge.  (The order of a form
equals the order of any potential walk to which it is
associated.)
\end{definition}

The ``form'' allows us to group together potential walks that determine
the same information on the graph; this can facilitate the task of studying
a walk sum.  A futher tool is the grouping of forms together by their
``type,'' which we now briefly motivate in rough terms to be made precise.
A type arises from a form with all its maximal
beaded paths collapsed to edges (by the appropriate supression).
Each collapsed edge inherits an edge length and $\Pi^+$ labelling from
the ``form,'' but in the type we forget this data.  As long as we collect
forms of a given type by
the first and last letters of all the collapsed edge labellings, we can
apply Lemma~\ref{lm:irdeigens} at each collapsed edge to study sums 
of the expansion polynomials in the $a_i$'s (and $v$)
over forms in such collections.

Before defining a ``type,''
we recall that for a form, $\Gamma=\Gamma(w;\vec t\>)$,
with specialization, $\iota$, the potential walk $(w;\vec t\>)$ pulls
back under $\iota^{-1}$ to a walk on $G_\Gamma$;
we remark that the vertex and edge numberings
of $\Gamma$
serve to remember in which order the vertices and edges were traversed
in the walk.

\begin{definition} A {\em type}\mytwoindex{type}{a graph representing a
number of forms, where we forget certain features of the form, such as
its $\Pi$-labelling and all or almost all its degree two vertices}, 
$T$, is a connected, 
oriented graph
$G_T=(V_T,E_T)$, with vertex and edge numberings
such that all vertices except possibly the first one are of degree
at least $3$.
A {\em labelling} of a type means a $\Pi^+$-labelling (recall
$\Pi^+$ is the
set of words over $\Pi$ of length at least $1$).
\end{definition}

To each form, $\Gamma$, we associate a type, $T=T(\Gamma)$, as follows.
Let $W$ be the set of beads of $G_\Gamma$ numbered greater than $1$.
We claim $W$ cannot contain a cycle, for then this cycle would be
disconnected from the vertex numbered $1$.
So we may form the supression of $W$
in $G_\Gamma$.  
This supression inherits a vertex and edge numbering from $G_\Gamma$
(ordering a vertex before another in the supression if it is numbered
less in $G_\Gamma$, and ordering edges in the supression by any
associated edge in $E_\Gamma$).
Of course, the $\Pi$-labelling of $\Gamma$ gives rise to a $\Pi^+$-labelling
of $G_{T(\Gamma)}$.

In other words, the beads of a form are ``less important'' features,
and the type is just the form with these ``less important'' features
supressed.

A $\Pi^+$-labelled type uniquely determines a form, and vice versa.  We
wish to group together forms corresponding to one type (inducing different
$\Pi^+$-labellings), the prototypical example being STSL's or SL's discussed
earlier in this section.  To do this it will be helpful to remember
a small part of the labelling, namely the starting and ending letter of
each $\Pi^+$-label.

\begin{definition} A {\em lettering} of a type, $T$, is the assignment to
each directed edge a {\em starting letter} in $\Pi$ and an {\em ending letter}
in $\Pi$ (such that opposite directed edges are lettered with the starting
letter of one being the inverse of the ending letter of the other).  
Given a form, $\Gamma$, with $T=T(\Gamma)$, or equivalently
a $\Pi^+$-labelling of $T$, the associated lettering assigns to each edge the
starting and ending letter of the $\Pi^+$-label assigned to it in its
orientation.
\end{definition}

It turns out that the notion of a type is too coarse to attack the
Alon conjecture (but sufficient for the results of
\cite{friedman_random_graphs}).  The problem is that some types, when
their edges take on certain $\Pi^+$-labellings, contain supercritical
tangles.  When such tangles occur (which they
can when their order is roughly $O(\sqrt{d})$), we must distinguish where
there tangles occur.  
This is what a ``new type'' does, where the type edges are partitioned into
``long'' and ``fixed'' (in length)
edges, and where supercritical tangles
lie on type
edges that are ``fixed.''
We then modify our walk sums to be ``selective''
(see Section~6), a notion which requires us to know where these tangles
occur in the type.

\begin{definition} 
\label{de:new_type}
A {\em $B$-new type}\mytwoindex{new type}{a type
with some additional information specified, such as a partition of
the types edges into a ``long'' and a ``fixed'' edge set}
is a collection, $\widetilde T=
(T;\El,\Ef;{\vec k}^{\rm fixed})$,
of (1) a lettered type,
$T$, (2) a partition of $E_T$ into two sets, $\El,\Ef$, (3) for each
$\ned_i\in \Ef$ an {\em edge length}, $\kf_i$, with $0<\kf_i<B$, and
(4) a $\Pi^+$-labelling of $\Ef$ with each $\ned_i\in \Ef$ 
labelled with a word
of length $\kf_i$.
A $\Pi^+$-labelling of $T$ (or, equivalently, a form, $\Gamma$), is said to be
{\em of $B$-new type $\widetilde T$} if each $\El$ label is of length
at least $B$, and each label corresponding to $\ned_i\in \Ef$ is of length
$\kf_i$ and agrees with the label specified by $\widetilde T$.
$\widetilde T$ is said to be {\em based on $T$}.
\end{definition}

\begin{theorem} For each $r>0$ there are finitely many types (up to
isomorphism) of order
at most $r$.  For each type, $T$, and each $B>0$ there are finitely many
$B$-new types based on $T$.
\end{theorem}
\proof The first statement is just Lemma~2.3 of \cite{friedman_random_graphs},
except that ``coincidence'' is used instead of ``order'' (and the
coincidence is the order plus one).  The second statement is clear
since there are finitely many (1) letterings, (2) partitions of $E_T$,
(3) choices of $\kf_i$ with $e_i\in \Ef$, and (4) labellings of each $\Ef$
edge, $e_i$, with a length $\kf_i<B$.
\proofbox

\subsection{Motivation of Types and New Types}

So far we have defined walk sums; we have seen that symmetric, size
invariant walk sums can be organized into forms, by
equation~(\ref{eq:fundamental}); we have seen that forms can be 
grouped by type and new type.  In this section we briefly explain
how and why we use types and new types.

Fix a symmetric, size invariant walk collection, $\cw$.
Organizing forms by type, we may write equation~(\ref{eq:fundamental}) as
$$
\walksum{\cw}{k,n} = 
\sum_T
\sum_{\Gamma\in T} W_\Gamma(\cw,k)\E{\Gamma}_n,
$$
where $\Gamma\in T$ means the form $\Gamma$ is of type $T$, and the
summation in $T$ ranges over all types.  Typically we need only sum over
$T$ of at most some order, so the sum in $T$ will effectively be a finite
sum.  

So fix a $T$ and define $\E{T}_{n,k}$ to mean
$$
\E{T}_{n,k} =
\sum_{\Gamma\in T} W_\Gamma(\cw,k)\E{\Gamma}_n.
$$
Let $T$'s edges be $E_T=\{e_1,\ldots,e_b\}$.
For a vector, $\vec k = (k_1,\ldots,k_{b})$, let $T(\vec k\>)$ denote
the set of forms, $\Gamma$,
of type $T$, such that for all $i$ the length of the beaded path in
$\Gamma$ corresponding to the edge $e_i$ in $T$ has length $k_i$.
For each $e_i\in E_T$ fix an integer $m_i\ge 1$.
Let $W_\Gamma(\vec m)=W_\Gamma(m_1,\ldots,m_b)$ 
denote the number of legal $\cw$ walks
in $\Gamma$ that traverse edge $e_i$ exactly $m_i$ times.
Clearly
$$
W_\Gamma(\vec m) = W_T(\vec m),
$$
i.e., $W_\Gamma(\vec m)$ depends only on $\vec m$ and the
type, $T$, of $\Gamma$.  This allows us to write 
$$
\E{T}_{n,k} = \sum_{\vec m} W_T(\vec m) 
\sum_{\vec k \cdot \vec m = k} \ \ 
\sum_{\Gamma\in T(\vec k\>)} \E{\Gamma}_n \;.
$$
Since each $\E{\Gamma}_n$ has a $1/n$ expansion, by adding expansions we
get an asymptotic expansion
$$
\sum_{\Gamma\in T(\vec k\>)} \E{\Gamma}_n =
n^{-{\rm ord}(T)} \bigl( q_0(\vec k\>) + q_1(\vec k\>)/n + \cdots \bigr),
$$
where we ignore error terms in this subsection;
of course,
$$
q_i(\vec k\>) = \sum_{\Gamma\in T(\vec k\>)} 
p_i\bigl( a_1(\Gamma),\ldots,a_{d/2}(\Gamma) \bigr).
$$

Perhaps the main technical point of Section~8 (see Theorem~\ref{th:f_vecm_est}
here, and Lemma~2.14 in \cite{friedman_random_graphs}) is that
the $P_{\vec m\>,i}$ defined by
$$
P_{\vec m\>,i}(k)
=\sum_{\vec k\cdot \vec m=k} q_i(\vec k\>)
$$
are {\dtreelike}, and
roughly speaking $P_{\vec m\>,i}$'s principal part and error term
decay at most like $(d-1)^{m/2}$ where
$m=m_1+\cdots+m_k$ (actually, the principal part is shown
in Lemma~2.14 in \cite{friedman_random_graphs} to decay at most like
$(d-1)^{-m}$).
The way this is done is very roughly as follows.
First, we fix a lettering of $T$ and apply Lemma~\ref{lm:irdeigens}
to each edge, $e_i$, of $T$, whose length is set to $k_i$.
This shows that the $q_i(\vec k\>)$ is a sum of {\dtreelike} functions
whose arguments are all sums of a subset of the $k_i$.  As an example,
consider
a function, $g(k_1+\cdots+k_b)$ with $g$ being {\dtreelike}.  The maximum
value of $k_1+\cdots+k_b$, given $\vec k\cdot\vec m=k$ with fixed
$\vec m$ and $k$, is $k-m+b$, which is achieved when and only when $k_i=1$
whenever $m_i\ge 2$ (assuming at least one $m_i=1$; otherwise
$g(k_1+\cdots+k_b)$ is bounded by 
$ck^c (d-1)^{k/2}$ since $k_1+\cdots+k_b\le k/2$).
This $k-m+b$ is where the decay exponential in $m$ comes from
(see Section~8 and/or Section~2 in \cite{friedman_random_graphs}
for details).

It follows that there is an asymptotic expansion
$$
\E{T}_{n,k} = n^{-{\rm ord}(T)} \bigl( Q_0(k) + Q_1(k)/n + \cdots \bigr),
$$
where
\begin{equation}\label{eq:key_problem}
Q_i(k) = \sum_{\vec m} W_T(\vec m\>) P_{\vec m\>,i}(k).
\end{equation}
Since the error term of the
$P_{\vec m\>,i}$ decay like $(d-1)^{-m/2}$ (roughly speaking),
the above sum for $Q_i(k)$ turns out to be {\dtreelike}
provided
that
\begin{equation}\label{eq:bound_W_T}
|W_T(\vec m\>)| \le c(d-1-\epsilon)^{m/2}
\end{equation}
for some $\epsilon>0$.

If all $Q_i$ were {\dtreelike} for all $T$, then we'd have a
fairly simple proof of the Alon conjecture.  The results of
\cite{friedman_random_graphs} are based on the fact that
equation~(\ref{eq:bound_W_T}) for all types up to a certain order
(of roughly $O(\sqrt{d})$); past order $O(\sqrt{d})$,
equation~(\ref{eq:bound_W_T}) generally fails to hold.
Furthermore, by Theorem~\ref{th:nottreelike} we see
for certain that at least one $Q_i$ fails to be {\dtreelike} for
an $i$ bounded by $O(\log d\sqrt{d})$.

The new approach in this paper is as follows.  Consider a
constant positive integer $B$, and consider
$$
Q_i(k) = \sum_{\vec m} W_T(\vec m\>) 
\sum_{\vec k\cdot \vec m=k} q_i(\vec k\>).
$$
We can divide this sum by fixing some of the $k_i$'s at fixed values
less than $B$ (call these $k_{t+1},\ldots,k_b$), and then summing
over the remaining $k_i$'s (namely $k_1,\ldots,k_t$) subject to
the remaining $k_i$'s being at least $B$ (or ``long'').  
This is where a ``new type'' comes from.
Next we fix a constant
$S<B$ and define a
``selective trace'' to be the sum of irreducible closed walks of a given
length that have no subpath of length $S$ tracing out a graph of
$\lambda_{\ird{}}$ at least $\sqrt{d-1}$.  If
$$
M_1=m_1+\ldots+m_t,\quad M_2=m_{t+1}+\ldots+m_b,
$$
then the corresponding ``selective'' version of $W_T(\vec m\>)$ 
(that depends on the new type, i.e., on knowing the particular $k_i$ that
are fixed and their values) is bounded by roughly
$(d-1-\epsilon)^{(BM_1+M_2)/2}$ (for appropriately large $S$ and $B$).
But consider
the $P_{\vec m\>,i}$ for the ``new type,'' i.e., the sum
$$
\sum_{\substack{\vec k\cdot \vec m=k \\ k_1,\ldots,k_t\ge B}} q_i(\vec k\>)
$$
with $k_{t+1},\ldots,k_b$ at their fixed values; it turns out they
decay like $(d-1)^{-(BM_1+M_2)/2}$.
Thus, after summing over all new types,
the corresponding selective analogues of the $Q_i(k)$ are {\dtreelike}.

Unfortunately, a selective trace does not generally equal the original
trace unless the graph in question is free of certain tangles.
Still, in Sections~9 and later we show how the asymptotic expansions
with {\dtreelike} coefficients of
selective traces can be used to control
a random graph's eigenvalues.

\section{The Selective Trace}

In this section we define a {\em selective trace}, and discuss some of
its properties.

\subsection{The General Selective Trace}

Fix a graph, $G=(V,E)$, coming from $\cgnd$, so that $V=\{1,\ldots,n\}$
and $G$ is $\Pi$-labelled.

By a {\em path\footnote{
  By a {\em path} one often means a sequence of vertices.  In case there
  are multiple edges in the graph, one needs to note also which edge is 
  traversed.  Finally, in the case of whole-loops in an undirected graph,
  one needs to remember in which ``direction'' each whole-loop is being
  traversed.  In the present situation, all the above information is contained
  simply in the
  initial vertex and the permutations, $\pi_i$ or $\pi_i^{-1}$, being taken
  on each step of the path.
} of length $k$} in $G$ we shall mean a vertex, 
$v\in V$, and a word of length $k$,
$w=\sigma_1\ldots\sigma_k$, over $\Pi$ (i.e., with each $\sigma_i\in\Pi$).
Such a path determines a subgraph in $G$ of those vertices and edges
traversed.  We say a path {\em traverses} a tangle, $\tang$, if the subgraph
traversed by the path contains the tangle, $\tang$.

\begin{definition} Let $\tset=\{\tang_1,\ldots\}$ be a (finite or infinite)
collection of
tangles.
For positive integer, $S$, 
the {\em set of $(S,\tset)$-selective closed walks (respectively, walks)} 
are those irreducible closed walks (respectively, walks)
that have no subpath of length at most $S$ that traverses a 
tangle in $\tset$.
The {\em $k$-th irreducible
$(S,\tset)$-selective trace of $G$}, 
$\irdsel_{S,\tset}(G;k)$\mythreeindex{IrSelTr}{$\irdsel_{S,\tset}(G;k)$}{
$k$-th irreducible $(S,\tset)$-selective trace of $G$, i.e., the number of
irreducible closed walks of length $k$ in $G$ such that no subpath of
length at most $S$ traces out a tangle in $\tset$}, is the
number of $(S,\tset)$-selective closed walks of length $k$.
\end{definition}
Intuitively, the selective trace modifies the standard irreducible
trace on those graphs
that have a tangle in $\tset$, and avoids those closed walks that
in some
short part trace out such a tangle.

\subsection{A Lemma on Selective Walks}

What is the point of the selective trace?  We can answer this question
in two ways.  First, since hypercritical 
tangles give large eigenvalues, any trace
with an arbitrarily long asymptotic expansion in $1/n$ with {\dtreelike}
coefficients must avoid hypercritical tangles (according to
Theorems~\ref{th:remarkable} and \ref{th:key_to_s});
a trace must be selective or its asymptotic expansion coefficients 
will not all be {\dtreelike}.
Second, there is a crucial technical theorem, 
Theorem~\ref{th:crucial_cycle_count}, about
counting irreducible contributions to a selective trace.
This lemma makes certain infinite sums converge for the selective trace
that would have to diverge for the standard trace---
for example, the infinite sum involving $W(T;\vec m)$ and $P_{i,T,\vec m}$
just above the middle
of page 351 in \cite{friedman_random_graphs}, for types of order $>d$;
for the same reason, this crucial theorem makes the $1/n$ expansion for
a selective trace have {\dtreelike} coefficients when they don't for a
trace that is not selective--- indeed, the
$(2d-1)^{k/2}$ bound in equation~(24) of \cite{friedman_random_graphs}
depends critically on $2i+2\le\sqrt{2d-1}$, and this equation
corresponds to the error in the $n^{-i}$ term in the expansion of the expected
value of the irreducible
trace (recall that $2d$ in \cite{friedman_random_graphs} corresponds to
our $d$).
We shall finish this subsection
with the crucial technical theorem, Theorem~\ref{th:crucial_cycle_count},
after setting up the necessary terminology.

A {\em relabelling} of a tangle, $\tang$, is a tangle, $\tang'$, that differs
from $\tang$ only in its edge labels.

\begin{definition} A set, $\tset$, of tangles is called {\em closed under
pruning (respectively, relabelling)}
if $\tang\in\tset$ implies $\tang'\in\tset$ for any 
pruning (respectively, relabelling), 
$\tang'$, of $\tang$.
\end{definition}

Note:  In the definition above, $\tang$ and $\tang'$ must be
$\cgnd$-tangles (or tangles in whatever model is discussed)---  a vertex
with two self-loops labelled both labelled $\pi_1$ is not a
$\cgnd$-tangle and is therefore not considered a relabelling of the
tangle where the self-loops are labelled $\pi_1$ and $\pi_2$.

\begin{definition} For a positive integer, $\tau$, let $\tset_\ord(\tau)$ be
the set of tangles whose order is at least $\tau$.
For positive integers $\tau_1\le \tau_2$, let $\tset_\ord(\tau_1,\tau_2)$
be the set of all tangles whose order is at least $\tau_1$ and at
most $\tau_2$.
We also write $\tset_\ord(\tau,\infty)$ for $\tset_\ord(\tau)$.
\end{definition}
Since pruning a tangle does not affect its order, $\tset_\ord(\tau_1,
\tau_2)$ is
closed under pruning; clearly $\tset_\ord(\tau_1,\tau_2)$
is closed under relabelling.

Consider a form, $\Gamma$,
of type $T$, in which $T$'s edges,
$e_i$, have length $k_i$ (as beaded paths arising from $\Gamma$).
For each $e_i\in E_T$ fix an integer $m_i\ge 1$.
Let $\tset$ be a set of tangles closed under relabelling.
Let $W_\Gamma(\vec m;S,\tset)$ be the number of legal closed walks (in particular,
beginning at the first vertex) in $\Gamma$ that traverse each $e_i$ 
exactly $m_i$ times (in either direction) and that are 
$(S,\tset)$-selective.  Since $\tset$ is closed under relabelling,
$W_\Gamma$ depends only on the length, $k_i$, of $e_i$ in $\Gamma$, not
on the particular $\Pi^+$ labels of length $k_i$.  So we may write
$$
W_\Gamma(\vec m;S,\tset)=W_T(\vec m,\vec k;S,\tset)\myindexW .
$$

Now given the above setting, call an edge, $\ned_i$, of $T$ {\em long} if
$k_i>S$, and {\em short} otherwise.
If a walk contains some $\tang\in\tset$
in any consecutive $S$ steps, then by possibly pruning these
consecutive steps along long edges at the beginning and end,
we get a consecutive walk over short edges that contains a pruning of $\tang$.
In particular, if $B>S$ and $\widetilde T=(T;\El,\Ef;\vec k^{\rm fixed})$ is a 
$B$-new type based on $T$, then $W_T$ depends on only $\widetilde T$,
$\vec m$, $S$, and $\tset$ provided that $k_i=\kf_i$; hence we may write
$$
W_T(\vec m,\vec k;S,\tset)=W_{\widetilde T}(\vec m;S,\tset).
$$

\begin{definition} 
We say that a collection of tangles, $\tset$, is {\em $r$-supercritical}
if it contains all supercritical tangles of order at most $r$.
\end{definition}
Next we give two examples of very natural $r$-supercritical tangle sets.
\begin{definition}
Let $\taufund$ be the smallest order of a supercritical tangle,
and let $\tset_{\rm fund}=\tsord(\taufund)$.  Let $\tseig$ the be set of all
supercritical tangles.
\end{definition}

$\tset_{\rm fund}$ and $\tseig$ are clearly $r$-supercritical for any $r$;
$\tsord(\taufund,r)$ and $\tseig[r]=\tseig\cap\tsord(\taufund,r)$ are also
clearly $r$-supercritical.
We arrive at our crucial technical theorem, that is the key to the
selective trace.

\begin{theorem}\label{th:crucial_cycle_count}
Let $T$ by any type, with specified edge set partition
$\El$,$\Ef$, and a $\Pi$-lettering specified.  
Let the edges be indexed so that
$$
\El=\{e_1,\ldots,e_t\}, \qquad \Ef=\{e_{t+1},\ldots,e_b\}.
$$
Then there is a $c$, an
$\epsilon>0$, and
an $S_0$ such
that the following is true for all $S\ge S_0$.  Let
$\tset$ be a set of tangles containing all supercritical tangles included
in a form of type $T$; e.g., by Lemma~\ref{lm:inclusion_order} we may
take $\tset$ to be any $r$-supercritical set for $r=\ord(T)$.
Let
$$
W_{\widetilde T,S}(M_1,M_2) = 
\sum_{\substack{ m_1+\ldots+m_t=M_1\\
m_{t+1}+\cdots+m_b=M_2\\ m_i\ge 1}}
W_{\widetilde T}(\vec m;S,\tset),
$$
for a $B$-new type, $\widetilde T$, with $B>S$ and with $\widetilde T$
having edge set partition $\El,\Ef$.  Then
$$
W_{\widetilde T,S}(M_1,M_2) \le c (\sqrt{d-1}-\epsilon)^{(S_0+1)M_1+M_2}
\le c (\sqrt{d-1}-\epsilon)^{BM_1+M_2}.
$$
\end{theorem}
\proof 
The general approach we take is based on the following crude estimate.
Let $\{f_i(z)\}_{i\in I}$ be a collection of non-negative power series,
and let $g(z)$ be a power series that majorizes each $f_i(z)$.
Suppose that
$g(z_0)$ converges for some $z_0\in(0,1)$.  Then the $z^k$ coefficient
of any of the $f_i(z)$'s is bounded by $g(z_0)z_0^{-k}$.

Specifically, we shall show that there is an $S_0$ such that the
following holds.
Let $B>S_0$, and let $\widetilde T$ be a $B$-new type subject to the
conditions of the theorem.  
Let $G$ be a VLG whose underlying graph is $T$, whose $\Ef$ edges take
their lengths from $\widetilde T$, and whose $\El$ edges all have length
$B$.  Let $f(z)=\sum c_k z^k$, where $c_k$ is the number of
$(S_0,\tset)$-selective irreducible walks of length $k$ in $G$.
We shall show that
there is a $z_0> (d-1)^{-1/2}$ and a $g=g(z)$ such that
(1) $g$ majorizes $f$, (2) $g$ depends only on $T$, and (3) $g(z_0)$
converges.  In that case
$$
W_{\widetilde T,S}(M_1,M_2) \le g(z_0) z_0^{-BM_1-M_2},
$$
which completes the theorem.

Let ${\cal G}_{\rm below}$ be the set of VLG's, $H$, whose underlying graph
is a subgraph of $T$ containing only $\Ef$ edges, with the property that
$\lambda_{\ird{}}(H)<\sqrt{d-1}$.  Let ${\cal G}_{\rm extreme}$ be the set
of elements
of ${\cal G}_{\rm below}$ that are not majorized by a different member
of ${\cal G}_{\rm below}$.  We claim that ${\cal G}_{\rm extreme}$ is
finite; indeed, if $H_1,H_2,\ldots$ were a distinct sequence of VLG's
in ${\cal G}_{\rm extreme}$, then by passing to a subsequence we could
assume that for every $e\in \Ef$ either the length of $e$ in $H_i$ is
constant or the length tends to infinity; but then $H_1$ would majorize
all $H_i$ with $i$ sufficiently large.

Let ${\cal G}_{\rm extreme}=\{H_1,\ldots,H_m\}$, and let
$h_i(z)=\sum c_{i,k} z^k$, where $c_{i,k}$ is the number
of irreducible walks of length $k$ in $H_i$.  These $c_{i,k}$ are
given as the number of walks of length $k-1$ in $(H_i)_{\ird{}}$; it follows
that $h_i(z)$ has radius of convergence greater than $(d-1)^{-1/2}$.  Also,
if $\widetilde c_{i,k}=c_{i,0}+\cdots+c_{i,k}$ is the number of irreducible
walks in $H_i$ of length at most $k$, then
$$
\widetilde h_i(z) = \sum_k \widetilde c_{i,k} z^k = \frac{1}{1-z}h_i(z)
$$
also has radius of convergence greater than $(d-1)^{-1/2}$.  Let
$z_0>(d-1)^{-1/2}$ be a value at which all $\widetilde h_i$ converge.

For any $S$ set
$$
\widetilde h_i^{S}(z) = \sum_{k>S} \widetilde c_{i,k} z^k,
$$
and set
$$
\widetilde h^{S}(z) = \sum_{i=1}^m \widetilde h_i^{S}(z) \quad\mbox{and}\quad
h(z) = \sum_{i=1}^m h_i(z).
$$
Consider an
$S_0$ sufficiently large so that
$\widetilde h^{S_0}(z_0)< 1/d$ (later we impose other lower bounds on
$S_0$'s value).

Let $B>S_0$.
Let $G$ be a VLG whose underlying graph is (the graph underlying) $T$, and
whose $\El$ edges have length at least $B$.  Let $d_k$ (respectively,
$\widehat d_k$) be the number of irreducible walks in $G$ of length $k$
that are $(S_0,\tset)$-selective (respectively, and never traverse an
$\El$ edge).  Let $f(z),\widehat f(z)$ be the generating functions
of $d_k,\widehat d_k$, respectively.

The functions $f(z),\widehat f(z)$ are clearly majorized by the
$f(z),\widehat f(z)$ in the case where $B=S_0+1$ and all $\El$ edges
have length $S_0+1$.  We shall assume this to be the case.

First, we claim that $f$ is majorized by
$$
\bigl(1 - \widehat f(z) dz^{S_0+1}\bigr)^{-1} \widehat f(z).
$$
This is because any walk in $G$ can be broken into alternating
subwalks that remain in
$\Ef$ and steps along $\El$ edges; each time an $\El$ edge is taken
its length is at least $S_0+1$, and there are at most $d$ such
$\El$ edges from which
to choose after finishing the $\Ef$ walk\footnote{Of course,
there are at most $d-1$ such $\El$ edges from which to choose except
possibly at the very first step of a walk.}.
To prove the theorem it therefore suffices to show that 
$\widehat f(z_0) dz_0^{S_0+1}$ is less than one for sufficiently large $S_0$.

Next, we claim that $\widehat f(z)$ is majorized by
$$
\bigl(1 - d\widetilde h^{S_0}(z)\bigr)^{-1} h(z).
$$
Assuming this, it is clear that for sufficently large $S_0$ we have
$\widehat f(z_0) dz_0^{S_0+1}<1$ and the theorem is proven.
For $k\ge S_0$, let $b_k$ be the number of $(S_0,\tset)$-selective
irreducible walks, $w=(v_0,e_1,v_1,\ldots,e_j,v_j)$,
in $G$ of length $k$ through only $\Ef$ edges such that
$w'=(v_0,e_1,v_1,\ldots,e_{j-1},v_{j-1})$ is of length less than $S_0$.
The walk $w'$ is contained in a subgraph of $G$ that is majorized by
one of the $H_i$.
Clearly
$$
b_{S_0} \le \sum_{i=1}^m c_{i,S_0}.
$$
When $k>S_0$, then from $v_{j-1}$ the walk has at most $d$ possible
directions to take (or at most $d-1$ possible directions if $j>1$),
and so
$$
b_k \le \sum_{i=1}^m \widetilde c_{i,k}.
$$
Thus $\sum b_k z^k$ is majorized by $d\widetilde h^{S_0}(z)$.  But
any walk over $\Ef$ edges can be broken into a series of walks of
the form $w$ as above, plus a final walk of length less than $S_0$.
The generating function for such walks is clearly majorized by $h(z)$.
\proofbox

\subsection{Determining $\taufund$ for $\cgnd$}

In order to use the selective trace, we must determine $\taufund$.
We begin by doing so for the model $\cgnd$; next subsection
we use similar
techniques for the models $\chnd$, $\cind$, and $\cjnd$.

More generally, for a given $\tau$, consider the task of finding the
tangle, $\tang$, in $\cgnd$, of order at most
$\tau$ with $\lambda_{\ird{}}(\tang)$
as large as possible.  To simplify this task, notice that pruning leaves
the order and $\lambda_{\ird{}}$ invariant (an irreducible closed
walk can never visit a leaf, so pruning a leaf doesn't affect the number
of irreducible closed walks); hence we may restrict our
search to those $\tang$'s that are completely pruned.

\begin{lemma}\label{lm:contract}
Let $G$ be a graph with edge $e=\{u,v\}$ with $u\ne v$.
Let $G_e$ be the {\em contraction} of $G$ along $e$, i.e. the graph
obtained by discarding $e$ and identifying $u$ with $v$.  Then
$\lambda_{\ird{}}(G)\le \lambda_{\ird{}}(G_e)$.
\end{lemma}
\proof Consider an irreducible closed walk, $c$, about $u$ in $G$.  Then we can
associate to this closed walk one in $G_e$, $\iota(c)$,
by discarding all occurrences
of $e$.  This association, $\iota$,
is an injection, since given a $G_e$ irreducible
closed walk about $u$ of the form $\iota(c)$,
we can infer when $e$ was taken (since 
$e=\{u,v\}$ with $u\ne v$) in the $G$ closed walk, giving rise to (at most) a
single $G$ closed walk.  Since this injection does not increase the length of
the closed walks, we conclude that the number of irreducible closed walks about $u$
in $G$ of length $\le k$ is no more than the number in $G_e$.  Hence
the conclusion of the lemma.
\proofbox

Since edge contraction reduces the number of vertices and of edges by
one each, edge contraction leaves the order invariant.  So in looking
for a $\lambda_{\ird{}}$ tangle of a given order, we may always assume
the tangle has no edge contraction that leaves it a tangle\footnote{
  In $\chnd$, two vertices joined by between $2$ and $d/2$ edges is
  a tangle (with apropriate $\Pi$-labelling), but contracting any edge
  gives self-loops, which are not feasible in $\chnd$.  Therefore 
  edge contraction can take graphs that can be tangles to graphs that
  cannot, at least for certain random graph models.}.

We now claim (by Lemma~\ref{lm:contract})
that for $\cgnd$ and $\tau\le (d/2)-1$, a vertex with $\tau+1$
whole-loops has the largest $\lambda_{\ird{}}$ of all tangles of order $\tau$
(recall that each self-loop is counted as one edge, according to
Definition~\ref{de:tangle_order}).
For this graph we clearly have $\lambda_{\ird{}}=2\tau+1$; hence
$\taufund$ is the smallest integer $\tau$ with $2\tau+1\ge\sqrt{d-1}$,
provided that this $\tau$ is $\le (d/2)-1$.
But we easily verify that this $\tau$,
$$
\taufund=\lceil (\sqrt{d-1}\;+1)/2 \rceil-1=\lceil (\sqrt{d-1}\;-1)/2 \rceil
$$
is indeed at most $(d/2)-1$ for all $d\ge 4$.
We have just established
the following theorem.

\begin{theorem} For the model $\cgnd$, we have 
$\taufund=\lceil (\sqrt{d-1}\;+1)/2 \rceil-1$.
\end{theorem}

\subsection{Determining $\taufund$ for $\chnd$, $\cind$, and $\cjnd$}

For $\chnd$ we have to remember that tangles can't have self-loops.
Thus contractions can only be done along non-multiple edges,
and $\taufund$ will not generally be the same for $\chnd$ and $\cgnd$.

\begin{lemma}\label{lm:contract_distance_two}
Let $u,v$ be vertices of distance two in a graph, $G$, 
i.e., there are no
edges joining $u$ and $v$, but there is a $w$ with edges to each of
$u,v$.  Let $G'$ be the graph obtained by identifying $u$ and $v$
and deleting one of the edges from $w$ to $u$ (or to $v$) (so that the order
of $G'$ is the same as that of $G$).
Then $\lambda_{\ird{}}(G)\le \lambda_{\ird{}}(G')$.
\end{lemma}
\proof Let $U$ be the vertex in $G'$ which is the identification of
$v$ and $u$.  Let the edges from $u$ to $w$ be enumerated
$e_1,\ldots,e_s$, and those from $v$ to $w$ enumerated $f_1,\ldots,f_t$.
The edges from $U$ to $w$ are $g_1,\ldots,g_r$, where $r=s+t-1$.

First consider the case when $V_G=\{u,v,w\}$, and consider the irreducible
closed walks about $w$ (which are necessary of even length).  Such a closed walk begins
in $w$ and takes two steps, visiting either $u$ or $v$, in, respectively,
$s(s-1)$ or $t(t-1)$ ways.  After coming back from a $u$ vertex, another
step of length $2$ can either (1) visit a $u$ vertex, in $(s-1)^2$ ways,
or (2) visit a $v$ vertex, in $t(t-1)$ ways; similarly for coming back from
a $v$ vertex.  Thus ``coming back from a $u$ vertex'' and
``coming back from a $v$ vertex'' forms a Markov chain, and the total number
of irreducible closed walks of length $k$ about $w$ is
\begin{equation}\label{eq:w_cycles}
I_1(k)=
\left[ \begin{array}{cc}s(s-1)&t(t-1)\end{array}\right]
\left[ \begin{array}{cc}(s-1)^2&t(t-1)\\s(s-1)&(t-1)^2\end{array}\right]^{(k-2)/2}
\left[ \begin{array}{c}1\\1\end{array}\right]
\end{equation}
We wish to compare this to the number of irreducible $G'$ closed walks about $w$,
of which there are clearly
$$
I_2(k)=r(r-1)^{k-1} = (s+t-1)(s+t-2)^{k-1}.
$$
For starters, we see
$$
I_2(2)-I_1(2)=2(s-1)(t-1)
$$
which is non-negative, since both $s,t\ge 1$.  Now since the maximum row sum
in the $2\times 2$ matrix of equation~(\ref{eq:w_cycles}) is
$$
s^2+t^2-2(s+t)+1+\max(s,t),
$$
we have
$$
I_1(k+2)\le I_1(k)m_1,\quad\mbox{where}\quad
m_1=s^2+t^2-2(s+t)+1+\max(s,t)
$$
for all $k$.  But
$$
I_2(k+2) = I_2(k)m_2,\quad\mbox{where}\quad m_2=(s+t-2)^2,
$$
and
$$
m_2-m_1 = 2st-2(s+t)+3-\max(s,t) = 1+2(s-1)(t-1)-\max(s,t),
$$
which is positive unless $s$ or $t$ is $1$.
Thus, provided that $s\ge 2$ and $t\ge 2$, we have
$$
\lambda_{\ird{}}(G)\le \sqrt{m_1}<\sqrt{m_2}=\lambda_{\ird{}}(G'),
$$
and
\begin{equation}\label{eq:compare_I}
I_1(k)\le I_1(2)m_1^{(k-2)/2}< I_2(2)m_2^{(k-2)/2} I_2(k) = I_2(k)
\end{equation}
for all even $k$.  If $t=1$ we calculate
\begin{equation}\label{eq:compare_I_t1}
I_1(k)=s(s-1)^{k-1}=I_2(k),
\end{equation}
and similarly when $s=1$.

We shall use the above calculation below.  We can now assume that
$V_G$ has a vertex, $x$, different from $u,v,w$.

There is a natural bijection of edges, $\iota$ from $E_{G}
\setminus(\{e_i\}\cup\{f_i\})$ to $E_{G'}\setminus\{g_i\}$.
Extend $\iota$ to a map on all of $E_G$ by defining $\iota(e_i)$ and
$\iota(f_i)$ to be a formal symbol $S$.  For any irreducible $G$ closed walk
about $x$ specified by its edges,
$c=(c_1,\ldots, c_k)$ with $c_i\in E_G$, we associate a sequence
$$
\iota(c) = \bigl( \iota(c_1),\ldots,\iota(c_\ell) \bigr).
$$
We claim that the number of $c$ with a given image $\iota(c)$ is no 
more than the number of $E_{G'}$ closed walks corresponding to $\iota(c)$
by changing all $g_i$ edges into $S$'s.  
Indeed, consider a block of consecutive $S$'s in $\iota(c)$, i.e.
$\iota(c_a)=\iota(c_{a+1})=\cdots=\iota(c_b)=S$, and $\iota(c_{a-1})\ne S$
and $\iota(c_{b+1})\ne S$;  $\iota(c)$ cannot begin or end with an $S$,
since the closed walk begins at $x$, and so we can assume $a\ge 2$ and $b\le \ell-1$.
By looking at $\iota(c_{a-1})$ and $\iota(c_{b+1})$ we can determine
whether or not the $S$-block begins in $u$, $v$, or $w$, and ends
in $u$, $v$, or $w$.  If the $S$-block begins in $w$ and ends in $w$, then
equations~(\ref{eq:compare_I}) and (\ref{eq:compare_I_t1})
show that there are no fewer $G'$ sequences
for the corresponding $S$-block than $G$ sequences.  Next compare those
$S$-blocks
that begin in a $u$ and end in a $w$.  The number of such sequences
in $G$ is
$$
\left[ \begin{array}{cc}s&0\end{array}\right]
\left[ \begin{array}{cc}(s-1)^2&t(t-1)\\s(s-1)&(t-1)^2\end{array}\right]^{(b-a)/2}
\left[ \begin{array}{c}1\\1\end{array}\right],
$$
whereas the number in $G'$ is $(s+t-1)(s+t-2)^{b-a}$ (since the non-$S$
edge $\iota(c_{a-1})$ can be followed by any $U$ to $w$ edge in $G'$);
so the $G'$ number is no less than the $G$ number for $b-a=0$ (since $t\ge 1$),
and each time $b-a$ is increased by $2$, the former number
gets multiplied by an
$m_2$, the latter gets multiplied by no more than $m_1$, where $m_1<m_2$,
provided that $s\ge 2$ and $t\ge 2$; the $s=1$ or $t=1$ case is easily
checked to result in equality.
The same arguement holds for $v$ to $w$ $S$-blocks.  For an $S$-block starting
and ending in $u$, we wish to compare
$$
\left[ \begin{array}{cc}s&0\end{array}\right]
\left[ \begin{array}{cc}(s-1)^2&t(t-1)\\s(s-1)&(t-1)^2\end{array}\right]^{(b-a-1)/2}
\left[ \begin{array}{c}s-1\\0\end{array}\right],
$$
with $(s+t-1)(s+t-2)^{b-a}$.
Again, it suffices to compare when $b-a=1$, which is immediate, and to check
$s=1$ or $t=1$ separately.
We argue for $S$-blocks starting in either $u$ or $v$ and ending in either
$u$ or $v$ similarly.
\proofbox

\begin{theorem}\label{th:taufund_chnd}
For the model $\chnd$, we have
$\taufund=\lceil \sqrt{d-1}\;\rceil-1$.
\end{theorem}
\proof As before, we consider a $\tau$ and search for those $\tang$ of order
at most $\tau$ with $\lambda_{\ird{}}(\tang)$ as large as possible.
By Lemma~\ref{lm:contract_distance_two}, 
and by contractions (in Lemma~\ref{lm:contract}),
we may restrict our search to those $\tang$ with
two or more edges between every pair of nodes.

First assume that $\tau+2\le d/2$.
If $\tang$ has two vertices, then $\tang$ has $\tau+2$ edges joining the
two vertices (since there are no self-loops in $\chnd$).
In this case
$\lambda_{\ird{}}(\tang)=\tau+1$.
We claim that this is as large
a $\lambda_{\ird{}}$ as possible (again, assuming $\tau+2\le d/2$).
Indeed, if $\tang$ has $r>2$ vertices,
then the maximum degree of a vertex is $|E|$ minus the edges not involved
with that particular vertex, which is at least $2$ for each pair of the
$r-1$ other vertices.  So the maximum degree is at most
$$
|E|-\binom{r-1}{2} 2 \le (|V|+\tau)-\binom{r-1}{2} 2 = \tau+r-(r-1)(r-2).
$$
Since $\lambda_{\ird{}}$ is no greater than the maximum degree minus $1$,
we have
$$
\lambda_{\ird{}} \le \tau+r-(r-1)(r-2)-1=\tau+1-(r-2)^2.
$$
It follows that if $r>2$, $\lambda_{\ird{}}$ is strictly less than
$\tau+1$.

To achieve $\lambda_{\ird{}}(\tang)=\tau+1$ with our $\tang$ having two vertices,
we required $\tau+2\le d/2$.  To get $\lambda_{\ird{}}(\tang)=\tau+1$ to equal or
exceed $\sqrt{d-1}$, we require $\tau+1=\bigl\lceil \sqrt{d-1} \;\bigr\rceil$,
for which we must have
$$
\bigl\lceil \sqrt{d-1} \;\bigr\rceil + 1 \le d/2.
$$
Since $d/2$ is an integer, this is equivalent to
$$
\sqrt{d-1}+1\le d/2,
$$
which we easily see holds for all even $d>2$ except $d=4,6$.

We conclude that $\taufund=\bigl\lceil \sqrt{d-1} \;\bigr\rceil-1$ 
for even $d\ge 8$.
It suffices to analyze the cases $d=4,6$.

For each order, $\tau$, and $d=4,6$, we must examine those tangles of
order $\tau$ and determine the largest possible $\lambda_{\ird{}}(\tang)$.
Let us note that if $\tang$ is a tangle of order $-1$, then it is a tree and
has $\lambda_{\ird{}}(\tang)=0$.  If $\tang$ is a completely pruned tangle of
order $0$, then $\tang$ is a closed walk and has $\lambda_{\ird{}}(\tang)=1$.

If $d=4$, then consider the tangle of order $1$ with three vertices,
consisting of one ``middle''
vertex joined by two edges to each of two vertices.
(This is a tangle by labelling the left to middle edges and the middle to
right edges $\pi_1,\pi_2$.)
We easily compute $\lambda_{\ird{}}=\sqrt{3}$, as this graph is
bipartite and the number of irreducible
walks of length $2m$ from the middle vertex,
all such walks being closed walks, is clearly $4\cdot 3^{m-1}$.
So for $d=4$, $\taufund=1$.

For $d=6$, consider the tangle, $\tang=(V,E)$ with $V=\{v_1,v_2,v_3\}$
with three edges connecting $v_1$ to $v_2$ (labelled $\pi_1,\pi_2,\pi_3$)
and two edges connecting
$v_2$ to $v_3$ (labelled $\pi_1,\pi_2$).
We claim that $\lambda_{\ird{}}(\tang)>\sqrt{5}$.  Say that a
closed walk about $v_2$ ends in ``state A'' if the last vertex before $v_2$
was $v_1$, and otherwise in ``state B'' (i.e. the second to last vertex
is $v_3$).  From state A, taking two additional irreducible steps, there
are 4 ways to reach another state A, and two ways to reach another state B.
From state B, taking two irreducible steps, there is one way to reach
another state B and six ways to reach another state A.  It easy follows
that $\lambda_{\ird{}}(\tang)$ is the square root of the largest eigenvalue
of
\begin{equation}\label{eq:d6tangle}
\left[ \begin{array}{cc}4&2\\6&1\end{array}\right].
\end{equation}
But $\lambda_1$ of this matrix is $(5+\sqrt{57})/2$, and this $\lambda_1$
is just
$\lambda_{\ird{}}(\tang)$.
It follows that $\lambda_{\ird{}}(\tang)>\sqrt{6}>\sqrt{5}$, and hence
$\taufund\le 2$.

We wish to rule out $\taufund=1$ when $d=6$.  Since we are considering only
completely pruned graphs, $\tang$,
each vertex has degree $\ge 2$.  Such a graph, $\tang$, of
order $1$ has all vertices of degree $2$ except for one of degree $4$ or
two of degree $3$.  In the case where there are vertices, $u,v$,
of degree $3$, therefore joined
by three disjoint beaded paths, then $\lambda_{\ird{}}(\tang)$ is greatest when
the beaded paths are each of length $1$ (by setting up an obvious map from
irreducible closed walks about $u$ from the general graph to the one with 
beaded paths
of length $1$); hence $\lambda_{\ird{}}(\tang)\le 2$ in this case,
since the graph of two vertices joined by three edges has 
$\lambda_{\ird{}}=2$.  Similarly,
in the case with $u$ of degree $4$, therefore having two beaded closed walks
from $u$, $\lambda_{\ird{}}(\tang)$ is greatest when the lengths of the two
closed walks are two (they cannot be one since $\chnd$ does not permit self-loops);
hence $\lambda_{\ird{}}(\tang)\le \sqrt{3}$ in this case.
Hence $\taufund>1$ and therefore $\taufund=2$.

We conclude that $\taufund=
\bigl\lceil \sqrt{d-1} \;\bigr\rceil-1$ also when $d=4,6$.
\proofbox

\begin{theorem} For the model $\cind$, we have
$\taufund=\bigl\lceil \sqrt{d-1}\;\bigr\rceil-1$ for all $d\ge 3$.
\end{theorem}
\proof We argue as with $\chnd$.  The only difference is that in 
$\cind$, two vertices can have as many as $d$ edges between them
in a tangle (as opposed to $d/2$ edges in an $\chnd$ tangle).  So
the argument in the previous thoerem shows that the two-vertex tangles
give that $\tau=\bigl\lceil \sqrt{d-1} \;\bigr\rceil-1$ equals $\taufund$ provided
that $\tau+2\le d$ (as opposed to $\tau+2\le d/2$ for $\chnd$).
But we easily verify that
$$
\bigl\lceil \sqrt{d-1} \;\bigr\rceil+1 \le d
$$
for all $d\ge 3$ (indeed, we have equality for $d=3$, and
each time we increase $d\ge 3$ by one, $\sqrt{d-1}$
increases by less than one).
\proofbox

\begin{theorem} For the model $\cjnd$, we have
$\taufund=\bigl\lceil \sqrt{d-1}\;\bigr\rceil-1$ for all $d\ge 3$.
\end{theorem}
\proof 
As in $\cind$, for any $\tau\le d-2$ there is a tangle $G_\tau$ that is
a pair of vertices 
with $\tau+2$ edges
joining them.  $G_\tau$ has order $\tau$ and $\lambda_{\ird{}}=\tau+1$; since
when $\tau+2=d$ we have $\lambda_{\ird{}}(G_\tau)
=d-1\ge \sqrt{d-1}$ giving a supercritical
tangle, we need worry only about whether or not there is a tangle
of order $\tau\le d-2$ that can beat the $\lambda_{\ird{}}$ of $G_\tau$.
Again, as with $\cind$,
Lemma~\ref{lm:contract_distance_two} can be applied to graphs with
half-loops, and so by the same argument as for $\cind$ we have that
only graphs on one or two vertices can possibly beat $G_\tau$.
So consider a graph on vertices $u,v$ with $a$ half-loops about $u$,
$c$ half-loops about $v$, and $b$ edges from $u$ to $v$.  An irreducible
path traverses edges of four different states: (1) half-loops about $u$,
(2) edges from $u$ to $v$, (3) edges from $v$ to $u$, and (4) half-loops
about $v$.  Now we write a transition matrix about the states:  for example
in state (1) we may either continue on one of $a-1$ half-loops in state (1)
or continue on one of $b$ edges in state (2).  We find the transition
matrix
$$
\left[ \begin{array}{cccc}a-1&b&0&0\\0&0&b-1&c\\a&b-1&0&0\\0&0&b&c-1
\end{array}\right],
$$
and $\lambda_{\ird{}}$ of our graph is this matrice's largest eigenvalue.
The order of the graph is $a+b+c-2$
(recall, each half-loop
contributes one to the order of a graph).
But the row sum is never greater than $a+b+c-1$ (and always less unless
$a$ or $c$ vanishes), and so if this graph
has order $\tau$ its $\lambda_{\ird{}}$ is no more than $\tau+1$.
Hence
no $\cjnd$ tangle of order $\tau$ beats $G_\tau$, provided that
$\tau\le d-2$.  Thus
$\taufund$ is the smallest number with $\taufund+1\ge \sqrt{d-1}$.
\proofbox

\section{Ramanujan Functions}

In this section we discuss Ramanujan functions in order to (1) explain
their significance, and (2) give some intuition on some very technical
issues surrounding the asymptotic expansion for irreducible traces
(as in Section~\ref{se:expansion}).

\begin{definition} A function, $f(k)$, on positive integers, $k$, is
said to be {\em {\dtreelike} of order $\alpha>0$}
if there is a polynomial $p=p(k)$ and a
constant $c>0$ such
that
$$
|f(k)-(d-1)^kp(k)|\le ck^c \alpha^k
$$
for all $k$.  We call $(d-1)^kp(k)$ the {\em principal term} of $f$,
and $f(k)-(d-1)^kp(k)$ the {\em error term} (both terms are uniquely
determined if $\alpha<d-1$).  A function is {\em super-{\dtreelike}}
if it is {\dtreelike} of order $1$.
\end{definition} 
A {\dtreelike} function as defined before, in Definition~\ref{de:ram},
is just a 
{\dtreelike} of order $\sqrt{d-1}$.

Let $N(k)$ be the number of irreducible cycles of length $k$
in a $d$-regular graph.  Then
in \cite{lubotzky} it is shown that if $N(k)$ is {\dtreelike}, then
any eigenvalue, $\lambda\ne \pm d$, of the graph satisfies 
$|\lambda|\le 2\sqrt{d-1}$.  The discussion there also shows that
in any case,
if $\lambda$ is the eigenvalue of largest absolute value $<d$, then
$N(k)$ is {\dtreelike} of order $\alpha$ with
$$
\alpha = \frac{ |\lambda|+\sqrt{\lambda^2-4(d-1)}}{2}
$$
(and not for any smaller an $\alpha$).
Any discussion of irreducible traces and eigenvalues is bound to be tied to
{\dtreelike} functions.

One important property of {\dtreelike} functions of order $\alpha$
is that they are closed under addition.  Another very important
property is that they are closed under {\em convolution}, which we
now formally explain.
This property
will be used in Section~\ref{se:finish}, and
refined versions of it will be used in Section~\ref{se:expansion}.
\begin{theorem}\label{th:baby_convolute}
Let $f_1,f_2$ be {\dtreelike} of order $\alpha$ with $\alpha<d-1$.  
Then their convolution,
$$
g(k)= (f_1*f_2)(k) = \sum_{j=1}^{k-1} f_1(j)f_2(k-j)
$$
is also {\dtreelike} of order $\alpha$.
\end{theorem}
The techniques in
Section~\ref{se:expansion} prove a more precise version of this
theorem (keeping track of the sizes of the
the error term and the coefficients of the principal part); for this
reason we keep the argument below concise.
\proof 
For $i=1,2$ let
$$
f_i(k) = (d-1)^k p_i(k) + e_i(k)
$$
where $p_i$ are polynomials and the $|e_i(k)|$ are bounded
by $ck^c\alpha^k$ for some $k$.  We may also write
$$
f_i(k) = (d-1)^k\bigl(p_i(k)+\tilde e_i(k)\bigr), \qquad
\mbox{where $\tilde e_i(k)=(d-1)^{-k}e_i(k)$.}
$$
Since convolution is bilinear, we easily see
$$
f*g= e_1*e_2+(d-1)^k(p_1*p_2 + p_1*\tilde e_2 + p_2*\tilde e_1).
$$
It suffices to show that 
$$
e_1*e_2,\quad (d-1)^k(p_1*p_2)(k), \quad (d-1)^k(p_1*\tilde e_2)(k),
\quad (d-1)^k(p_2*\tilde e_1)(k)
$$
are {\dtreelike} of order $\alpha$.

According to Sublemma~2.15 of \cite{friedman_random_graphs},
$p_1*p_2$ is a polynomial.  Next
$$
(p_1*\tilde e_2)(k)=\sum_{j=1}^{k-1} (d-1)^{-j} p_1(k-j)e_2(j)
= \Sigma_1-\Sigma_2
$$
where
\begin{eqnarray*}
\Sigma_1& =& \sum_{j=1}^\infty p_1(k-j)(d-1)^{-j}e_2(j),\\
\Sigma_2 &=& \sum_{j=k}^\infty p_1(k-j)(d-1)^{-j}e_2(j).
\end{eqnarray*}
Writing 
$$
p_1(k-j) = \sum a_{r,s} k^r j^s,
$$
we see that
$\Sigma_1$ is a polynomial, and $\Sigma_2$ is bounded by
$ck^c\alpha^k(d-1)^{-k}$ (see Section~8, especially Lemma~\ref{lm:useful_aux},
for details).
This shows $(d-1)^k(p_1*\tilde e_2)$ is {\dtreelike} of order $\alpha$.
Similarly, so is $(d-1)^k(p_2*\tilde e_1)$; $e_1*e_2$ is easily also seen
to be so (with zero principal term).
\proofbox

\section{An Expansion for Some Selective Traces}
\label{se:expansion}

In this section we prove the first crucial expansion theorem.
Our second such theorem, Theorem~\ref{th:with}, will extend these ideas.

\begin{theorem}\label{th:main_expansion}
Let $r$ be a positive integer, and let $\tset$ be a set of
tangles containing all supercritical tangles of order less than $r$.
Then there is an $S_0=S_0(r)$
such that for all $S\ge S_0$ the following holds.  
We have
$$
\E{\irdsel_{S,\tset}(G;k)} = f_0(k)+\frac{  f_1(k)}{n}+\cdots+
\frac{  f_{r-1}(k)}{n^{r-1}}+\frac{{\rm error}}{n^r},
$$
where 
the $f_i$ are {\dtreelike} and the error term satisfies the bound given in
Theorem~\ref{th:SSIICexpansion}.
\end{theorem}

We begin by explaining why this theorem is an easy consequence of
Theorem~\ref{th:SSIICexpansion}
and the following theorem.

\begin{theorem}\label{th:reduce1}
Fix a lettering, $\cl$, of type $T$, and fixed non-negative
integers 
$\ell_1,\ldots,\ell_{d/2}$.  Let
$$
R_{T,\cl}(k_1,\ldots,k_b) = 
\sum_{(w_1,\ldots,w_b)} 
\prod_{j=1}^{d/2}
\bigl( a_j(w_1)+\ldots+a_j(w_b) \bigr)^{\ell_j}
$$
where the sum is over all tuples of words $(w_1,\ldots,w_b)$ such that each
$w_i$ is irreducible and of length $k_i$ and is compatible with $\cl$.
Let $\widetilde T$ be a $B$-new type based on $T$, and let
\begin{equation}\label{eq:define_f}
f(k) = \sum_{m_i\ge 1}\;\;
\sum_{\substack{ k_1m_1+\ldots+k_bm_b=k\\ k_i\ge B\;{\rm if}\;e_i\in
\El \\ k_i=\kf_i\;{\rm if}\;e_i\in \Ef}}
W_{\widetilde T}(\vec m;S,\tset)R_{T,\cl}(k_1,\ldots,k_b)
\end{equation}
(with $W$ as in Theorem~\ref{th:crucial_cycle_count}).  Then
$f$ is {\dtreelike} for all $B\ge B_0=B_0(T)$.
\end{theorem}

Assume Theorem~\ref{th:reduce1} for the moment.  Let $\cw$ 
(respectively, $\cw_T$ and $\cw_{\widetilde T}$) be the walk collections
corresponding to irreducible, $(S,\tset)$-selective walks
(respectively, and that are associated to the type $T$ and $\widetilde T$).
These walk collections are all SSIIC.  According to 
Theorem~\ref{th:SSIICexpansion}
$$
\walksum{\cw}{k,n} =   f_0(k)+\frac{  f_1(k)}{n}+\cdots+
\frac{  f_{r-1}(k)}{n^{r-1}}+\frac{{\rm error}}{n^r},
$$
where 
\begin{equation}\label{eq:want_to_be_ram}
f_i(k) = \sum_{j=0}^{r-1}\;\; \sum_{(w;[\vec t\>]){\rm \;order\;}j,\in\cw(k)}
p_{i-j}\bigl( a_1(w;[\vec t\>]),\ldots,a_d(w;[\vec t\>])\bigr).
\end{equation}
It suffices to show that these $f_i$ are {\dtreelike}.  
We know there are finitely many types of order at most $r-1$.
Now fix
$S_0$ to be the max over $B_0(T)$ over types, $T$, of order at most $r-1$
(and $B_0$ as in Theorem~\ref{th:reduce1}).  For any $S\ge S_0$,
choose $B=S+1$; of course, $B\ge B_0(T)$ for any type, $T$, of order
at most $r-1$.
We know there are finitely many $B$-new types based on any type.
So the sum involving $\cw$ in equation~(\ref{eq:want_to_be_ram}) decomposes
as a finite sum over $\cw_T$'s or $\cw_{\widetilde T}$'s.
Furthermore, each expansion polynomial $p_{i-j}$, involving a fixed new type,
$\widetilde T$, is just
a function of $a_1,\ldots,a_{d/2}$ over appropriate forms, 
and each $a_i$ of the form is just the sum of
the $a_i$ along each edge of the form.  Therefore
Theorem~\ref{th:reduce1} just says that each $f_i$, when summing
over a $\cw_{\widetilde T}$, is {\dtreelike}.  Summing over all
$\cw_{\widetilde T}$ shows that the $f_i$ corresponding to $\cw$ are
also {\dtreelike}.

\noindent{\bf Proof (of Theorem~\ref{th:reduce1})\ \ }
Clearly it suffices to prove the following theorem with 
$R_{T,\cl}$ replaced by
$$
\sum_{(w_1,\ldots,w_b)} 
\prod_{i=1}^{b} 
\prod_{j=1}^{d/2}
a_j^{\ell_{ij}}(w_i),
$$
with $\ell_{ij}$ any set of non-negative integers.

Our Lemma~\ref{lm:irdeigens}
reduces the above theorem to the following.

\begin{theorem}\label{th:reduce3}
With notation as in Theorem~\ref{th:reduce1}, let
$K_1,K_2,K_3$ be a partition of $k_1,\ldots,k_b$, and let $|K_i|$ for
$i=1,2,3$ denote the sum of the $k_j$ in $K_i$.  Then for fixed non-negative
integers $\ell_1,\ldots,\ell_b$, Theorem~\ref{th:reduce1} holds with
$R_{T,\cl}$ replaced by
\begin{equation}\label{eq:specific_R}
R_{T,\cl}(k_1,\ldots,k_b) = 
(d-1)^{|K_1|}(-1)^{|K_2|}k_1^{\ell_1}\cdots
k_b^{\ell_b}.
\end{equation}
More generally, Theorem~\ref{th:reduce1} holds with
$R_{T,\cl}$ replaced by
\begin{equation}\label{eq:general_R}
R_{T,\cl}(k_1,\ldots,k_b) = 
(d-1)^{|K_1|} k_1^{\ell_1}\cdots
k_u^{\ell_u} \beta(k_{u+1},\ldots,k_b),
\end{equation}
where the edges are ordered so that
$$
\{ i | \mbox{$e_i\in \El$ and $k_i\in K_1$} \} = \{1,\ldots,u\},
$$
and where $\beta$ is a function such
that
$$
|\beta(k_{u+1},\ldots,k_b)| \le c(|k_{u+1}|+\cdots+|k_b|)^c
$$
for some constant $c$.
\end{theorem}
The $R$ of equation~(\ref{eq:specific_R}) is all that is needed for $\cgnd$;
it will be convenient (if not necessary) to use the $R$ of
equation~(\ref{eq:general_R}) for $\cjnd$ (see Section~14).
\proof
It suffices to deal with the $R$ of
equation~(\ref{eq:general_R}).
Let 
$$
\widetilde K_i = \{ k_j\in K_i \mid \ned_j\in \El \} .
$$
We may assume $\El=\{\ned_1,\ldots,
\ned_t\}$
and $\widetilde K_1=\{k_1,\ldots,k_u\}$;
set 
$$
\begin{array}{lll}
M_1=m_1+\cdots+m_t, & \qquad & M_2=m_{t+1}+\cdots+m_b, \\
j_1=k_1m_1+\cdots+k_tm_t, & & 
\mbox{and  $j_2=k_{t+1}m_{t+1}+\cdots+k_bm_b$.} \end{array}
$$
Clearly it suffices to prove the theorem for
$$
f(k)=\sum_{\vec m} W_{\widetilde T}(\vec m;S,\tset) 
\sum_{\substack{k_1m_1+\cdots+k_b m_b=k\\ k_i\ge B\;{\rm for}\; i\le t}}
(d-1)^{|\widetilde K_1|}
k_1^{\ell_1}\ldots k_u^{\ell_u}\beta(k_{u+1},\ldots,k_t),
$$
understanding that $k_{t+1},\ldots,k_t$ are fixed by $\veckf$.

\begin{definition} The {\em (coefficient) norm}, $|p|$, of a polynomial, $p$
(which is possibly multivariate), is the largest absolute value among
its coefficients.
\end{definition}
Working with this notion of a norm is a bit ``weak,'' (i.e., sometimes
much stronger statements would hold with other norms), 
but this notion is sufficient for our
purposes.

Let 
$$
f_{\vec m}(k)=\sum_{\vec k\cdot\vec m=k}(d-1)^{|\widetilde K_1|}
k_1^{\ell_1}\ldots k_u^{\ell_u}\beta(k_{u+1},\ldots,k_t).
$$
\begin{theorem}\label{th:f_vecm_est}
For any vector of positive integers, $\vec m$, we have
$f_{\vec m}$ is {\dtreelike} with principal term $(d-1)^kp_{\vec m}(k)$
and error term $e_{\vec m}(k)$ satisfying
$$
|p_{\vec m}| \le c(d-1)^{(-BM_1-M_2+c)/2},
$$
and 
$$
|e_{\vec m}(k)| \le ck^c (d-1)^{(k-BM_1-M_2+Bc)/2},
$$
where $c$ depends only on the $\ell_i$ and $\beta$.
\end{theorem}
\proof Fix a value of $\vec m$.  Without loss of generality we may assume
$m_1=\cdots=m_s=1$ and $m_{s+1},\ldots,m_u\ge 2$.  For now assume that
$s\ge 1$; we will later indicate the minor changes needed for the situation
$s=0$ (i.e. when there are no $m_i$ belonging to $\widetilde K_1$ that
equal one).
Let
$$
g_{\vec m}(r)= \sum_{\substack{ k_{s+1},\ldots,k_b{\rm \;s.t.}\\
k_{s+1}m_{s+1}+\cdots+k_bm_b=r\\  k_i\ge B\;{\rm for}\; i\le t }}
(d-1)^{k_{s+1}+\cdots+k_u}\beta(k_{u+1},\ldots,k_t).
$$
\begin{lemma}\label{lm:crucial_B_type_est}
If $\widetilde T$ is a $B$-new type, then
$$
|g_{\vec m}(r)|\le cr^c (d-1)^{(r-BM_1-M_2+Bc)/2}
$$
for some constant $c$ depending only on $\widetilde T$.
\end{lemma}
\proof
Since each $k_i$ is at most $r$,
$$
\beta(k_{u+1},\ldots,k_t)
$$
is bounded by $cr^c$, and
it suffices to prove the 
estimate for $g_{\vec m}$ replaced with
$$
\sum_{\substack{k_{s+1}m_{s+1}+\cdots+k_bm_b=r\\ 
k_i\ge B\;{\rm for}\; i\le t }}
(d-1)^{k_{s+1}+\cdots+k_u}.
$$
But there are only $\binom{r+b-s-1}{b-s-1}$ ways of writing
$r$ as the sum of $b-s$ positive integers.  So it suffices to show
$$
(d-1)^{k_{s+1}+\cdots+k_u} \le (d-1)^{(r-BM_1-M_2-Bc)/2}.
$$
Now we have 
$$
r = k_{s+1}m_{s+1}+\cdots+k_bm_b,
$$
so
$$
2k_{s+1}+\cdots+2k_u =
r - (k_{s+1}m_{s+1}+\cdots+k_bm_b)+(2k_{s+1}+\cdots+2k_u)
$$
$$
=r - (m_{s+1}-2)k_{s+1} - \cdots - (m_u-2)k_u -
m_{u+1}k_{u+1} - \cdots - m_bk_b.
$$
Since $m_i\ge 2$ for $i$ between $s+1$ and $u$, and since $k_i\ge B$ for
$i\le t$ (and $k_i\ge 1$ for all $i$), we conclude
$$
2k_{s+1}+\cdots+2k_u \le r - (m_{s+1}-2)B - \cdots - (m_u-2)B
$$
$$
-m_{u+1}B - \cdots - m_tB-m_{t+1}-\cdots-m_b
$$
$$
= r-(m_{s+1}+\cdots+m_t)B+2(u-s)B-(m_{t+1}+\cdots+m_b)
$$
$$
= r - \bigl(M_1-s-2(u-s)\bigr)B-M_2.
$$
Hence
$$
k_{s+1}+\cdots+k_u\le (2k_{s+1}+\cdots+2k_u)/2
\le \Bigl(r - \bigl(M_1-s-2(u-s)\bigr)B-M_2\Bigr)/2.
$$
But $u,s$ are bounded by the number of edges in $\widetilde T$.
\proofbox

We will need another lemma.
\begin{lemma}\label{lm:Q}
For any non-negative integers $\ell_1,\ldots,\ell_s$ there is
a polynomial $Q$ such that for
all $k\ge s$ we have
$$
\sum_{\substack{k_1+\cdots+k_s=k\\ {\rm integers}\;k_i\ge 1}} 
k_1^{\ell_1}\ldots k_s^{\ell_s} = Q(k).
$$
\end{lemma}
\proof This is a special case of Sublemma~2.15 of \cite{friedman_random_graphs}
(proven in a straightforward induction on $s$).
\proofbox

Now let 
$$
j_{11}=k_1m_1+\cdots+k_sm_s=k_1+\cdots+k_s, \qquad
j_{12}=k_{s+1}m_{s+1}+\cdots+k_tm_t
$$
(so that $j_{11}+j_{12}=j_1$).  In the notation of the above two lemmas,
letting $j'=j_{12}+j_2$,
we have
$$
f_{\vec m}(k) = \sum_{j_{11}+j'=k} (d-1)^{j_{11}} Q(j_{11})
g_{\vec m}(j')
$$
\begin{equation}\label{eq:Qgsum}
= \sum_{r=1}^{k-s} (d-1)^{k-r} Q(k-r) g_{\vec m}(r);
\end{equation}
here we sum until $r=k-s$ since Lemma~\ref{lm:Q} requires $k\ge s$ (and
$Q(k-r)$ vanishes for $k<s$), and we
sum from $r=1$ to simplify the expression,
despite the fact that $g_{\vec m}(r)$ clearly vanishes
for 
$$
r<B(m_{s+1}+\cdots+m_t)+m_{t+1}k_{t+1}+\cdots+m_bk_b.
$$
The
sum in equation~(\ref{eq:Qgsum}) is clearly $\Sigma_1(k)-\Sigma_2(k)$, where
\begin{eqnarray*}
\Sigma_1(x) =& \sum_{r=1}^{\infty} &(d-1)^{k-r} Q(x-r) g_{\vec m}(r), \\
\Sigma_2(x) =& \sum_{r=k-s+1}^{\infty}& (d-1)^{k-r} Q(x-r) g_{\vec m}(r),
\end{eqnarray*}
assuming these sums converge.

We claim $\Sigma_1(k)$ will be the principal part of $f_{\vec m}(k)$,
and $\Sigma_2(k)$ will be the error term.  First we need the following
lemma.

\begin{lemma}\label{lm:useful_aux}
For any positive integer, $D$, there is a $C_2$ such that the following
holds.
Let $g(r)$ be a function defined on non-negative integers, $r$,
such that $|g(r)|\le C_1r^D\rho^r$, with $\rho<1$.  Let $Q=Q(x)$ 
be any polynomial
of degree at most $D$.
Then (1) the infinite sum
$$
h(x) = \sum_{r=1}^\infty Q(x-r) g(r)
$$
is convergent (in coefficient norm), (2) the degree of $h$ is that of $Q$, and
(3) we have
$$
|h|\le C_1C_2(1-\rho)^{-2D} |Q|.
$$
The same is true for the
sum
$$
h_u(x) = \sum_{r=u+1}^\infty Q(x-r) g(r),
$$
for any positive integer $u$, except that we replace the last claim with
the estimate
$$
|h_u|\le C_1C_2u^{2D}(1-\rho)^{-2D}\rho^u|Q|
$$
(however, the $C_2$ in this equation might need to be larger than that
in the estimate for $h$).
\end{lemma}
\proof 
First we observe that since
$$
\sum_{r=0}^\infty \binom{r}{j}\rho^r = \frac{\rho^{j+1}}{(1-\rho)^j},
$$
we have
$$
\sum_{r=0}^\infty r^j\rho^r = \frac{q_j(\rho)}{(1-\rho)^j},
$$
where $q_j$ is some polynomial of degree at most $j+1$.

We first prove the claims on $h$.
By the binomial theorem, and the fact that $Q$'s degree is bounded, it
suffices to examine only the cases where 
$Q(x-r)$ is replaced by $x^ir^j$ for $i+j\le D$.
In this case $h(x)$ becomes
$$
\sum_{r=1}^\infty x^ir^jg(r) =
x^i\sum_{r=1}^\infty r^jg(r),
$$
and we have
$$
\sum_{r=1}^\infty r^j |g(r)| \le
\sum_{r=0}^\infty r^j C_1r^D\rho^r = \frac{C_1\rho^{j+D+1}}{(1-\rho)^{j+D}}.
$$
This establishes the claim on $h$.  The claim on
$h_u$ is reduced to $h$ via
$$
h_u(x) = \sum_{r=1}^\infty \tilde Q(x-r) \tilde g(r),
$$
where $\tilde Q(x)=Q(x-u)$ and $\tilde g(r)=g(r+u)$.  So $\tilde g$
satisfies the same estimate as does $g$, except with an extra factor 
of $(r+u)^Dr^{-D}\rho^u\le Cu^D\rho^u$; 
the binomial theorem implies that $|\tilde Q|$ is at most
$|Q|u^D$ times a constant depending on $D$.
\proofbox

We continue with the proof of Theorem~\ref{th:f_vecm_est}.
We have
$$
(d-1)^{-k}\Sigma_1(k) = \sum_{r=1}^\infty Q(k-r) [(d-1)^{-r}g_{\vec m}(r)]
= \sum_{r=1}^\infty Q(k-r) \tilde g(r),
$$
where $\tilde g(r)=(d-1)^{-r}g_{\vec m}(r)$.  Now $Q$ is fixed in the
theorem, so $|Q|$ can be regarded as a constant.  Also, since
$$
|g_{\vec m}(r)| \le cr^c(d-1)^{(r-BM_1-M_2+Bc)/2},
$$
according to Lemma~\ref{lm:crucial_B_type_est}, we have
$$
|\tilde g(r)| \le cr^c(d-1)^{(-BM_1-M_2+Bc)/2}.
$$
It follows that $(d-1)^{-k}\Sigma_1(k)=h(k)$, where $h$ is a polynomial with
$$
|h|\le c (d-1)^{(-BM_1-M_2+Bc)/2},
$$
assuming that $d>2$ (so that $1-\rho$ with $\rho=(d-1)^{-1/2}$ is strictly
positive).

Furthermore, Lemma~\ref{lm:useful_aux} also implies that
$$
|\Sigma_2|\le c(d-1)^k (k-s+1)^{2D}(d-1)^{-(k-s)/2}(d-1)^{(-BM_1-M_2+Bc)/2}
$$
$$
\le c'k^{2D}(d-1)^{(-BM_1-M_2+Bc+k)/2}.
$$
Now we see that $\Sigma_1(k)+\Sigma_2(k)$ is the decomposition of 
$f_{\vec m}(k)$ into principal and error terms, as claimed before, and
that these terms satisfy the bounds stated in Theorem~\ref{th:f_vecm_est}.

Finally we indicate the minor changes when $s=0$.  In this case we take
$Q$ to be the function $Q(0)=1$ and $Q$ vanishing elsewhere.
Then $f_{\vec m}(k)=g_{\vec m}(k)$, so
Lemma~\ref{lm:crucial_B_type_est} shows that $f_{\vec m}$ is {\dtreelike}
with zero principal part.
\proofbox

We continue with the proof of Theorem~\ref{th:reduce3}.
We are studying
$$
f(k) = \sum_{\vec m} W_{\widetilde T}(\vec m;S,\tset)f_{\vec m}(k).
$$
Set 
$$
F(M_1,M_2;k) = \sum_{\substack{m_1+\ldots+m_t=M_1\\
\scriptstyle m_{t+1}+\cdots+m_b=M_2}}
W_{\widetilde T}(\vec m;S,\tset)f_{\vec m}(k),
$$
so that
$$
f(k)=\sum_{M_1,M_2>0}
F(M_1,M_2;k).
$$
Theorem~\ref{th:crucial_cycle_count} combined with 
Theorem~\ref{th:f_vecm_est} gives that for fixed $M_1,M_2$ we have
that $F(M_1,M_2;k)$ is {\dtreelike} with principle term
$(d-1)^k P_{M_1,M_2}(k)$ and error term $E_{M_1,M_2}(k)$ where
$$
|P_{M_1,M_2}| \le (d-1)^{(-BM_1-M_2+c)/2}
c (\sqrt{d-1}-\epsilon)^{BM_1+M_2},
$$
$$
|E_{M_1,M_2}(k)| \le
ck^c (d-1)^{(k-BM_1-M_2+Bc)/2}
c (\sqrt{d-1}-\epsilon)^{BM_1+M_2},
$$
and the degree of $P_{M_1,M_2}$ is bounded independent of $M_1,M_2$.
So we sum over all $M_1,M_2$
to conclude that
$f$ is {\dtreelike}.
\proofbox
\section{Selective Traces In Graphs With (Without) Tangles}

Let us review our general approach to the Alon conjecture.
We are interested in expansions in $1/n$
of the expected value of the $k$-th
irreducible trace.  Unfortunately these expansions have some coefficients
that fail to be {\dtreelike}, and, as explained in Section~2, this
prevents us from proving the Alon conjecture.
Replacing irreducible traces with
selective traces gives
{\dtreelike} coefficients up to any desired power of $1/n$.  However,
selectivity, in the presence of appropriate tangles, modifies the
irreducible trace in a way that seems hard to control.  
Thus we don't know how to use the results of the last section to
conclude the Alon conjecture.

The last main
idea of the proof is to get an expansion for the expected value
of selective traces counted only when appropriate tangles are present
(i.e., the selective trace multiplied by a characteristic function
over those graphs in $\cgnd$ with appropriate tangles).
The methods of the last section generalize, rather tediously, to
such expansions.  These expansions will also have 
{\dtreelike} coefficients up to any desired power of $1/n$.
It follows that we also get such expansions for the expected value
of the selective trace counted only when appropriate tangles are
{\em not} present; for this count, the selective trace and irreducible
trace are the same.  This information turns out to be enough to prove
the Alon conjecture (with an auxilliary lemma proven in
Section~11).

Before doing the above, 
it is crucial to know that a certain set of tangles
is finite.

\begin{definition}
Recall that $\tseig$ is the set of supercritical tangles, i.e.,
whose $\lambda_{\ird{}}$ is at least
$\sqrt{d-1}$.  Recall that $\tseig[r]$ is the subset of elements of 
$\tseig$ of
order at most $r$.  Let $\tsmin[r]$ be the set of tangles of
$\tseig[r]$ that are minimal with respect to inclusions, i.e., that
don't have another element of $\tseig[r]$ properly included in it.
\end{definition}

\begin{lemma}\label{lm:finiteness}
The set $\tsmin[r]$ is finite.
\end{lemma}
This means that containing a supercritical tangle of order at most $r$
is equivalent to containing
one of a finite set of tangles.
\proof
Assume that $\tsmin[r]$ is not finite.  With each tangle we associate
a type which is the labelled graph obtained by supressing the degree
two vertices.  Since there are finitely many types of order at most $r$,
there must be an infinite number of $\tsmin[r]$ tangles of some type,
$T$.  By passing to a subsequence we may assume there is an infinite
sequence of $\tsmin[r]$ elements, $\{\tang_i\}$, such that for each edge
of $T$ the associated labelling is either constant or has length
tending to infinity; furthermore, the length must tend to infinity 
along at least one $T$ edge.  Let $\tang_\infty$ be the limiting tangle,
where we discard all edges with length tending to infinity.  

We claim $\lambda_{\ird{}}(\tang_i)=\lambda_1(T_{\ird{}}^i)$, where
$T_{\ird{}}^i$ is the VLG with underlying directed graph $T_{\ird{}}$,
and where $e=(v_1,v_2)\in E_{T_{\ird{}}}$ has length equal to the length,
$\ell(v_1)$,
of $v_1$ in $\tang_i$ (recall $v_1$ can be viewed as a directed edge of
$T$); indeed, with a vertex path $v_1,\ldots,v_r$ in $T_{\ird{}}$ with
$v_j=(u_j,u_{j+1})$ for $u_j\in V_T$, we associate the walk
$u_1,\ldots,u_{r+1}$, which has $\tang_i$ length equal to the sum of
the $\ell(v_i)$.
For a closed walk, where $u_{r+1}=u_1$, its length (in $T_{\ird{}}^i$)
$\ell(v_1)+\cdots+\ell(v_r)$, which corresponds to a unique 
$(\tang_i)_{\ird{}}$ closed walk of the same length, arising from
the subdivided $v_j$.  This correspondence is clearly a length preserving
bijection between $T_{\ird{}}^i$ closed walks and $(\tang_i)_{\ird{}}$ closed walks.
Hence $\lambda_{\ird{}}(\tang_i)=\lambda_1(T_{\ird{}}^i)$.

Similarly $\lambda_{\ird{}}(\tang_\infty)=\lambda_1(T_{\ird{}}^\infty)$
with $T_{\ird{}}^\infty$ defined similarly.  Now by 
Theorem~\ref{th:infinite_edge}, $\lambda_1(T_{\ird{}}^i)\to
\lambda_1(T_{\ird{}}^\infty)$, and so
$\lambda_{\ird{}}(\tang_\infty)\ge \sqrt{d-1}$.  Also $\tang_\infty$'s
order is less than that of the $\tang_i$ (because of the edge removal(s)).
Hence
$\tang_\infty$ is again a $\tseig[r]$ tangle.
But $\tang_\infty$ properly
contains (all) $\tang_i$, which contradicts the supposed
minimality of
the $\tang_i$.
Hence $\tsmin[r]$ is finite.
\proofbox
We illustrate the above lemma with an example.  Let $\tang_i$ be a
sequence of tangles whose underlying graph is the same except for one
beaded cycle of length $i$ about some vertex.  
This infinite collection of tangles would prove troublesome
to the methods of this section.
However either (1) $\lambda_{\ird{}}(\tang_i)<\sqrt{d-1}$ for some $i$,
at which point only finitely many of the $\tang_i$ are relevant,
or (2) the limiting tangle, $\tang_\infty$, has
$\lambda_{\ird{}}(\tang_\infty)\ge\sqrt{d-1}$, in which case a $\tang_i$
inclusion implies a $\tang_\infty$ inclusion.

\begin{theorem}\label{th:with}
Let $\tset$ be a finite set of pruned (nonempty)
tangles of order at least $1$.  Let $\chi_\tset$ be the
indicator function of the event that $G\in\cgnd$ contains a (i.e., at 
least one) tangle
from $\tset$, i.e.,
$$
\chi_\tset(G)=\left\{ \begin{array}{ll} 1 & \mbox{if $G$ contains a tangle
from $\tset$,} \\ 0 & \mbox{if not.}\end{array} \right. 
$$
Let $\tset'$ be a set of tangles including all
supercritical tangles of order less than $r$.  
Then for any $r$ there is an $S_0=S_0(r,\tset,\tset')$ such
that for all $S\ge S_0$ we have an expansion
\begin{equation}\label{eq:with_expansion}
\E{\chi_\tset\irdsel_{S,\tset'}(G;k)} = f_0(k)+\frac{  f_1(k)}{n}+\cdots+
\frac{  f_{r-1}(k)}{n^{r-1}}+\frac{{\rm error}}{n^r},
\end{equation}
where 
the $f_i$ are {\dtreelike} and the error term satisfies 
$$
|{\rm error}| \le c k^{\widetilde r}(d-1)^k
$$
with $c$ and $\widetilde r$ depending only on $r$, $\tset$, and $\tset'$.
\end{theorem}

We believe this theorem is true even if $\tset$ contains cycles, i.e.,
tangles of order $0$.  But to prove this would be more difficult,
since the last part of Lemma~\ref{lm:truncate} would not be true
(see the remark that follows this lemma).  

Before giving the proof of Theorem~\ref{th:with}, we give an important
corollary of it and Theorem~\ref{th:main_expansion}.
\begin{corollary} With notation and conditions
as in Theorem~\ref{th:with}, we have
that
$$
\E{(1-\chi_\tset)\irdsel_{S,\tset'}(G;k)}
$$
also has an expansion of the form given by the right-hand-side of
equation~(\ref{eq:with_expansion}).
\end{corollary}

\proof {\bf (of Theorem~\ref{th:with})}\ \ 
We want to generalize walk sums, forms, types, etc. into structures
that incorporate the presence of a $\tset$ tangle.  We shall first
explain why we run into inclusion/exclusion and tangle automophisms
(as in Theorem~\ref{th:tangle_count}).

Define a
{\em potential tangle specialization} as a pair, $(\Omega,\sigma)$,
of a tangle, $\Omega$, and an inclusion $\sigma\from V_\Omega\to\intn$.
Let $\chi_{\Omega,\sigma}$ denote the indicator function of the event
that $\Omega$ is contained in $G\in\cgnd$ via the map $\sigma$.
Let $\chi_{(w;\vec t\;)}$ be the indicator function of the event,
${\cal E}(w;\vec t\;)$, of the
potential
walk $(w;\vec t\;)$.
We plan to study sums involving various 
$\chi_{(w;\vec t\;)}\chi_{\Omega,\sigma}$.

For now consider $\chi_{\Omega,\sigma}$ alone.  Since the word ``form''
was used earlier to mean ``look at the graph traced out and
forget the specific values of the vertices,''
it makes sense to view $\Omega$ itself as the ``form'' of $(\Omega,\sigma)$,
and let
$$
\E{\Omega}_n = \sum_\sigma \E{\chi_{\Omega,\sigma}}
=\frac{n!}{(n-v)!}\prod_{i=1}^{d/2} \frac{(n-a_i)!}{n!},
$$
where, as usual, the $a_i$ are the number of $\Omega$ edges labelled $\pi_i$
and $v$ is the number of vertices in $\Omega$.  The problem is that
$\E{\Omega}_n$ gives the expected number of times $\Omega$ is included
into a random graph;  when computing traces and walk sums, we do wish
to count; but when looking at tangle inclusions, when don't wish to count---
we want to compute only the function $\chi_\tset$, i.e., we want a $1$ when
a $\tset$ tangle is present, and a $0$ otherwise.

Given two tangles, $\tang,\tang'$, let $N(\tang,\tang')$ denote the
number of inclusions of $\tang$ into $\tang'$.  Of course,
$N(\tang,\tang)$ is the number of automorphisms of $\tang$.  Also,
$\E{\Omega}_n$ is just the expected value of $N(\Omega,G)$ for a 
$G\in\cgnd$ (viewing $G$ as a tangle).  Let $\tang\le\tang'$
(respectively, $\tang<\tang'$) denote that $\tang$ has an inclusion
(respectively, proper inclusion) into $\tang'$.
By a $\tset$-tangle we mean any tangle isomorphic to an element of
$\tset$.
A {\em derived tangle of $\tset$} is a tangle that is the 
nonempty union of $\tset$-tangles.  Let
$\tset^+$ be a collection of one derived tangle of $\tset$ in
every tangle isomorphism class.

\begin{proposition} There exist reals numbers $\{\mob_\Omega\}_{\Omega\in
\tset^+}$ such that for any $\Omega'\in\tset^+$ we have
$$
\sum_{\Omega\le\Omega'} N(\Omega,\Omega')\mob_\Omega = 1.
$$
\end{proposition}
\proof This is just generalized M\"obius inversion:  $\tset^+$ is a
partially ordered set, and given $\Omega'$
there are only finitely many $\Omega$
with $\Omega<\Omega'$.  Furthermore $N(\Omega',\Omega')$ is positive for
all $\Omega'$.  So we can inductively solve for $\mob_\Omega$.
\proofbox

It follows that
\begin{eqnarray}\label{eq:mother1}
\E{\chi_\tset\irdsel_{S,\tset'}(G;k)} 
&=& \sum_{(w,\vec t\>)\in\cw} \E{\chi_{(w;\vec t\;)}\chi_\tset} \\
\label{eq:mother2}
&=& \sum_{(w,\vec t\>)\in\cw}
\quad\sum_{(\Omega,\sigma),\Omega\in\tset^+}  
\E{\chi_{(w;\vec t\;)}\chi_{\Omega,\sigma}} \mob_\Omega,
\end{eqnarray}
where $\cw$ is the walk collection corresponding to the irreducible
$(S,\tset')$ selective walks.

\begin{lemma}\label{lm:truncate}
Let $\cw_{<r}$  be those
elements of $\cw$ of order less than $r$,
and similarly for $\tset^+_{<r}$.
In equation~(\ref{eq:mother2}), by replacing the summation over $\cw$
and $\tset^+$ by summation over $\cw_{<r}$ and $\tset^+_{<r}$, the
difference is at most $C(d-1)^k n^{-r}k^{2r}$.
Furthermore, the set $\tset^+_{<r}$ is finite for each $r$.
\end{lemma}
We remark that if $\tset$ could contain a tangle of order $0$, i.e.,
one whose underlying graph is a cycle, then $\tset^+_{<1}$ would be
infinite, containing arbitrarily large disjoint unions of tangles
whose underlying graph is a cycle.
\proof
We know that we can ignore $(w;\vec t\;)$'s of order at least $r$ by
using
Theorem~\ref{th:SSIICexpansion} and the fact that $|\chi_\tset|\le 1$
in equation~(\ref{eq:mother1}).
This leaves us with a sum over $\cw_{<r}$ and $\tset^+$, and it is left
to see what happens when $\tset^+$ is replaced by $\tset^+_{<r}$.

For the $(\Omega,\sigma)$'s, notice that every tangle in $\tset^+$
of order at least $r$ must contain a tangle of order between
$r$ and $r+s-1$, where $s$ is the maximum number of edges in a $\tset$-tangle;
this follows because taking the union of any graph with a $\tset$-tangle
increases the number of edges by at most $s$.
Lemma~\ref{lm:order_increases} shows that there are only finitely many
elements of $\tset^+$ of order at most $r+s-1$ (and that any such element
is the union of at most $r+s-1$ elements of $\tset$).
Theorem~\ref{th:tangle_count} now shows that there is a $C=C(r)$
such that a graph contains a $\tset^+$-tangle of order at least $r$
with probability at most $Cn^{-r}$.  Let $\tset^+_{<r}$ be the subset
of tangles of order less than $r$ in $\tset^+$.  Then
$$
\chi_\tset(G) = h(G) + \sum_{\Omega\in \tset^+_{<r}} N(\Omega,G) \mob_\Omega,
$$
where $h(G)$ is a function bounded by a $1$ plus the finite sum
of all $|\mob_\Omega|$ over $\Omega\in \tset^+_{<r}$.
Truncating the sum
in equation~(\ref{eq:mother2}) to $\Omega\in \tset^+_{<r}$ therefore
introduces an error of at most
$$
\sum_{(w,\vec t\>)\in\cw} \E{\chi_{(w;\vec t\;)}h(G)}\le
d(d-1)^{k-1} C n^{-r} \max(h).
$$
\proofbox

According to Lemma~\ref{lm:truncate}, it suffices to fix an
$\Omega\in\tset^+_{<r}$ and to show that
\begin{equation}\label{eq:Omega_aim}
\sum_{(w,\vec t\>)\in\cw_{<r}}\sum_\sigma 
\E{\chi_{(w;\vec t\;)}\chi_{\Omega,\sigma}}
\end{equation}
has an expansion like the right-hand-side of 
equation~(\ref{eq:with_expansion}).
So fix an $\Omega\in\tset^+_{<r}$; we now define the ``form'' of
$(w;\vec t\>)$, but we incorporate into the form
the information of how $\Omega,\sigma$
overlaps with $(w;\vec t\>)$ by allowing $\Omega$ and $\Gamma$ to share
vertices and edges.

\begin{definition}
\label{de:Omega_isom}
An $\Omega$-specialization of a form, $\Gamma$,
is an inclusion $\iota\from V_\Gamma\cup V_\Omega\to\intn$.  An
$\Omega$-isomorphism of forms is an isomophism of forms $\Gamma_1\to\Gamma_2$
which is the identity from $V_{\Gamma_1}\cap  V_\Omega$ to
$V_{\Gamma_2}\cap V_\Omega$.  We introduce the notation
$$
\E{\Gamma;\Omega}_n
= \frac{n!}{(n-v)!}\prod_{i=1}^{d/2} \frac{(n-a_i)!}{n!},
$$
where $v=|V_\Gamma\cup V_\Omega|$ 
and $a_i$ is the number of $\pi_i$ labels occurring
in $\Gamma\cup\Omega$.
\end{definition}
Given $(w;\vec t\;)$ and $\Omega,\sigma$ as above, we set
$[\vec t;\sigma]$ to be the set of all pairs, 
$(\vec s,\tau)$, such that there is a permutation of
the integers taking $\vec t$ to $\vec s$ and $\sigma$ to $\tau$;
we set $[\vec t;\sigma]_n$ to the same, with the additional requirement that
the components of $\vec s$ and the image of $\tau$
lie in $\intn$.
If we let $\Gamma$ be the form of $w$, sharing vertices and edges
with $\Omega$
where $\sigma$ and $\vec t$ ``overlap,'' then
$$
\E{\Gamma;\Omega}_n = \sum_{ (\vec s,\tau)\in [\vec t;\sigma]_n}
\E{ \chi_{(w;\vec s)}
\chi_{\Omega;\tau} }.
$$
It follows that the expression in
equation~(\ref{eq:Omega_aim}) is just
\begin{equation}\label{eq:Omega_fundamental}
\sum_{\Gamma\in \cw_{<r}} W_\Gamma(\cw,k)\E{\Gamma;\Omega}_n,
\end{equation}
summed over one $\Gamma$ from each $\Omega$-isomophism class
(compare equation~(\ref{eq:fundamental})).

We now go through the rest of Sections~5, 6, and 8, indicating how to
modify our results to allow deal with sums in 
equation~(\ref{eq:Omega_fundamental}).
Let us begin by remarking that the expansion of Theorem~\ref{th:exp_polys}
generalizes easily here, in that
$$
\E{\Gamma;\Omega}_n = n^{v-e} \biggl( p_0+\frac{p_1}{n}+\cdots+\frac{p_q}{n^q}
+ \frac{{\rm error}}{n^{q+1}} \biggr)
$$
where
the $p_i=p_i(a_1,\ldots,a_{d/2},v)$ are the expansion polynomials, and
$$
|{\rm error}| \le e^{qk/(n-k)}\bigl( v(v-1)/2 + a_1(a_1-1)/2 + \cdots
+ a_{d/2}(a_{d/2}-1)/2 \bigr)^q
$$
(by equation~(\ref{eq:error_term}) ).  Since our potential walk is of
length $k$, we have
$$
v \le k + |V_\Omega|,
$$
and
$$
\sum a_i \le k + |E_\Omega|.
$$
Since $\Omega$ is fixed, it follows that the error is bounded
by $ck^{2r-2}$ in the range $k\le n/2$, 
since the order of $\Gamma\cup\Omega$ is at least $1$
(since $\Gamma\cup\Omega$ is pruned and contains an element of $\tset$).

We next claim that given $w$ (and our fixed
$\Omega$), the number of equivalence 
classes $[\vec t;\sigma]$ corresponding to $w$ and $\Omega$ is
bounded by $Ck^{2r+|V_\Omega|}$;
indeed, Lemma~\ref{lm:equiv_classes} shows there at most $Ck^{2r}$
$\vec t$ classes.  The additional $|V_\Omega|$ choices of $\sigma$
values can be chosen from at most $k+|V_\Omega|-1$ ``old'' values plus
one ``new'' value each time, for a total number of equivalence classes
of at most $Ck^{2r}$ times $(k+|V_\Omega|)^{|V_\Omega|}$, which is
the claimed bound.

We obtain that the sum in equation~(\ref{eq:Omega_fundamental}) is 
\begin{equation}\label{eq:Omega_Gamma_exp}
\sum_{\Gamma\in \cw_{<r}}  W_\Gamma(\cw,k)\biggl[ 
\frac{p_0(\Gamma;\Omega)}{n^{\ord(\Gamma;\Omega)}} + \cdots + 
\frac{p_{r-1-\ord(\Gamma;\Omega)}(\Gamma;\Omega)}{n^{r-1}}
+ \frac{{\rm error}(\Gamma;\Omega)}{n^r} \biggr]
\end{equation}
with $p_i(\Gamma)=p_i\bigl(a_1(\Gamma;\Omega),\ldots,v(\Gamma;\Omega)\bigr)$ the
expansion polynomials,
with
$$
\sum_{\Gamma\in \cw_{<r}}  W_\Gamma(\cw,k) {\rm error}(\Gamma;\Omega) = 
O(k^{4r-2+|V_\Omega|})(d-1)^k.
$$

To attack the expansion polynomial sums in equation~(\ref{eq:Omega_Gamma_exp}),
we introduce straightforward generalizations of types and new types.
\begin{definition} An $\Omega$-type\mythreeindex{Omega-type}{$\Omega$-type}{
a generalization of type that incorporates information of an included tangle,
$\Omega$} is an oriented graph 
$G_T=(V_T,E_T)$, with vertex and edge partial numberings
such that (1) $\Omega$ is a subgraph of $G_T$,
(2) all vertices of $G_T$
except possibly the first one and possibly $V_\Omega$ vertices
are of degree
at least $3$, and (3) all vertices and edges not in $\Omega$ are numbered.
\end{definition}
To a form, $\Gamma$, we associate its $\Omega$-type, $T$, by taking
$\Gamma\cup\Omega$ and supressing the degree $2$ vertices in
$V_\Gamma\setminus V_\Omega$.  $T$'s edges are partially numbered since
we number the edges in they order they must be traversed by a 
corresponding walk, and we don't number an edge if the walk doesn't
traverse the edge (such edges are the edges of $E_\Omega\setminus E_\Gamma$).

\begin{definition} A $B$-new $\Omega$-type is a collection
$\widetilde T=
(T;\El,\Ef;{\vec k}^{\rm fixed})$,
(1) a lettered $\Omega$-type,
$T$, (2) a partition of $E_T$ into two sets, $\El,\Ef$, (3) for each
$\ned_i\in \Ef$ an {\em edge length}, $\kf_i$, with $0<\kf_i<B$, and
(4) a $\Pi^+$-labelling of $\Ef$ with each $\ned_i\in \Ef$ 
labelled with a word
of length $\kf_i$.  Furthermore, we require that $E_\Omega$ is contained
in $\Ef$, and that each $E_\Omega$ edge has length $1$ in $\widetilde T$
and is labelled with its $\Omega$ label in $\widetilde T$.
\end{definition}
So conditions (2)--(4) are just as for a new type
(See Definition~\ref{de:new_type}), and condition (1) here involves an
$\Omega$-type instead of a type.

Again, it is easily seen that there are finitely many (isomorphism
classes) of $\Omega$-types of a given order, and finitely many
$B$-new $\Omega$-types belonging to a type, $T$,
for a given $B$ (and $T$).  It suffices to show that for $B$ and $S$
sufficiently large with $B>S$ we have that for any $B$-new $\Omega$-type,
$\widetilde T$, and any polynomial, $p$, in the $a_i$'s we have
$$
\sum_{\Gamma\in \widetilde T}  W_\Gamma(\cw,k)  
p\bigl( a_1(\Gamma;\Omega),\ldots,a_{d/2}(\Gamma;\Omega) \bigr)
$$
is {\dtreelike}.

For integers $\{m_i\}$ indexed over $e_i$ edges of $E_T$,
set
$$
W_{\widetilde T}(\vec m;S,\tset')
$$
to the number of walk classes of new type $\widetilde T$,
traversing the edge $e_i$ $m_i$ times, that are
irreducible $(S,\tset')$ selective cycles and that respect the
$T$ partial numbering and orientation; we assume $m_i\ge 1$ or $m_i=0$
according to whether or not $e_i$ is numbered (i.e., $e_i$ is to be
traversed on our walk); assuming $\widetilde T$
is a $B$-new type with $B>S$, this number does not depend on the
``long'' (i.e., $\El$) edge lengths.
Index the edges so that $\El=\{e_1,\ldots,e_t\}$
and $\Ef=\{e_{t+1},\ldots,e_{b'}\}$, with $\{e_{t+1},\ldots,e_b\}$ the
numbered edges (so $m_{b+1}=\cdots=m_{b'}=0$).
We claim that the proof of Theorem~\ref{th:crucial_cycle_count} shows that
$$
W_{\widetilde T,S}(M_1,M_2) =
\sum_{\substack{ m_1+\ldots+m_t=M_1\\
\scriptstyle m_{t+1}+\cdots+m_b=M_2} }
W_{\widetilde T}(\vec m;S,\tset'),
$$
satisfies the bound
$$
W_{\widetilde T,S}(M_1,M_2) \le c B (\sqrt{d-1}-\epsilon)^{cM_1+M_2}.
$$
for $S\ge S_0$
for some $c$, $S_0$, and $\epsilon$ depending only on $\widetilde T$.
This is because the argument of Theorem~\ref{th:crucial_cycle_count}
is unaffected by the two essential new features that
$T$ has over types, which are the possible presence of
(1) some more degree $2$ vertices (other than just the vertex numbered
$1$), and (2) some edges whose lengths are
fixed at $1$ (namely $E_\Omega$ edges).
But each $a_i=a_i(\Gamma;\Omega)$,
which is the number of edges labelled $\pi_i$ in $\Gamma\cup\Omega$, has
$$
a_i(\Gamma;\Omega)=a_i(\Gamma)+a_i(\Omega\setminus\Gamma),
$$
and $a_i(\Omega\setminus\Gamma)$ is fixed in
a new $\Omega$-type, $\widetilde T$ (by the edge partial numbering).  
Thus for any polynomial $p$ and new
typoid, $\widetilde T$, there is a polynomial $\widetilde p$ such that
\begin{equation}\label{eq:almost_done}
p\bigl( a_1(\Gamma;\Omega),\ldots,a_{d/2}(\Gamma;\Omega) \bigr) =
\widetilde p\bigl( 
a_1(\Gamma),\ldots,a_{d/2}(\Gamma) \bigr).
\end{equation}
But it is easy to see that Theorem~\ref{th:reduce1} holds if we extend
the definition of type and new type to allow any fixed number of vertices
of degree $2$.  Furthermore
the condition that $\Gamma$ belongs to $\widetilde T$ is the same as 
$\Gamma$ belonging to this extended notion of new type.
We conclude that either side of equation~(\ref{eq:almost_done}) is
{\dtreelike}.
\proofbox

\section{Strongly Irreducible Traces}

We wish to use Theorem~\ref{th:with} to estimate eigenvalues.  However,
it is easier to use {\em strongly irreducible traces} rather than 
irreducible traces.  We explain this and develop the properties of
the strongly irreducible trace in this section.

\begin{definition} A word $w\in\Pi^*$ is {\em strongly irreducible} if
$w$ is irreducible and $w=\sigma_1\ldots\sigma_k$ with $\sigma_1\ne
\sigma_k^{-1}$.
\end{definition}
For any of our irreducible traces, selective irreducible traces,
irreducible walk sums, etc., we can form its ``strongly irreducible''
version where we discard contributions from words that are not
strongly irreducible.
In any graph, labelled or not, one can speak of strongly
irreducible closed walks as those closed walks that are irreducible and whose last
step is not the opposite of its first step.
\begin{definition} The {$k$-th strongly irreducible trace} of a graph,
$G$, is the number of strongly irreducible closed walks of length $k$ for a
positive integer $k$; we
denote it $\sit(G,k)$ or $\sit(A,k)$ if $A$ is the adjacency matrix of $G$.
If $G$ has half-loops, we consider each half-loop to be a strongly
irreducible closed walk (of length 1) and include it in our count for
$\sit(G,1)$ or $\sit(A,1)$.
\end{definition}
Half-loops are only a concern for us in the model $\cjnd$; the reason
that we count half-loops as strongly irreducible is to make
Lemma~\ref{lm:main_sit} hold.

With a closed walk of length $k$ in $G_{\ird{}}$ about a directed edge, $e$, of 
$G$, we may associate the strongly irreducible closed walk in $G$ about the vertex
in which $e$ originates.  It follows that if $\mu_i$ are the eigenvalues
of $G_{\ird{}}$, we have 
\begin{equation}\label{eq:SIT_sum}
\sit(G,k) = \sum_{i=1}^{nd} \mu_i^k
\end{equation}
for all $k\ge 2$; for $k=1$ we must add the number of half-loops to the
right-hand-side of the above equation, since there is no edge in
$G_{\ird{}}$ from a vertex to itself when the vertex corresponds to a
half-loop.

We will study the relationship between $\irdtr{G}{k}$ and $\sit(G,k)$
and its consequences.  The most important consequence is the following
theorem.
\begin{theorem}\label{th:strongtrace}
For $|\lambda|\le d$, let
$$
\mu_{1,2}(\lambda) = \frac{\lambda\pm\sqrt{\lambda^2-4(d-1)}}{2},
$$
and set
\begin{equation}\label{eq:tilde_q}
\widetilde q_k(\lambda)=\mu_1^k(\lambda)+\mu_2^k(\lambda)+
\bigl( 1+(-1)^k \bigr)(d-2)/2.
\end{equation}
If $G$ is a $d$-regular graph with no half-loops and
adjacency matrix eigenvalues $\lambda_1\ge\cdots\ge\lambda_n$,
then
\begin{equation}\label{eq:q_k_sum}
\sit(G,k) = \sum_{i=1}^n \widetilde q_k(\lambda_i).
\end{equation}
Furthermore, if instead $G$ has no whole-loops, then the same is
true with $\widetilde q_k$ replaced by $\widehat q_k$ where
$$
\widehat q_k(x) = \left\{ \begin{array}{ll}\widetilde q_k(x)&
\mbox{if $k$ is even or $k=1$,} \\ \widetilde q_k(x)-x&
\mbox{if $k\ge 3$ is odd.}\end{array}\right.
$$
\end{theorem}
Furthermore, we shall see that $\widetilde q_k$, like the $q_k$
of Lemma~\ref{lm:chebyshev}, are polynomials of degree $k$ that may
alternatively be expressed as a simple linear combination of
Chebyshev polynomials (plus the $\pm 1$ eigenvalue contribution for
the $\widetilde q_k$).
(In a sense, equation~(\ref{eq:q_k_sum}) says that
to each eigenvalue, $\lambda$, of $G$, there correspond
eigenvalues $\mu_{1,2}(\lambda)$ of multiplicity one each and eigenvalues
$1$ and $-1$ of multiplicity $(d-2)/2$ each in $G_{\ird{}}$.)

\begin{lemma}\label{lm:main_sit}
Let $G$ be a $d$-regular graph on $n$ vertices with $h$ half-loops.
Then for all
integers $k\ge 2$ we have\footnote{
	For $k=2$ the summation in the formula to follow is ignored, since it
	ranges from $i=1$ to $i=\lfloor(k-1)/2\rfloor=0$.
}
$$
\irdtr{G}{k} = \sit(G,k)+(d-2)\sum_{i=1}^{\lfloor(k-1)/2\rfloor}(d-1)^{i-1}
\sit(G,k-2i)
$$
$$
+ \left\{\begin{array}{ll} 0 & \mbox{if $k$ is even,}\\ (d-1)^{(k-3)/2}h &
\mbox{if $k$ is odd.} \end{array}\right.
$$
Therefore for all $k\ge 4$ (and for $k=3$ when $h=0$) we have
\begin{equation}\label{eq:sit_recur}
\irdtr{G}{k}-(d-1)\irdtr{G}{k-2} = \sit(G,k)-\sit(G,k-2)
\end{equation}
\end{lemma}
\proof
The last equation follows from the previous one, with a little care
when $k=3$; indeed, for $k=3$ we have $\sit(G,k-2)=\sit(G,1)=2w+h$,
where $w,h$ are the number of whole- and half-loops.
So
$$
\irdtr{G}{3}=\sit(G,3)+(d-2)\sit(G,1)+h,
$$
and
$$
\irdtr{G}{1}=\sit(G,1)=\Trace{A} = 2w+h.
$$
Hence
$$
\irdtr{G}{3}-(d-1)\irdtr{G}{1}-\sit(G,3)+\sit(G,1)=h.
$$
Thus equation~(\ref{eq:sit_recur}) holds with $k=3$ if $h=0$.
We similarly show that this equation holds regardless of $h$ for $k\ge 4$.

So it suffices to prove the first equation of the lemma.
Each irreducible closed walk about a vertex, $v$, begins by traversing a
path, $p$, to a vertex, $w$,
then follows a strongly irreducible (nonempty)
closed walk about $w$, and then backtracks
over $p$ (and this statement is only true
if we count half-loops as strongly irreducible); each irreducible closed walk
has a uniquely determined such $p$ and $w$.
So we may count irreducible closed walks of length $k$ in $G$ by counting how
many paths of length $i$ there are from $w$ that when combined with
a strongly irreducible closed walk about $w$
of length $k-2i$ yield an irreducible closed walk, $C$, of length $k$.  
The strongly irreducible closed walk's length, $k-2i$, must be positive, or
else $C$ isn't irreducible.
If $C$
is of length at least $2$ or is a whole-loop, then there are
$d-2$ possibilities for the first edge of the path (since two edges are
ruled out by $C$ in either case), and $d-1$ possibilities for all edge
choices thereafter.  For a half-loop there are $d-1$ possibilities for 
the first edge (since only the single half-loop edge is ruled out).
So the contribution per strongly irreducible closed walk of length $k-2i$
is $1$ for $i=0$, $(d-2)(d-1)^{i-1}$ for $i\ge 1$, except in a half-loop,
where the contribution per half-loop is $(d-1)^i$, i.e., an additional
$(d-1)^{i-1}$ beyond the standard contribution.  Since $k$ is odd and
$i=(k-1)/2$ in the case of a half-loop, the additional amount beyond
the standard contribution is $(d-1)^{(k-3)/2}$ per half-loop.
\proofbox

We return to the proof of the theorem.  
First assume that $G$ has no half-loops.
We will prove by induction on $k\ge1$
that
\begin{equation}\label{eq:tilde_trace}
\sit(G,k)=\sum_{i=1}^n \widetilde q_k(\lambda_i),
\end{equation}
where $\widetilde q_k$ are polynomials of degree $k$.
Clearly
$$
\sit(G,1)={\rm Trace}(A),
$$
and so $\widetilde q_1$ exists as desired with
$\widetilde q_1(\lambda)=\lambda$.  Of the closed walks of length $2$, all
irreducible closed walks are strongly irreducible, so $\widetilde q_2(\lambda)
=\lambda^2-d$.
Lemmas~\ref{lm:main_sit} and \ref{lm:chebyshev} now imply (by induction on
$k$) that
polynomials $\widetilde q_k(\lambda)$ exist of degree $k$ satisfying
equation~(\ref{eq:tilde_trace}), and that the $\widetilde q_k$, for 
$k\ge 2$, are
annihilated by
\begin{equation}\label{eq:recurrence}
(\sigma_k^2-1)\bigl(\sigma_k^2-\lambda\sigma_k+(d-1)\bigr),
\end{equation}
where $\sigma_k$ is the ``shift in $k$'' operator, i.e., 
$\sigma_k\bigl( f(k) \bigr)=f(k+1)$ (here we use the fact that the $q_k$
are annihilated by $\sigma_k^2-\lambda\sigma_k+(d-1)$, mentioned below
Lemma~\ref{lm:chebyshev}).
Since $\mu_{1,2}(\lambda),\pm 1$ are the four roots in $\sigma_k$ 
of equation~(\ref{eq:recurrence}), we have
$$
\widetilde q_k(\lambda) = c_1\mu_1^k+c_2\mu_2^k+c_3+c_4(-1)^k,
$$
where the $c_i=c_i(\lambda)$ assuming that the four roots are distinct.
There are now two ways to finish the theorem.

The first method to finish the theorem is to calculate
$\widetilde q_3,\widetilde q_4$ and verify equation~(\ref{eq:tilde_q}) 
holds for $k=2,3,4,5$, i.e.,
that $c_1=c_2=1$ and $c_3=c_4=(d-2)/2$ work in those cases.  Then by
uniqueness (i.e., the nonvanishing of a Vandermonde determinant),
those $c_i$'s must be the unique $c_i$'s that work for all $k$.

Another way to finish the theorem is to use 
the fact that $c_1(d)=1$ (see the remark after
Lemma~\ref{lm:irdeigens}), and then argue that $c_1(\lambda)=c_2(\lambda)=1$
by analytic continuation.  First, the $c_i$'s are the unique solutions to a
$4\times 4$ system of equations with coefficient analytic in $\lambda$;
hence the $c_i$ are, indeed, analytic in $\lambda$.  Next,
notice that $\mu_1(\lambda)$ at
$\lambda=d$ analytically continues to $\mu_2(\lambda)$ at $\lambda=d$
by one loop about $2\sqrt{d-1}$; thus is suffices to prove that $c_1=1$
near $d$.  
Next notice that different $\lambda$'s give different $\mu$'s
(indeed, $r^2-\lambda_1r
+(d-1)=0$ and $r^2-\lambda_2 r+(d-1)=0$ for the same $r$ implies
$r(\lambda_2-\lambda_1)=0$), so if $\lambda\ne d$ is an eigenvalue of 
multiplicity $k$ in a graph, then $c_1(\lambda)$ times $k$ must be 
an integer.
But there is a sequence, $z_n\to 1$, with $c_1(z_n)$ an integer
or half-integer (namely a cycle of length $n$
where each edge has multiplicity $d/2$
has $z_n=(d/2)\cos(2\pi/n)$ as an eigenvalue of multiplicty two).
So by continuity $c_1(z_n)=1$
for sufficiently large $n$, and thus $c_1$ is identically $1$.

It suffices to determine
$c_3,c_4$, which from $\widetilde q_1,\widetilde q_2$ we find are
(the constant functions) $c_3=c_4=(d-2)/2$.

Finally, if $G$ has only half-loops (no whole-loops),
then the $k$ even formula and $k=1$ formula
are unchanged.
We easily see by induction on odd $k\ge 3$ that the polynomial $\widehat q_k
=\widetilde q_k(x)-x$ works (we use the fact that $h$, the number of
half-loops, is the trace of $A_G$, i.e., $h=\sum\lambda_i$).
\proofbox

\begin{theorem} Fix an integer $d>2$ and a real
$\epsilon>0$.  There
is an $\eta>0$ such that if $G$ is a $d$-regular graph with
$|\lambda_i|\le d-\epsilon$ for all $i>1$, then the $\mu_i$ of
equation~(\ref{eq:SIT_sum}) satisfy $\mu_1=d-1$ and
$|\mu_i|\le d-\eta$ for all $i>1$.
\end{theorem}
\proof The $\mu_i$ must be $\pm 1$ or roots of the equation in $\mu$
$$
\mu^2-\mu\lambda_i+(d-1)=0,
$$
or
$$
\mu =  \frac{\lambda_i \pm \sqrt{\lambda_i^2-4(d-1)}}{2}.
$$
For $i=1$ we have $\lambda_1=d$ and the corresponding $\mu$'s are
$\mu=d-1,1$.  
The other $\mu$'s are either $1$ or come from $\lambda_i$ with
$i>1$.  But for $|\lambda_i|\le d-\epsilon$ it is easy to see that
the corresponding $\mu$'s are bounded away from $d-1$.
\proofbox

We can form a selective, strongly irreducible trace by taking
$$
\ssit_{S,\tset'}(G;k)
$$
to be the number of strongly irreducible closed walks of length $k$ that
are $(S,\tset')$-selective.  Notice that there are no more strongly
irreducible closed walks than irreducible closed walks in any lettered type,
and the strong irreducibility of a potential walk can
be determined from its image in the corresponding lettered type.
Hence the expansion theorems of Sections 6--9, especially
Theorem~\ref{th:with}, carry over to $\ssit$ replacing $\irdsel$,
by simply replacing
$$
W_{\widetilde T,S}(M_1,M_2)
$$
by the same number of walk classes of the new $\Omega$-type, $\widetilde T$, 
except requiring that the walks are strongly irreducible.
\section{A Sidestepping Lemma}

\begin{lemma}\label{lm:ugly}
Fix integers $r,\widetilde r,d$ with $d>2$,
polynomials $p_0,\ldots,p_r$, a constant, $c$, and
an integer $D$.  
Assume that for each $n$ we have complex-valued random variables
$\theta_1,\ldots,\theta_m$ such that $m=Dn$.
Assume that $1-\theta_i$ is of absolute value at most $1$, and
is purely real if its absolute value is greater than $(d-1)^{-1/2}$.
Furthermore assume
that for all integers $k\ge 1$ we have
\begin{equation}\label{eq:ugly_start}
\E{ \sum_i (1-\theta_i)^k } = \sum_{j=0}^{r-1} p_j(k)n^{-j} \; + \;
O\bigl(k^{\widetilde r}n^{-r} + k^c (d-1)^{-k/2}\bigr).
\end{equation}
Then for sufficiently large $n$ we have
\begin{equation}\label{eq:ugly_finish}
\E{ \sum_i \chi_{\{|\theta_i|>\log^{-2}n\}} (1-\theta_i)^k }
= O(Dn^{1-(r/3)}+k^c(d-1)^{-k/2})
\end{equation}
for all $k$ with $1\le k\le n^\gamma$ for some constant $\gamma>0$, where the
constant $\gamma$ and the constant in the $O(\,\cdot\,)$ notation depends
only on $r,\widetilde r,d$ and the maximum degree of the $p_i$.
\end{lemma}

After this section we will apply this lemma with the $(d-1)(1-\theta_i)$
being the eigenvalues of $G_{\ird{}}$,
with $k$ proportional (for fixed $r$) to $\log n$.

If $\sigma_k$ denotes the ``shift with respect to $k$,'' i.e., 
$\sigma_k\bigl( f(k) \bigr) = f(k+1)$, then some fixed power of 
$\sigma_k-1$ annihilates the $p_j(k)$, and also
$$
(\sigma_k-1)^i (1-\theta)^k = (-\theta)^i (1-\theta)^k.
$$
This allows us to say a lot about the $\theta_i$ by applying some power
of $\sigma_k-1$ to equation~(\ref{eq:ugly_start}).
For the irreducible trace, the $(1-\theta_i)^k$ are replaced by certain
Chebyshev polynomials of $\theta_i$, and applying powers of $\sigma_k-1$
to them seems more awkward; this is why we have introduced strongly
irreducible traces.

In \cite{friedman_random_graphs}, we worked with irreducible traces, not
strongly irreducible traces.  There we had the $(1-\theta_i)^k$ replaced
by Chebyshev polynomials of $1-\theta_i$; however (1) we knew that
$\theta_1=0$ and $\theta_i$ was bounded away from $0$ with probability
$1-O(n^{1-d})$, and (2) we could only prove the asymptotic expansion up to
$r$ which was roughly propotional to $d^{1/2}$.  So we could directly
apply the analogue of equation~(\ref{eq:ugly_start}) with $k$ roughly
$\log^2 n$ to determine that $p_0(k)=1$ and the higher $p_j$ vanish
(up to $j$ roughly proportional to $d^{1/2}$).  In this paper the
arbitrary length of the asymptotic expansion for a type of trace comes
at the cost of having far less control over the $\theta_i$, and we have
no ability to determine the $p_j$ exactly.  Fortunately
Lemma~\ref{lm:ugly} allows us to control the $\theta_i$ bounded away
from $0$, and fortunately we will see that
the $\theta_i$ for $i>1$ are bounded away
from $0$ with probability $1-O(n^{-\taufund})$.

We wish to comment that one expects polynomials $p_j=p_j(k)$ to arise from
the binomial expansion.  Namely (by Taylor's theorem),
\begin{equation}\label{binom_expansion}
(1-\theta)^k = \sum_{i=0}^{s-1} \binom{k}{i}(-\theta)^i \; + \;
\binom{k}{s}(-\theta)^s(1-\xi)^{k-s},
\end{equation}
for some $\xi\in[0,\theta]$.  So for those $\theta\le n^{-\beta}$ for some
constant $\beta>0$, we may take $s$ roughly $r/\beta$ and get an error term
in $\xi$ bounded by roughly $O(n^{-r})$; we get a similarly bounded error
term when taking expected
values of $(1-\theta)^k\chi_E$ where $E$ is
the event that $\theta\le n^{-\beta}$.
In this way, equation~(\ref{binom_expansion}) could give rise to the terms
of an asymptotic expansion.
\proof (of Lemma~\ref{lm:ugly}).
Let $s$ be a fixed even integer such that
the maximum degree of the $p_j$ is at most $s-1$.  We apply $(\sigma_k-1)^s$ to
equation~(\ref{eq:ugly_start}) to conclude that
$$
\E{ \sum_i (-\theta_i)^s(1-\theta_i)^k } = 
O\bigl(k^{\widetilde r}n^{-r} + k^c (d-1)^{-k/2}\bigr)
$$
for any $k$, where the constant in the $O(\,\cdot\,)$ notation depends
on $d$ and $s$.  We conclude that for $\log^2 n\le k\le n^{r/(2\widetilde r)}$
we have
$$
\E{ \sum_i \theta_i^s(1-\theta_i)^k } = O(n^{-r/2}).
$$
Applying this for $k=\lfloor n^\gamma\rfloor$ and 
$k=\lfloor\log^2 n\rfloor$, where
$\gamma=r/(2\widetilde r)$, and subtracting
we conclude
$$
\E{ \sum_i \theta_i^s \bigl( (1-\theta_i)^{\lfloor\log^2n\rfloor} - (1-\theta_i)^{\lfloor n^\gamma\rfloor}
 \bigr)}
 = O(n^{-r/2}).
$$
Since $\theta_i$ is real unless $1-\theta_i=(d-1)^{-1/2}$, we conclude
that
$$
\E{ \sum_i |\theta_i|^s \bigl( |1-\theta_i|^{\lfloor\log^2n\rfloor} - |1-\theta_i|^{\lfloor n^\gamma\rfloor}
 \bigr)}
$$
$$
 = O(n^{-r/2})+O\bigl( n(d-1)^{-(\log^2 n)/2} \bigr)= O(n^{-r/2}).
$$
Now for any $\theta_i\le \log^{-2}n$ we have $(1-\theta_i)^{\log^2n}$
is at least roughly $1/e$ for large $n$; also for $\theta_i\ge n^{-\alpha}$
for a constant $\alpha>0$ we have $(1-\theta_i)^{n^\gamma}$ is near $0$
for large $n$ provided $\alpha<\gamma$, and also $\theta_i^s\ge n^{-s\alpha}$.
We conclude that
$$
\E{ \sum_i (\chi_{\{n^{-\alpha}\le\theta_i\le\log^{-2}n\}}) n^{-s\alpha} }
=O(n^{-r/2}),
$$
and hence, since $\theta_i$ is real for $|\theta_i|<\log^{-2}n$ for
$n$ large,
\begin{equation}\label{eq:midrange}
\E{ \chi_{n^{-\alpha}\le\theta_i\le\log^{-2}n} } = O(n^{s\alpha-(r/2)}).
\end{equation}

Let $\alpha>0$ be fixed with $s\alpha<r/6$, and set
\begin{eqnarray*}
A_1[i,n] &=& \;\mbox{The event that $\theta_i<n^{-\alpha}$},\\
A_2[i,n] &=& \;\mbox{The event that $n^{-\alpha}\le \theta_i\le\log^{-2}n$},\\
A_3[i,n] &=& \;\mbox{The event that $\log^{-2}n<|\theta_i|$}.
\end{eqnarray*}
Equation~(\ref{eq:midrange})
implies that
$$
\prob{ A_2[i,n]} = O(n^{-r/3}).
$$
Since the number of
$\theta_i$ is linear in $n$, we conclude that
\begin{equation}\label{eq:big_expected}
\E{ \sum_i \chi_{A_2[i,n]} (1-\theta_i)^k } = O(n^{1-(r/3)}).
\end{equation}

By the comment just before the proof, we have (using Taylor's theorem)
$$
\E{ \chi_{A_1[i,n]} (1-\theta_i)^k } = 
\sum_{j=0}^{j\le 2r/\alpha} \binom{k}{j} \E{ \chi_{A_1[i,n]}(-\theta_i)^j}
\;+\;O(k^{1+(2r/\alpha)}n^{-2r}).
$$
Summing over $i$ in the above involves summing over $i$ in the expected
values and multiplying the error term by a number linear in $n$.
So let
$$
q(k,n) = \sum_i\sum_{j=0}^{j\le r/\alpha} \binom{k}{j} 
\E{ \chi_{A_1[i,n]}(-\theta_i)^j},
$$
which is a polynomial of fixed degree in $k$ whose coefficients depend on
$n$ (and the $\theta_i$ which are given for each value of $n$).
Fix a $\gamma$ for which
$$
O(k^{1+(2r/\alpha)}n^{1-2r}) = O(n^{-r/3})
$$
for all $k\le n^\gamma$, i.e., fix a $\gamma$ with
$$
\gamma\bigl( 1 + (r/\alpha) \bigr) \le 5r/3-1.
$$
Then for all $k\le n^\gamma$ we have
\begin{equation}\label{eq:A1}
\E{ \sum_i \chi_{A_1[i,n]} (1-\theta_i)^k } = q(k,n) + O(n^{-r/3}).
\end{equation}

Now combine
$$
\E{ \sum_i (1-\theta_i)^k } = \sum_{j=1}^3 \E{ \sum_i \chi_{A_j[i,n]}
(1-\theta_i)^k }
$$
with equations~(\ref{eq:A1}) and (\ref{eq:big_expected}) to conclude
that for $k\le n^\gamma$ we have
$$
\E{ \sum_i (1-\theta_i)^k } = q(k,n) + \E{ \sum_i \chi_{A_3[i,n]}
(1-\theta_i)^k } + O(n^{1-(r/3)}).
$$
On the other hand, equation~(\ref{eq:ugly_start}) just says
$$
\E{ \sum_i (1-\theta_i)^k } = p(k,n) +
O\bigl(k^{\widetilde r}n^{-r} + k^c (d-1)^{-k/2}\bigr),
$$
where $p(k,n)$ is the polynomial in $k$ given as the sum of the $p_j(k)/n^j$.
Therefore
$$
\E{ \sum_i \chi_{A_3[i,n]}(1-\theta_i)^k } 
$$
\begin{equation}\label{eq:A_three_sum}
= p(k,n) -q(k,n)+
O(n^{1-(r/3)}) + O\bigl(k^{\widetilde r}n^{-r} + k^c (d-1)^{-k/2}\bigr)
\end{equation}
for all $k\le n^\gamma$.
Since $A_3[i,n]$ implies $|1-\theta_i|\le (1-\log^{-2}n)$ and thus
$|1-\theta_i|^k\le e^{O(\log^{-2}n)}$ for $k\ge \log^4n$, we have
\begin{equation}\label{eq:for_the_end}
\E{ \sum_i \chi_{A_3[i,n]}
|1-\theta_i|^k }  = O(n^{-r/3})
\end{equation}
for $k\ge\log^4n$ for $n$ sufficiently large.  We conclude that
\begin{equation}\label{eq:pq_close}
p(k,n)-q(k,n) = O(n^{1-r/3})
\end{equation}
for all $k$ with $\log^4 n\le k\le n^\gamma$.

\begin{sublemma} Let $g(k)$ be a polynomial in $k$ of degree $\le s-1$
such that $|g(i)|\le 1$ for integers $i=a,a+1,\ldots,b$ for some
integers $a,b$ with $a\le b$.  Then $|g(i)|\le 2^s-1$ for integers $i$
with
$$
a-\frac{b-a}{s-1} \le i \le a.
$$
\end{sublemma}
\proof We have $(\sigma_k-1)^s g =0$, and therefore
$$
g(x) = \sum_{i=1}^d \binom{s}{i}(-1)^{i-1} g(x+hi)
$$
for any $x$ and $h$.  Given $i<a$, let $h=a-i$ and $x=i$ in the above;
$x+h,x+2h,\ldots,x+sh$ are integers between $a$ and $b$ provided that
$$
i+(a-i)s \le b,
$$
so $i\ge (as-b)/(s-1)$ or $i\ge a-(b-a)/(s-1)$.  If so, then
$$
|g(i)| \le \sum_{i=1}^d \binom{s}{i} |g(x+hi)| \le 
\sum_{i=1}^d \binom{s}{i} = 2^s-1.
$$
\proofbox

Recall equation~(\ref{eq:pq_close}) and the fact that $p$ and $q$ are
polynomials in $k$ (for fixed $n$) of bounded degree.  So applying the
above sublemma for $a=2\lceil(\log^4 n)/2\rceil$ 
and $b=2\lfloor n^\gamma/2\rfloor$
implies
that $p(k,n)-q(k,n)=O(n^{1-(r/3)})$ for $1\le k\le\log^4n$.
Equation~(\ref{eq:pq_close}) now holds for all $1\le k\le n^\gamma$.
We conlude that equation~(\ref{eq:for_the_end}) holds 
for all $1\le k\le n^\gamma$.
Adding this to
equation~(\ref{eq:big_expected}) yields the desired
equation~(\ref{eq:ugly_finish}).
\proofbox
\section{Magnification Theorems}

In this section we use standard counting arguments to prove theorems
implying
``magnification'' or ``expansion'' for ``most'' random graphs;
here
``most'' means all graphs excepting a set of probability $O(n^{-\taufund})$.
These theorems will then be used with Lemma~\ref{lm:ugly} to
prove Theorems~\ref{th:main}, \ref{th:mainh}, and
\ref{th:maini}.

A graph, $G$, with $n$ vertices is said to be a $\gamma$-magnifier
if for all subsets of vertices, $A$, of size at most $n/2$ we have
$$
|\Gamma(A)-A| \ge \gamma|A|,
$$
where $\Gamma(A)$ denotes those vertices connected to some member of $A$
by an edge.  Alon has shown that any $d$ regular $\gamma$-magnifier has
$$
\lambda_2(G) \le d-\frac{\gamma^2}{4+2\gamma^2}
$$
(see \cite{alon_eigenvalues}; see \cite{dodziuk,jerrum1,jerrum2} for related
``edge magnification'' results).

\begin{definition}
Say that a $d$-regular graph on $n$ vertices is a {\em $\gamma$-spreader}
if for every subset,
$A$, of at most $n/2$ vertices we have
$$
|\Gamma(A)|\ge (1+\gamma)|A|.
$$
\end{definition}
\begin{theorem}\label{th:separation}
Let $G$ be a $d$-regular $\gamma$-spreader.
Then for all $i>1$ we have 
$$
\lambda_i^2(G) \le d^2 -\frac{\gamma^2}{4+2\gamma^2}.
$$
\end{theorem}
\proof 
Since the graph is $d$-regular, we have $|\Gamma(B)|\ge |B|$ for all subsets
of vertices, $B$.  Taking $B=\Gamma(A)$ yields
$$
|\Gamma^2(A)|\ge |\Gamma(A)| \ge (1+\gamma)|A|.
$$
Hence $G^2$, the graph on $V_G$ whose edges are paths in $G$ of length $2$
(and whose adjacency matrix is $A_G^2$), is a $d^2$-regular $\gamma$-magnifier.
Now apply Alon's result on magnification and eigenvalues to $G^2$, whose
eigenvalues are $\lambda_i^2(G)$.
\proofbox

We now establish that for all our models, a graph will be a
$\gamma$-spreader for some fixed $\gamma=\gamma(d)>0$ with probability
$1-O(n^{-\taufund})$.

\begin{theorem}\label{th:magnify_cgnd}
For any $\epsilon>0$ and even $d\ge 4$ there is a $\gamma>0$
such that $G\in\cgnd$ is a $\gamma$-spreader
with probability
$1-O(n^{-\taufund})$.
\end{theorem}
Later we shall prove this theorem for other models of random graphs, by
very similar calculations.  This theorem is easy for $d$ sufficiently
large; but when $d=4$ (or later possibly $d=3$) one has to calculate
fairly carefully.
\proof 
Fix $A,B\subset\intn$, and consider the event that $\Gamma(A)\subset B$.
We will impose the condition that $a=|A|\le n/2$ and
$|B|=a+\lfloor \gamma a\rfloor$.

Fix constants, $\gamma,C$ with $0<\gamma<1/C$.  Consider the situation
where
$|A|<C$.  In this case
$|B|=|A|=a$.
But since $G$ is $d$-regular and we have $d|A|$ edges
leaving $A$,
these edges comprise all edges incident upon $B$ (since $|B|=|A|$).
Thus $A\cup B$ is a union of connected components of $G$.  But this
cannot occur if $G$ has no supercritical
tangles of size at most $2C$ (since each connected component of $G$ has
$\lambda_{\ird{}}=d-1$).  For a constant
$C$ there are only a constant number of tangles of size at most $2C$.
Thus, by forsaking a probability of $O(n^{-\taufund})$, we may assume
that $a=|A|\ge C$ for any fixed constant, $C$ (provided that we then
take $\gamma<1/C$) for sufficiently large $n$.

So consider a random
permutation, $\pi=\pi_i$, and consider the event that $\pi$
and $\pi^{-1}$ map $A$ to $B$.
Let $C_1=A\cap B$, $C_2=A\setminus B$, $C_3=B\setminus A$, and let
$c_i=|C_i|$.  
We view $\pi$ as determined by a perfect matching of a bipartite graph
on inputs, $I$, and outputs, $O$, with $I,O$ being copies of $\intn$
(and $i\in I$ mapped to $\pi(i)\in O$).
Viewing $\pi$ as a bipartite matching, it consists of
(1) $r$ edges from $C_1$ to $C_1$ (i.e., the $I$ vertices corresponding
to $C_1$ to those $O$ vertices corresponding to $C_1$),
(2) $c_1-r$ edges from $C_1$ to $C_3$, (3) $c_1-r$ edges from $C_3$ to
$C_1$, (4) $c_2$ edges from $C_2$ to $C_3$, and (5) $c_2$ edges from
$C_3$ to $C_2$.  (This is true since a $C_2$ vertex, either input or output,
must be paired with a $C_3$ vertex, and a $C_1$ vertex must be paired with
either a $C_1$ or $C_3$ vertex.)  So the event that $\pi$ and $\pi^{-1}$
map $A$ to $B$ with $c_1,c_2,c_3,r,A,B$ all held fixed has probability
$$
p(c_1,c_2,c_3,r)=\left[ \binom{c_1}{r}^2 r! \right]
\left[ \binom{c_3}{c_1-r} (c_1-r)! \right]^2 \times
$$
$$
\left[ \binom{c_3-c_1+r}{c_2} c_2! \right]^2
\left[ n(n-1)\cdots(n-2c_1-2c_2+r+1)\right]^{-1}
$$
(The first expression in square brackets corresponds to choosing $r$
$C_1$ to $C_1$ edges; the second expression corresponds
to choosing $c_1-r$ $C_1$ to
$C_3$ edges, and is squared to include choosing the $C_3$ to $C_1$ edges;
etc.)
The probability taken over all $A,B$ of a given $c_1,c_2,c_3$ (and with
$r$ fixed) is therefore at most
\begin{equation}\label{eq:bimon_and_p}
\binom{n}{c_1,c_2,c_3,n-c_1-c_2-c_3} p^{d/2}(c_1,c_2,c_3,r).
\end{equation}
It suffices to show that this expression is $O(n^{-s})$ with
$s=\taufund+4$,
provided that $a$ is sufficiently large (and at most $n/2$), since then
we can sum equation~(\ref{eq:bimon_and_p}) over the at most $n^4$ relevant
values of $c_1,c_2,c_3,r$.
We should remind
ourselves that $c_1,c_2,c_3,r$ range over integers with
$$
c_1+c_2=a,\quad
c_1+c_3=a+\lfloor \gamma a\rfloor,\quad
r \le c_1.
$$
Furthermore, considering the expression defining $p$, we have 
$c_3-c_1+r\ge c_2$.

We now write
\begin{equation}\label{eq:b_factorials}
b=b(c_1,c_2,c_3,r,n) = \binom{n}{c_1,c_2,c_3,n-c_1-c_2-c_3}
\end{equation}
$$
= \frac{n!}{c_1!\;c_2!\;c_3!\;(n-c_1-c_2-c_3)!}
$$
and 
\begin{equation}\label{eq:p_factorials}
p=p(c_1,c_2,c_3,r,n) = \frac{(c_1!\;c_3!)^2\;(n-2c_1-2c_2+r)!}{
\bigl( (c_1-r)!\;(c_3-c_1-c_2+r)!\bigr)^2\; r!\;n!}.
\end{equation}
We make some general remarks about analyzing the factorials in the
above two equations:
\begin{enumerate}
\item All factorials in the above equations are of the form $(\mu n)!$
for some $\mu\in[0,1]$.  Stirling's formula $m!\sim (m/e)^m\sqrt{2\pi m}$
implies that
\begin{equation}\label{eq:Stirling}
\frac{1}{n}\log[ (\mu n)! ] = \mu\log(n/e) + \mu\log\mu + O\left(\frac{\log n}{
n}\right),
\end{equation}
where the constant in the $O(\,\cdot\,)$ is independent of $n$ and 
$\mu\in[0,1]$.
\item In analyzing $b$ and $p$ above, we may ignore the $\mu\log(n/e)$ term
in equation~(\ref{eq:Stirling}).  This is because $b,p$ are {\em balanced}
in that the sum of the numbers to which factorials are applied is the same
in the numerator and denominator; in other words, the $\mu\log(n/e)$ terms
in the numerator will exactly cancel those in the denominator.
\item Let $f(\theta)=-\theta\log\theta$.
We claim that for $\theta_1,\theta_2\in[0,1]$ we have
$$
|f(\theta_1)-f(\theta_2)| \le \max\bigl( f(|\Delta\theta|),f(1-|\Delta\theta|)
\bigr),\qquad\mbox{with}\quad \Delta\theta=\theta_2-\theta_1.
$$
Indeed, since $f''(\theta)=-1/\theta<0$ for $\theta>0$, $f$ is concave
in $[0,1]$, and so $g(\theta)=f(\theta+\Delta\theta)-f(\theta)$ is decreasing
in $\theta$ for $\Delta\theta$ fixed; so $|g|$'s maximum over an interval is
taken at its endpoints, and since $f(0)=f(1)=0$, the above claim is 
established.

Next, a Taylor expansion shows that $-\epsilon\log\epsilon\ge
-(1-\epsilon)\log(1-\epsilon)$ for sufficiently small $\epsilon>0$.  Hence
there is an $\epsilon_0$ such that 
\begin{equation}\label{eq:difference}
|f(\theta_1)-f(\theta_2)| \le f(|\theta_1-\theta_2|)
\end{equation}
for all $\theta_1,\theta_2\in[0,1]$ with $|\theta_1-\theta_2|\le \epsilon_0$.
\end{enumerate}

Let $\nu_i,\rho,\alpha,\delta$ ($i=1,2,3$) be the non-negative reals given by
$$
c_i=\nu_i n,\quad r=\rho n,\quad a=\alpha n,\quad \lfloor \gamma a\rfloor
=\delta n.
$$

We have that
$$
\nu_1+\nu_2=\alpha,\quad \nu_1+\nu_3=\alpha+\delta,
\quad \rho\le\nu_1,\quad \rho\ge \nu_1+\nu_2-\nu_3.
$$
We conclude that
$$
|\nu_2-\nu_3| \le \delta,\qquad |\nu_1-\rho|\le \delta.
$$
It follows 
from equation~(\ref{eq:Stirling}), remark~(2) below it, and
equation~(\ref{eq:difference}),
that we may replace $\nu_3$ with $\mu_2$ and $\rho$ with
$\nu_1$ in calculating $(\log b)/n$ and incur an additive error term
of at most $O(\delta\log\delta)$.  Thus we get
$$
\frac{\log b}{n} = h(\nu_1,\nu_2)+
O\left(|\delta\log\delta|+\frac{\log n}{n}\right),
$$
where
\begin{equation}\label{eq:ubiqu_h}
h(\nu_1,\nu_2) = -\nu_1\log\nu_1-2\nu_2\log\nu_2
-(1-\nu_1-2\nu_2)
\log(1-\nu_1-2\nu_2).
\end{equation}
Similarly we calculate
$$
\frac{-\log p}{n} = h(\nu_1,\nu_2)+
O\left(|\delta\log\delta|+\frac{\log n}{n}\right),
$$
i.e., we have
the exact same equation (!)
for $\log b$ replaced by $-\log p$
(this ``coincidence'' happens for the other models as well).
Hence
$$
\frac{\log(bp^2)}{n} = -h(\nu_1,\nu_2)
+ O\left(|\delta\log\delta|+\frac{\log n}{n}\right).
$$
Since $\nu_1+\nu_2=\alpha$, we have either (or both) $\nu_i$ are
$\ge \alpha/2$.  Hence
$$
h(\nu_1,\nu_2) \ge -(\alpha/2)\log(\alpha/2).
$$

Now we claim that for any constant $C>0$ there is a constant $\gamma>0$
such that for all $\alpha\in[0,1/2]$ we have
\begin{equation}\label{eq:alphalog}
-\alpha\log\alpha \ge -C (\gamma\alpha)\log(\gamma\alpha).
\end{equation}
Indeed, for $\gamma<1$ fixed we have
\begin{equation}\label{eq:g_alpha}
g(\alpha) = \frac{(\gamma\alpha)\log(\gamma\alpha)}{\alpha\log\alpha}
=\gamma+\frac{\gamma\log\gamma}{\log\alpha}
\end{equation}
is increasing for $\alpha\in[0,1/2]$.  Hence it suffices to choose a
$\gamma>0$ sufficiently small so that
$$
g(1/2)=\frac{(1/2)\log(1/2)}{(\gamma/2)\log(\gamma/2)} \ge C,
$$
so that
$$
g(\alpha)\ge g(1/2)\ge C,
$$
which along with equation~(\ref{eq:g_alpha}) yields 
equation~(\ref{eq:alphalog}).

It follows that for sufficiently small $\gamma>0$ we have
$$
\frac{\log(bp^2)}{n} \ge -(\alpha/2)\log(\alpha/2) + 
O\left(|\delta\log\delta|+\frac{\log n}{n}\right),
$$
and, since $\delta\le\gamma a/n=\gamma\alpha$, this expression is
$$
\ge -(\alpha/4)\log(\alpha/2) + 
O\left(\frac{\log n}{n}\right).
$$
Hence for any constant, $C_1$, there is a $C_2$ such that if $a\ge C_2$
(i.e., $\alpha\ge C_2/n$) then 
$$
\frac{\log(bp^2)}{n} \le \frac{-C_1\log n}{n}
$$
for all $n$ sufficiently large.  In other words $bp^{d/2}$,
i.e., the expression in equation~(\ref{eq:bimon_and_p}),
is at most $n^{-C_1}$; this, by the discussion after
equation~(\ref{eq:bimon_and_p}),
completes the proof.
\proofbox

\begin{theorem} Theorem~\ref{th:magnify_cgnd} holds in the models
$\chnd$, $\cind$, and $\cjnd$.
\end{theorem}
\proof
In $\chnd$ each permutation occurs with probability at most $n$ times
its probability in $\cgnd$.  Therefore the same analysis goes through,
except that $p$ is multiplied by at most a factor of $n$.  This changes
the expression for $n^{-1}\log(bp^2)$ by an $O(n^{-1}\log n)$ factor,
so the same proof carries over.

For $\cind$ we again set $C_i$ and $c_i$ as before.  A perfect matching
in $\intn$ will have
(1) $r$ vertices of $C_1$ paired amongst themselves,
(2) $c_1-r$ vertices of $C_1$ paired with $C_3$ vertices, and
(3) $c_2$ vertices of $C_2$ paired with $C_3$ vertices.
This data determines the pairing for $r+2(c_1-r)+2c_2$ vertices.
The expression for $b$, representing the number of ways the $C_i$ can
be chosen, is the same as before.  We now derive an
expression for $p$, the probability that a single perfect matching
matches all $A$ vertices to those in $B$.

For an even integer, $m$, let {\em $m$ odd factorial} be
$$
m\oddf = (m-1)(m-3)\cdots 3 = \frac{m!}{2^{m/2}(m/2)!},
$$
which is just
the number of perfect matchings of $m$ elements.  
Stirling's formula yields
$$
m\oddf \sim \sqrt{2}\;(m/e)^{m/2}
$$
(so that for our purposes $m\oddf$ can be regarded as replacable by
the square root of $m!$).

We have
$$
p=p(\{c_i\},r,n)=
\left[ \binom{c_1}{r} r\oddf \right]\;
\left[ \binom{c_3}{c_1-r} (c_1-r)! \right]\;
$$
$$
\times
\left[ \binom{c_3-c_1+r}{c_2} c_2! \right]\;
\frac{ (n-2c_1-2c_2-r)\oddf }{n\oddf}.
$$
We get that $-\log p$ is
$$
\frac{h(\nu_1,\nu_2)}{2} + 
O\left(|\delta\log\delta|+\frac{\log n}{n}\right),
$$
with $h$ as in equation~(\ref{eq:ubiqu_h}).
Since $b$ is unchanged,
by analyzing as before we see that there is a fixed $\gamma>0$ such that
for any constant $C_1$ there is a $C_2$ such that
$bp^3=O(n^{-C_1})$ provided that $a\ge C_2$.

Next we consider $\cjnd$.  Let $G$ be a random graph in $\cjnd$,
so $V_G=\intn$.
Consider the graph $G'$ formed by adding one new
vertex, $w=n+1$, to $G$ and replacing
each half-loop about a vertex, $v$, in $G$ by an edge from $v$ to $w$.
Then $G'$ is precisely distributed as an element of $\cinpd$; indeed,
a perfect matching on $V_{G'}$ matches $w$ to some element of $V_G=\intn$
and then randomly matches the remaining $n-1$ elements of $V_G$.

Now we know that $G'$ is a $\gamma$-spreader with probability
$1-O(n^{-\taufund})$.  But for any $A\subset V_G$, $\Gamma_{G'}(A)$
consists of at most one more vertex than $\Gamma_G(A)$.  Hence
for $|A|\le |V_G|/2$ and $G'$ being a $\gamma$-spreader, we have
$$
|\Gamma_G(A)\setminus A| \ge \gamma|A|-1 \ge \gamma'|A|,
$$
where $\gamma'=\gamma-(1/c')$, provided that $|A|\ge c'$.  Hence $G$
is a $\gamma$-spreader on sets, $A$, of size $\max(c,c')\le |A|\le n/2$.
On smaller sets, $A$, we have $G$ is a $\gamma'$-spreader with probability
$1-O(n^{-\taufund})$, assuming $\gamma'\le 1/\max(c,c')$, by the
argument given before for $\cgnd$.  (Notice that the $\taufund$
for $\cind$ and $\cjnd$ are the same.)  Hence a random graph in $\cjnd$ is a
$\gamma'$-spreader with probability $1-O(n^{-\taufund})$.
\proofbox

\section{Finishing the $\cgnd$ Proof}

Here we quickly finish the proof of Theorem~\ref{th:main}, which
proves Alon's conjecture for $\cgnd$.

Fix a value of $r$ to
be specified later.
Let $\tset'=\tseig[r-1]$ be the set of supercritical tangles of order less
than $r$,
and let
$\tset=\tsmin[r-1]$ (which we recall is the set of minimal $\tset'=
\tseig[r-1]$ elements with respect to inclusion);
we know that $\tset$ is finite by
Lemma~\ref{lm:finiteness}, and we recall that if $G$ contains a
$\tset'$ tangle then it contains an element of $\tset$.
The probability that $\chi_\tset(G)=1$, i.e., that
$G$ contains at least
one element of $\tset$, is at most $O(n^{-\taufund})$, since 
every one of the finitely many tangles in $\tset$ occurs with probability
proportional to $1/n$ to the order of tangle
(Theorem~\ref{th:tangle_count}), and each tangle order is
at least $\taufund$.  Given that $\chi_\tset(G)=0$, we have that $G$
contains no supercritical tangle of order less than
$r$, and hence no irreducible
closed walk
can fail to
be $(S,\tset')$-selective for any $S$.  Hence for all $S$ and $k$ we have
$$
\chi_\tset(G)=0 \quad\mbox{implies} \quad\ssit_{S,\tset'}(G;k)=\sit(G;k).
$$
Thus
$$
\E{ (1-\chi_\tset)\ssit_{S,\tset'}(G;k)} =
\E{ (1-\chi_\tset) \sit(G;k)}.
$$

Now according to Theorem~\ref{th:strongtrace} we have
$$
\sit(G;k) = \sum_{i=1}^n \mu_1^k(\lambda_i)+\mu_2^k(\lambda_i)+
\bigl( 1+(-1)^k \bigr)(d-2)/2,
$$
for even $k\ge 2$, where
$$
\mu_{1,2}(\lambda) = \frac{\lambda\pm\sqrt{\lambda^2-4(d-1)}}{2}.
$$
In other words, there are $nd$ numbers $\nu_i$,
such that $\sit(G;k)$ is the sum of
the $k$-th powers of these numbers.  
Also for each $i$ we have $\nu_i$ is not real only if
it is of absolute value $\sqrt{d-1}$.
Combining this and Theorem~\ref{th:with} we see that
$$
\theta_i=1-(1-\chi_\tset)\nu_i/(d-1)
$$
are random variables that
satisfy
the conditions of Lemma~\ref{lm:ugly} for each $i$ (and $G$).
It follows that
\begin{equation}\label{eq:verybig}
\E{ (1-\chi_\tset)\sum_{i=1}^n
\sum_{\substack{j\;\rm such\;that\cr |\mu_j(\lambda_i)|\le (d-1)(1-\log^{-2}n)}}
\mu_j(\lambda_i)^k } 
= O(Dn^{1-(r/3)}+k^c(d-1)^{-k/2})
\end{equation}
for all $k$ with $1\le k\le n^\gamma$ for some constant $\gamma>0$
depending only on $r$.

According to Theorems~\ref{th:separation} and \ref{th:magnify_cgnd}
there is an $\epsilon>0$ such that with probability
$1-O(n^{-\taufund})$ we have $|\lambda_i|\le d-\epsilon$ for all
$i\ne 1$; in this case there is an $\epsilon'=\epsilon'(\epsilon)>0$
such that $|\mu_j(\lambda_i)|\le (d-1)-\epsilon'$
for all $j=1,2$ and $i\ne 1$.

We claim that for any $G$ and an even integer $k$ we have
$$
\sum_{i\;{\rm s.t.}\;\mu_{1,2}(\lambda_i)\;{\rm not\;real}}\quad \sum_{j=1}^2 
\mu_j(\lambda_i)^k \ge  - 2(n-1)(d-1)^{k/2}
$$
indeed, if $\mu_j(\lambda_i)$ is not real, it
is of absolute value $\sqrt{d-1}$; if $\mu_j(\lambda_i)$ is real then
its $k$-th power is non-negative.

Now let $A$ be the event that $\chi_\tset=0$ and that
$|\mu_j(\lambda_i)|\le (d-1)-\epsilon'$
for all $j=1,2$ and $i\ne 1$.
Let $B=B(\eta)$ be the event that
for some $j$ and some $i\ne 1$ we have
$|\mu_j(\lambda_i)|\ge e^\eta\sqrt{d-1}$ for
an arbitrary fixed $\eta>0$.
$A\cap B$ implies that for even integer $k$ we have
$$
\sum_{i=2}^n\sum_{j=1}^2 
\mu_j(\lambda_i)^k \ge \left( e^\eta\sqrt{d-1}\right)^k - 2(n-2)(d-1)^{k/2}.
$$
It follows, using equation~(\ref{eq:verybig}), that for even $k$,
$$
\prob{ A\cap B } \left( e^\eta\sqrt{d-1}\right)^k \le
\E{ \sum_{i\;{\rm s.t.}\;\mu_{1,2}(\lambda_i)\;{\rm real}} \quad\sum_{j=1}^2 
\mu_j(\lambda_i)^k}
$$
$$
= \E{ \sum_{i=2}^n \sum_{j=1}^2 
\mu_j(\lambda_i)^k} -
\E{ \sum_{i\;{\rm s.t.}\;\mu_{1,2}(\lambda_i)\;{\rm not\;real}}\quad
\sum_{j=1}^2 
\mu_j(\lambda_i)^k}
$$
$$
\le O(Dn^{1-(r/3)}(d-1)^k+k^c(d-1)^{k/2}) + 2(n-1)(d-1)^{k/2}.
$$
We now take
$$
k = 2 \left\lceil \frac{r\log n}{3\log(d-1)} \right\rceil.
$$
We have
$$
(k/2)-1 \le  \frac{r\log n}{3\log(d-1)} \le k/2.
$$
Hence
$$
n^{-r/3}\le (d-1)^{-(k/2)+1},
$$
and so
$$
\prob{ A\cap B } \le
c\max(k^c,n) e^{-k\eta}
$$
$$
\le cn e^{-k\eta}
=cn n^{-\alpha r},
$$
where $\alpha=(2/3)\eta/\log(d-1)$, i.e. $\alpha$ is a positive constant
(depending only on $\eta$ and $d$).  Choosing $r$ so that
$\alpha r -1 > \taufund$, we have
$$
\prob{ A\cap B } = O(n^{-\taufund}).
$$
But we have already seen 
(Theorems~\ref{th:separation} and \ref{th:magnify_cgnd})
that
$$
\prob{ \complement A } = O(n^{-\taufund}),
$$
where $\complement A$ is the complement of $A$.
Hence
$$
\prob{ B} = \prob{B\cap A} + \prob{B\cap\complement A} = O(n^{-\taufund}).
$$
For any $\epsilon>0$ there is an $\eta>0$ such that $|\lambda|\ge
2\sqrt{d-1}\;+\epsilon$ implies $|\mu_i(\lambda)|\ge e^\eta\sqrt{d-1}$
for at least one $i$, which is the event
$B=B(\eta)$ above.  It follows that for any $\epsilon>0$ we have
$$
\prob{ \mbox{$|\lambda_i|\ge 2\sqrt{d-1}\;+\epsilon$ for some $i>1$}}
= O(n^{-\taufund}).
$$
This (and Theorem~\ref{th:improved_bound}) proves Theorem~\ref{th:main}.
\section{Finishing the Proofs of the Main Theorems}
\label{se:finish}

We now complete the proofs of Theorems~\ref{th:mainh}
and \ref{th:maini}, i.e., we establish the Alon
conjecture for $\chnd$, $\cind$, and $\cjnd$.

The proofs of the theorems are as the proof for $\cgnd$.
We only need to establish the following results for the different models
of random graph:
\begin{enumerate}
\item Labelling:
The model comes with edges labelled from a set $\Pi$ such that
to each $\pi\in\Pi$ we associate a $\pi^{-1}\in\Pi$ such that
$(\pi^{-1})^{-1}=\pi$ (in other words,
the elements of $\Pi$ are paired, with the possibility that an element
is paired with itself).
\item Coincidence: If $k$ of the random edges have been determined,
and if we fix any two vertices, $v,w$,
in the graph, then the probability that an
edge of a given label takes $v$ to $w$ is at most $c/(n-ck)$ for some
constant $c$.  We have only briefly mentioned coincidences in this paper, but
our Lemmas~\ref{lm:big_order} and \ref{lm:equiv_classes}, proven in
\cite{friedman_random_graphs}, require a property like this.
\item Expansion with Error:
Consider a $\Pi$-labelled graph, $H$, with vertices a subset of $\intn$,
that can occur as a subgraph of a graph in our model.
The probability that $H$ occurs must 
depend only on the
number of edges, $a_\pi$, of each label, $\pi$ (of course, 
$a_\pi=a_{\pi^{-1}}$).  Furthermore, this probability
times the number subsets of $\intn$ of size $V_H$ is, for every positive 
integer $r$,
$$
\Esymm(H)_n = \left(\sum_{i=0}^{r-1} \frac{p_i(\vec a)}{n^i}\right)
+ \frac{{\rm error}}{n^r},
$$
where $p_i$ are polynomials in $\vec a$ (where $\vec a$ is the collection
of all $a_\pi$) and where 
$$
|{\rm error}| \le c k^{r'}
$$
for all $k\le n/c$, where $c_1,r'$ depend only on $r$.
Furthermore, $p_i=0$
if $i$ is less than the order of $H$.
\item Simple Word Sum:
Let $\ird{k,\sigma,\tau}$ be those words that begin with $\sigma$, end
in $\tau$, and are {\em irreducible} (meaning no consecutive occurrence of
$\pi$ and $\pi^{-1}$).  Then for any polynomial, $p=p(\vec a)$ (with
$\vec a$ as above), we require
\begin{equation}\label{eq:simple_word}
\sum_{w\in\ird{k,\sigma,\tau}} p\bigl( a_1(w),\ldots,a_{d/2}(w),k\bigr)
= (d-1)^k Q_1(k)+ E(k)
\end{equation}
for a polynomial, $Q_1$, and a function $E$ with $|E(k)|\le ck^c$ for
some constant $c$ (i.e., the above sum is super-{\dtreelike}).
\item $\taufund$ determination:
We must determine $\taufund$ for the model.
\item Spreading: 
There is a constant $\gamma>0$ such that the probability that a
random graph has $|\lambda_i|\ge d-\gamma$ for some $i\ne 1$ is of order
at most $n^{-\taufund}$.
\end{enumerate}

We have already shown spreading and determined $\taufund$ for all
three models.  The labelling of the models is quite simple:
$\chnd$ is labelled like $\cgnd$; $\cind$ is labelled with
$\Sigma=\{\sigma_1,\ldots,\sigma_d\}$ where $\sigma_i^{-1}=\sigma_i$
(each $\sigma_i$ represents a perfect matching);
$\cjnd$ is labelled with $\Sigma\cap T$ with $\Sigma$ as before and
$T=\{\tau_1,\ldots,\tau_d\}$ with $\tau_i^{-1}=\tau_i$, and where
the $\sigma_i$ represent the near perfect matching and the $\tau_i$
represents the single completing half-loop for $\sigma_i$.

Coincidence is easily checked for all three models.

We address the issue of Simple Word Sum.  The word sum for $\chnd$
is the same as for $\cgnd$.  For $\cind$, the technique of Lemma~2.11
of \cite{friedman_random_graphs} reduces the matter to the irreducible
eigenvalues
of a vertex with $d$ half-loops; since these eigenvalues are the
eigenvalues of a $d\times d$ matrix which is $0$ on the diagonal and
$1$'s elsewhere, the eigenvalues are $d-1$ with multiplicity $1$ and
$-1$ with multiplicity $d-1$.  Hence the simple word sum of
equation~(\ref{eq:simple_word}) is given by 
\begin{equation}\label{eq:matching_walk}
(d-1)^k Q_1(k)+(-1)^k Q_2(k)
\end{equation}
where $Q_i$ are polynomials.
For $\cjnd$ we can break the sum by how many half-loops are involved.
For a fixed set of half-loops involved in the irreducible word,
the sum is a convolution of functions of the form in
equation~(\ref{eq:matching_walk}), which by Theorem~\ref{th:baby_convolute}
is again super-{\dtreelike}.

We now establish Expansion with Error for the three models.
Equation~(\ref{eq:probability}) has the $\chnd$ analogue
$$
P(w;\vec t\>) = \prod_{i=1}^{d/2} \frac{(n-a_i-1)!}{(n-1)!}.
$$
Now recall the proof of Theorem~\ref{th:exp_polys}, especially
equations~(\ref{eq:g_rational}) and (\ref{eq:error_term}).
For $\chnd$, we have
\begin{equation}\label{eq:chnd_form}
\Esymm(H)_n= n(n-1)\cdots(n-v+1)
\prod_{i=1}^{d/2} \frac{(n-a_i-1)!}{(n-1)!}.
\end{equation}
$$
= n^{v-e} g(1/n),
$$
with $g$ as in equation~(\ref{eq:g_rational})
with $b_1,\ldots,b_v$ being $0,1,\ldots,v-1$
and $c_1,\ldots,c_e$ being the collection of the sequences 
$1,2,\ldots,a_i$.  Hence for a walk of length at most $k$ we have
$$
\sum b_j+\sum c_j \le \binom{k}{2}+\binom{k+1}{2}=k^2.
$$
Accordingly Expansion with Error holds for $\chnd$ with expansion polynomials
determined by equation~(\ref{eq:chnd_form}), and with
error term bounded by
$$
e^{rk/(n-k)}k^{2r};
$$
this bound is $\le ck^{r'}$ for all $k\le n$ with $r'=2r$ and $c=e^r$.

Similarly for $\cind$ we have the analogue
$$
P(w;\vec t\>) = \prod_{i=1}^{d} \frac{(n-a_i)\oddf}{n\oddf}.
$$
The analysis goes through essentially as before; in the error bound
we have $\sum b_j$ is again $\binom{k}{2}$, but this time the
$\sum c_j$ is as large as
$$
1+3+5+\cdots+(2k-1) = k^2
$$
(taking one $a_i=2k$ and the rest $0$).
So Expansion with Error holds for $\cind$ with
error term bounded by
$$
e^{rk/n}\left(k^2+\binom{k}{2}\right)^{r}\le
e^{rk/n}(2k)^{2r}.
$$

For $\cjnd$, consider a random $1$-regular graph, $G'$, consisting of a near
perfect matching plus one complementing half-loop on the vertex set
$\intn$.  Notice that the
number of such graphs is $n(n-1)\oddf$.  Hence the probability of
occurrence of a specified half-loop and $a$ other matchings in $G'$ is
$$
\frac{(n-1-2a)\oddf}{n(n-1)\oddf} = \frac{1}{n(n-2)\cdots(n-2a)},
$$
and the probability of $a$ specified matchings (with no specified half-loop)
is
$$
\frac{(n-2a)(n-1-2a)\oddf}{n(n-1)\oddf} = \frac{1}{n(n-2)\cdots(n-2a-2)}.
$$
So for any specification of half-loops in $H$, i.e., any fixing of each
$a_{\tau_i}$ to $0$ or $1$,
$\Esymm(H)_n$ is a polynomial in the $a_{\sigma_i}$'s; this makes 
$\Esymm(H)_n$ a polynomial
in the $\vec a$, namely
$$
\sum_{I\subset\{1,\ldots,d\}} \biggl(p_I(a_{\sigma_1},\ldots,a_{\sigma_d})
\prod_{i\in I} a_{\tau_i}
\prod_{i\notin I} (1-a_{\tau_i})\biggr).
$$
We also see that, in the terminology above, $\sum b_j = \binom{k}{2}$ 
and $\sum c_j \le 2\binom{k-1}{2}$.  Hence Expansion with Error holds
for $\cjnd$ as well.

This establishes the six required results mentioned at the beginning of
this section for the models $\chnd$, $\cind$, and $\cjnd$.
Theorems~\ref{th:mainh} and \ref{th:maini} follow.

\section{Closing Remarks}

We make a number of final remarks.

\paragraph{Stronger conjectures:}
As mentioned before, numerical experiments indicate that the average
(and median) $\lambda_2$ for a random graph is $2\sqrt{d-1}+\epsilon(n)$,
where $\epsilon(n)$ is a negative function (tending to $0$ as $n\to\infty$).
By the results of Friedman and Kahale (extending the Alon-Boppana result),
$-\epsilon(n)\le O(\log^{-2}n)$ (see \cite{friedman_geometric_aspects}).
However, the trace method, even with selective traces, seems to require
some fundamental new idea in order to have any hope of achieving
$\epsilon(n)$ that is zero or negative.

\paragraph{Critical $d$:}
As mentioned before, when there is a critical tangle of order
strictly less than that of any hypercritical tangle, then our techniques
leave a gap in that we can only prove $\lambda_2>2\sqrt{d-1}$ with
probability at least $c/n^s$ where $s>\taufund$.  This case is extremely
interesting, since it seems that there should be a theorem that closes
this gap, and such a theorem would either get around a poorly bounded
$W_{\widetilde T,S}(M_1,M_2)$ or improve the very interesting
Theorem~\ref{th:remarkable} (or do something else).

\paragraph{Relative Alon Conjecture:}
Following \cite{friedman_relative}, it seems quite possible to
relativize the main theorems in this paper.  Namely, fix a 
``base'' graph, $B$, (or, more generally, a ``base''
pregraph, in the sense of 
\cite{friedman_geometric_aspects}).  Fix an $\epsilon>0$.  Then we
believe that most
random coverings of $B$ of degree $n$ have all ``new'' eigenvalue
$\le \epsilon+\rho$, where $\rho$ is the spectral radius of the
universal cover of $B$.  Similarly, we can ask for $\epsilon$
to be zero or even a negative function of $n$.  See
\cite{friedman_relative,friedman_tillich_generalized} for further
discussion and a result in this direction.

\paragraph{Alternate Proof with Trace (see the end of Section~2):}
It may be possible to analyze the expected irreducible trace over all
of $\cgnd$.  As remarked in Theorem~\ref{th:nottreelike} and the
discussion thereafter, the coefficients of $g_i(k)$ there could no
longer be {\dtreelike}.
It may be possible to analyze selective traces without discarding 
contributions from tangled graphs.  In other words, if we better 
understood how selectivity affected irreducible traces, we might
not need Section~9 (and certain parts of our understanding of
these traces might improve).  Clearly selectivity in $G$ can be
expressed in terms of walks in an induced subgraph of a ``higher block
presentation'' of $G$ (see \cite{marcus,kitchens}).
However, it is not clear what can be said about the eigenvalues
of induced subgraphs of a higher block presentation; the author has
only some weak results in this directions
(see \cite{friedman_SVD}).

\section*{Glossary}
{\em This glossary contains a term or a piece of notation, followed
by a colon (:), followed by a brief description, followed by the page
number(s) where the term/notation is explained.}
\medskip
\renewenvironment{theindex}{\begin{description}}{\end{description}}
\input all.ind 


\end{document}